RUHR-UNIVERSITÄT BOCHUM

RUB

# Untersuchung von Effekten maßgeschneiderter Elektroden auf kapazitiv gekoppelte Radio Frequenz Mikroplasmajets mittels Emissionsspektroskopie und Laserdiagnostik

## Lena Bischoff

**Masterarbeit**

Bochum, 12.03.2021

Lehrstuhl für Allgemeine Elektrotechnik und Plasmatechnik

AEPT

# Untersuchung von Effekten maßgeschneiderter Elektroden auf kapazitiv gekoppelte Radio Frequenz Mikroplasmajets mittels Emissionsspektroskopie und Laserdiagnostik

Lena Bischoff

Bochum, 12.03.2021

# Eidesstattliche Erklärung

Ich erkläre, dass ich keine Arbeit in gleicher oder ähnlicher Fassung bereits für eine andere Prüfung an der Ruhr-Universität Bochum oder einer anderen Hochschule eingereicht habe.

Ich versichere, dass ich diese Arbeit selbstständig verfasst und keine anderen als die angegebenen Quellen benutzt habe. Die Stellen, die anderen Quellen dem Wortlaut oder dem Sinn nach entnommen sind, habe ich unter Angabe der Quellen kenntlich gemacht. Dies gilt sinngemäß auch für verwendete Zeichnungen, Skizzen, bildliche Darstellungen und dergleichen.

Ich versichere auch, dass die von mir eingereichte schriftliche Version mit der digitalen Version übereinstimmt. Ich erkläre mich damit einverstanden, dass die digitale Version dieser Arbeit zwecks Plagiatsprüfung verwendet wird.

12.03.2024
Datum/Date

*Luna Bischoff*
Unterschrift/Signature

# Inhaltsverzeichnis

# Abbildungsverzeichnis

# 1 Einleitung

Mit der sogenannten Plasmamedizin erhalten Mediziner eine neue Perspektive in Bereichen wie der therapeutischen Wundheilung. Bei dem interdisziplinären Wissenschaftsgebiet an der Schnittstelle zwischen Physik, Medizin und Biologie werden nichtthermische Atmosphärendruckplasmen mit Hilfe spezieller Plasmageräte erzeugt, um medizinische Effekte direkt am Patienten zu erzielen [1].

Klinische Studien belegen, dass nichtthermische Atmosphärendruckplasmen antibakteriell wirken und die natürliche Wundheilung durch die Stimulation der Gewebeneubildung unterstützen [2]. Die nach dem aktuellen Forschungsstand ermittelten wirksamen Komponenten eines Plasmas sind dabei unter anderem reaktive Radikale wie zum Beispiel reaktive Sauerstoff- und Stickstoffspezies (RONS) [3]. Diese Eigenschaften bieten einen innovativen Therapieansatz wie etwa bei der Inaktivierung multiresistenter Bakterien oder der Behandlung von chronischen Wunden [2].

Eine weitere Anwendungsmöglichkeit von nichtthermischen Atmosphärendruckplasmen ist die schonende Sterilisation und Desinfektion von medizinischen Produkten wie beispielsweise chirurgischen Instrumenten oder die zusätzliche Erhöhung der Biokompatibilität von Implantaten in der Orthopädie [2]. Zudem werden erste Überlegungen hinsichtlich einer möglichen plasmabasierten Desinfektion von mit dem Coronavirus kontaminierten Oberflächen angeregt, um mit der Plasmamedizin auch während der weltweiten Pandemie einen wichtigen Forschungsbeitrag zu leisten [4]. Ein weiteres junges und aussichtsreiches Forschungsgebiet ist die Wirkung von Plasma auf Krebszellen [5].

Die Steuerung und Optimierung der physikalischen und chemischen Eigenschaften von nichtthermischen Atmosphärendruckplasmen durch externe Kontrollparameter ermöglichen es, Plasmageräte mit einer speziellen Plasmachemie zu entwerfen, die beispielsweise auf eine bestmögliche Wundheilung beim Menschen abgestimmt sind. Zur Optimierung müssen demnach im Plasma gezielt reaktive Radikale durch Ionisation und Dissoziation des Neutralgases erzeugt werden. Diese Stoßprozesse erfordern energetische Elektronen, die durch Heizungsmechanismen im Plasma erzeugt werden und abhängig von der Energieverteilungsfunktion der Elektronen sind [6].

Diese Arbeit wird innerhalb eines Forschungsprojektes des Sonderforschungsbereiches 1316 der Ruhr-Universität angefertigt. Das Teilprojekt A4 beschäftigt sich mit der Prozesskontrolle eines kapazitiv gekoppelten Radio Frequenz Mikroplasmajets (COST-Jet), welcher zu den nichtthermischen Atmosphärendruckplasmen gehört und als eine Referenzquelle in der Plasmamedizin dient. Standardmäßig handelt sich hierbei um ein kapazitiv gekoppeltes Plasma zwischen zwei aus Edelstahl bestehenden planparallelen Elektroden im Abstand von 1 mm.

Mittels maßgeschneiderter Spannungsformen (Voltage Waveform Tailoring) und ausgewählter Elektrodenmaterialien und -strukturen soll die Elektronenheizungsdynamik und die Energieverteilungsfunktion der Elektronen kontrolliert werden.

Ziel dieser Arbeit ist die Untersuchung von Effekten unterschiedlicher maßgeschneiderter Elektroden auf die Elektronenheizungsdynamik des COST-Jets. Basis dieser Überlegungen sind frühere Forschungsergebnisse aus dem Bereich der kapazitiv gekoppelten Niederdruckplasmen (CCP), die sich mit diesem Effekt beschäftigen. Da sich jedoch die Elektronenheizungsdynamik der Radio Frequenz Mikroplasmajets von denen der kapazitiv gekoppelten Niederdruckplasmen fundamental unterscheidet, bedarf es hier ebenfalls einer ausführlichen Untersuchung. Der COST-Jet wird in dieser Arbeit mit einer angelegten sinusförmigen Wechselspannung von 13,56 MHz betrieben. Mit Hilfe der phasenaufgelösten optischen Emissionsspektroskopie (PROES) wird die Elektronenheizungsdynamik in der Plasmaentladung für verschiedene maßgeschneiderte Elektrodenanordnungen, Gasmischungen und angelegte Spannungsamplituden untersucht. Die absoluten Dichten von atomarem Sauerstoff werden anschließend mittels der Zweiphotonen laserinduzierten Fluoreszenzspektroskopie (TALIF) untersucht, um Informationen über die Produktion reaktiver Teilchen in Abhängigkeit der verschiedenen Elektrodenmaterialien und -strukturen zu erhalten.

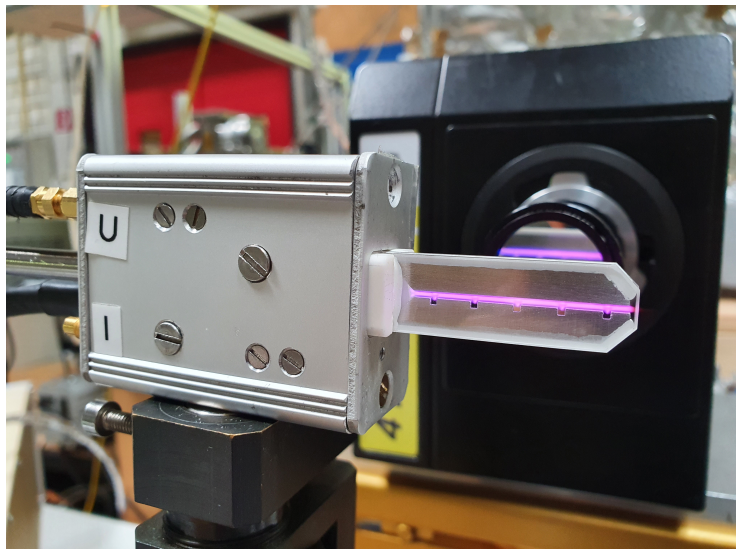

Abbildung 1.1: Foto des COST-Jets mit einer rechteckig strukturierten Elektrode während einer PROES-Messung.

# 2 Grundlagen

## 2.1 Atmosphärendruckplasmen

### 2.1.1 Charakterisierung

Eine wesentliche und charakterisierende Größe zur Unterscheidung von Plasmen ist der Druck. Dieser hat einen großen Einfluss auf ein Plasma und bestimmt die Elektronenheizungsdynamik im Plasmaprozess. Man unterscheidet dabei grundsätzlich zwischen Niederdruckplasmen, Hochdruckplasmen und Atmosphärendruckplasmen [7].

Niederdruckplasmen (< $10^3$ Pa) werden in der Industrie häufig für Beschichtungsprozesse und Ätzprozesse eingesetzt [7]. Durch Ätzprozesse in der Mikroelektronikherstellung werden beispielsweise filigrane Siliziumstrukturen in mikroelektromechanischen Systemen (MEMS) realisiert, welche sich in Sensoren von Fahrzeugsystemen und Smartphones befinden [8]. Dünnschichtsolarzellen oder Kratzschutzschichten von Displays werden durch plasmagestützte Beschichtungsprozesse hergestellt. Neben den technisch realisierten Niederdruckplasmen gibt es auch natürliche Niederdruckplasmen wie zum Beispiel interstellare Gasnebel im Weltall. Bei Hochdruckplasmen (> $10^5$ Pa) ist der Druck deutlich höher als in der umgebenden Atmosphäre. Diese Art von Plasma findet man häufig in Gasentladungslampen [9].

Atmosphärendruckplasmen werden hingegen bei einem Druck gezündet, welcher dem der umgebenden Atmosphäre entspricht. Im Gegensatz zu einem Niederdruckplasma benötigt man somit keine Vakuumkammer, die zur Aufrechterhaltung des unterschiedlichen Druckniveaus zum Atmosphärendruck dient. Dies ermöglicht eine kostengünstige Herstellung von Atmosphärendruckplasmen. Außerdem gilt nach dem vereinfachten Paschen-Gesetz, dass bei einer niedrigen Zündspannung der Elektrodenabstand eines Atmosphärendruckplasmas wenige Millimeter beträgt. Somit kann die Plasmaentladung auf relativ geringen räumlichen Dimensionen realisiert werden, was somit Vorteile bei der praktischen Handhabung bietet [2].

Atmosphärendruckplasmen lassen sich in thermische und nichtthermische Plasmen klassifizieren. Bei den thermischen Atmosphärendruckplasmen befinden sich die Elektronen, Ionen und Neutralteilchen in einem thermischen Gleichgewicht ($T_\mathrm{e} \approx T_\mathrm{i} \approx T_\mathrm{g}$). Dabei bezeichnet $T_\mathrm{e}$ die Elektronentemperatur, $T_\mathrm{i}$ die Ionentemperatur und $T_\mathrm{g}$ die Schwerteilchentemperatur. Aufgrund der hohen Temperatur der Teilchen finden thermische Atmosphärendruckplasmen beispielsweise Anwendung beim Lichtbogenschweißen. Bei nichtthermischen Atmosphärendruckplasmen, die mit einer Wechselspannung, die typischerweise im MHz-Bereich liegt, betrieben werden, unterscheiden sich die Temperaturen der Teilchen ($T_\mathrm{e} \gg T_\mathrm{i} \geq T_\mathrm{g}$) [2]. Diese

Eigenschaft lässt sich durch die Betrachtung der Plasmafrequenz genauer erklären, welche neben dem Druck ebenfalls eine wesentliche und charakterisierende Größe eines Plasmas ist [7]:

$$\omega_{\mathrm{pe,i}} = \sqrt{\frac{n_{\mathrm{e,i}} \mathrm{e}^2}{\varepsilon_0 \, m_{\mathrm{e,i}}}} \quad . \tag{2.1}$$

Dabei steht der Index e für Elektronen und der Index i für Ionen als Ladungsträger. Außerdem bezeichnet $\varepsilon_0$ die elektrische Feldkonstante, $n$ die Ladungsträgerdichte, e die Elementarladung und m die Masse der jeweiligen Ladungsträger. Die Plasmafrequenz beschreibt die Frequenz, mit der die verschiedenen Ladungsträger nach einer Auslenkung aus ihrer Gleichgewichtsposition im Plasma oszillieren. Elektronen und Ionen können hierbei nur auf Potentialschwankungen auf oder unterhalb der Zeitskala der jeweiligen Plasmafrequenz reagieren. Da die Masse $m_i$ der Ionen im Vergleich zu der Masse $m_e$ der Elektronen deutlich größer ist, ist die Elektronenplasmafrequenz viel größer als die Ionenplasmafrequenz [10]. Das in dieser Arbeit verwendete Atmosphärendruckplasma hat eine typische Elektronendichte von circa $n_e \approx 10^{16} \, cm^{-3}$, womit sich eine Plasmafrequenz der Elektronen von $\omega_{\mathrm{pe}} \approx 6\,\mathrm{GHz}$ und für die Ionen von $\omega_{\mathrm{pi}} \approx 70\,\mathrm{MHz}$ ergibt [11]. Somit können die Elektronen der Potentialschwankung folgen und geheizt werden, wohingegen die Ionen nur sehr träge folgen können. Da die Elektronenmasse deutlich kleiner als die Ionenmasse ist, findet kein thermischer Energietransfer statt und die Ionentemperatur liegt bei Raumtemperatur. Das Neutralgas bleibt ebenfalls kalt, da ausschließlich die Ionen durch ihr Masseverhältnis ihre thermische Energie an das Neutralgas übertragen können. Es handelt sich dementsprechend um ein Plasma im nichtthermischen Gleichgewicht. Aus dieser Eigenschaft ergibt sich die biomedizinische Anwendung von nichtthermischen Atmosphärendruckplasmen in der Plasmamedizin und somit die schonende Behandlung von temperatursensitiven Oberflächen bei zeitgleicher kalter Wirktemperatur [2, 12].

### 2.1.2 Zündungsmechanismen

Der Zündungsmechanismus bei einem Atmosphärendruckplasma basiert auf dem Townsend-Effekt [13]. Zur Veranschaulichung wird ein Aufbau mit zwei Elektroden betrachtet, an denen eine Spannung anliegt und zwischen denen sich ein Gas befindet. Schon vor Beginn des Plasmas liegen freie Elektronen vor, die zum Beispiel durch die kosmische Höhenstrahlung erzeugt werden. Diese Elektronen bezeichnet man als sogenannte Primärelektronen. Wird ein äußeres elektrisches Feld angelegt, werden die Primärelektronen durch das angelegte elektrische Feld zur Anode hin beschleunigt. Hat ein Primärelektron die benötigte Ionisationsenergie aufgenommen, kann es die Atome und Moleküle auf seinem Weg zur Anode ionisieren, dissoziieren oder anregen. Es entstehen weitere freie Elektronen, die wiederum durch das elektrische Feld beschleunigt werden und sich dann an Stoßionisationsprozessen beteiligen [13]. Die erzeugten Ionen werden durch das elektrische Feld in Richtung der Kathode beschleunigt. Treffen die Ionen auf die Kathode, können diese zusätzlich Sekundärelektronen herauslösen. Der Townsend-Effekt führt schließlich zur Ausbildung einer Elektronenlawine im Umfeld der Kathode, die dann im elektrischen Feld beschleunigt wird [13].

Mit verschiedenen technischen Möglichkeiten wird bei einem nichtthermischen Atmosphärendruckplasma der sogenannte Streamer-zu-Bogen-Übergang unterbunden. Beim Streamermechanismus unterscheidet man zwischen einem positiven und negativen Streamer, die sich in der Richtung der Fortbewegung unterscheiden. Anschaulich besteht ein Streamer aus einem Streamerkopf, der von einer Elektronenlawine nach Townsend gebildet wird. Dahinter folgt ein quasineutraler Plasmakanal, in dem ein ausgeglichenes Verhältnis an positiven und negativ geladenen Teilchen herrscht [14].

Bei der Ausbildung eines negativen Streamers wird die Elektronenlawine durch das äußere elektrische Feld von der Kathode zur Anode beschleunigt und durch weitere Elektronen verstärkt. Die hohe Ladungsdichte innerhalb der Elektronenlawine führt ebenso zu einem lokal erhöhten elektrischen Feld, das ausreicht, um weitere Teilchen zu ionisieren [14]. Die Trägheit der so erzeugten Ionen führt dazu, dass sich die positive Ladung am Ort der Ionisation weniger bewegt und die leichteren Elektronen vom elektrischen Feld fortbewegt werden. So entsteht hinter dem Streamerkopf ein quasineutraler Plasmakanal. Erreicht der Streamerkopf die Anode, entsteht ein leitfähiger Kanal und es kommt zu einem Kurzschluss. Zusätzlich fließen Ladungsträger aus der Kathode über den leitfähigen Kanal an der Anode ab und es kommt zu einem Zusammenbruch des elektrischen Feldes im Spalt. Die Spannung sinkt herab und insgesamt bildet sich ein sehr hoher Strom aus [15]. In diesem Fall bildet sich eine Bogenentladung aus, bei der sich die Temperaturen der im Plasma enthaltenen Teilchen angleichen. Somit würde es sich um ein thermisches Plasma handeln, welches für eine biomedizinische Anwendung nicht mehr geeignet wäre [16].

Der Streamer-zu-Bogen-Übergang kann beispielsweise durch die Aufladung eines Dielektrikums zwischen den Elektroden verhindert werden. Dieses Konzept kommt bei der sogenannten dielektrisch behinderten Entladung zum Einsatz. Zudem lässt sich die Entstehung von Streamern durch das Anlegen hochfrequenter oder gepulster Spannungen verhindern, was bei kapazitiv gekoppelten Mikro-Atmosphärendruckplasmajets eingesetzt wird. Dabei wird die Spannung so schnell variiert, dass der Stromfluss behindert wird. Erfahrungsgemäß kommt es bei der Verwendung von Edelgasen, insbesondere Helium, ebenfalls weniger zu einer Streamerbildung, da diese eine thermische Instabilität unterdrücken [16].

## 2.2 Betriebsmodi von Mikro-Atmosphärendruckplasmajets

Mikro-Atmosphärendruckplasmajets können in drei verschiedenen Modi betrieben werden, welche jeweils einen Einfluss auf die Elektronenheizung und der damit zusammenhängenden Elektronenenergieverteilungsfunktion (engl. Electron energy distribution function, EEDF) haben. Dabei unterscheidet man zwischen dem Ω-Modus, dem Penning-Modus und dem kontrahierten Modus. Die Abbildungen 2.1 a) - c) zeigen jeweils die Plasmaentladung eines mit Helium betriebenen COST-Jets in den verschiedenen Betriebsmodi. Die Betriebsmodi sind abhängig von externen Kontrollparametern wie zum Beispiel der Eingangsleistung oder der reaktiven Gasbeimischung [6].

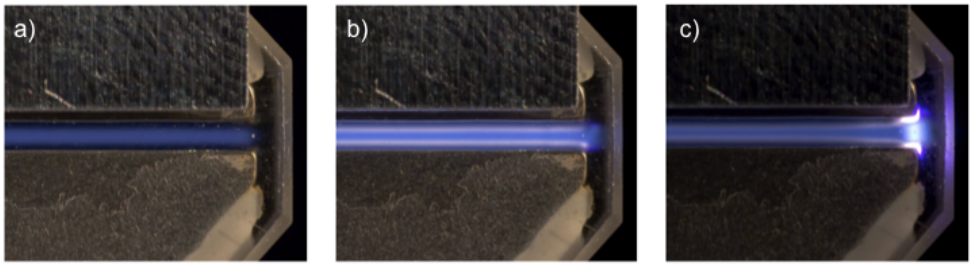

Abbildung 2.1: Foto der Plasmaentladung des mit Helium betriebenen COST-Jets im a) Ω-Modus, b) Penning-Modus und c) kontrahierten Modus [17].

**Ω−Modus**

Der Ω-Modus zeichnet sich durch eine homogene Intensitätsverteilung aus und kann bei einer geringen Ausgangsleistung beobachtet werden. Abbildung 2.1 a) zeigt die Plasmaentladung im Ω-Modus eines mit Helium betriebenen COST-Jets. Hierbei ähnelt die Leuchtemission des Ω-Modus der Leuchtemission des $\alpha$−Modus im Niederdruck, jedoch ergeben sich aufgrund des höheren Drucks fundamentale Unterschiede in der Heizungsdynamik [6]. Der dominierende Heizungsmechanismus im Ω-Modus basiert auf einem hohen Driftfeld innerhalb des Plasmabulks und im Gegensatz zum $\alpha$−Modus somit nicht auf der stochastischen Heizung von Elektronen. Bei der stochastischen Heizung werden die Elektronen stoßfrei in Richtung des Plasmabulks durch die Wechselwirkung mit der expandierenden Randschicht beschleunigt oder durch die Wechselwirkung mit der kollabierten Randschicht gebremst [18]. Der Ω-Modus eines Atmosphärendruckplasmas basiert hingegen auf der Beschleunigung von Elektronen im Plasmabulk durch ein hohes Driftfeld [6]. Diese hochenergetischen Elektronen (über 22,7 eV) können dann zu einer Ionisierung der Neutralteilchen innerhalb des Plasmabulks führen. Die Ursache für diesen Heizungsmechanismus ist der deutlich höhere Druck im Vergleich zu einem Plasma im Niederdruckbereich. Die somit höhere Stoßfrequenz führt insgesamt zu einer geringeren DC-Leitfähigkeit $\sigma_{dc}$, die wie folgt definiert ist:

$$\sigma_{dc} = \frac{n_e e^2}{m_e \nu_m}. \tag{2.2}$$

Dabei bezeichnet $n_e$ die Elektronendichte, e die Elementarladung, $m_e$ die Elektronenmasse und $\nu_m$ die Stoßfrequenz. Der schematische Verlauf der DC-Leitfähigkeit eines Atmosphärendruckplasmas in Abhängigkeit des Ortes ist in Abbildung 2.2 a) vereinfacht dargestellt und sinkt im Randschichtbereich aufgrund der dort sinkenden Elektronendichte. Außerdem ist nach dem Ampère'schen Gesetz der Gesamtstrom innerhalb einer kapazitiven Hochfrequenzentladung ortsunabhängig [19]. Der Gesamtstrom $\vec{j}_{tot}$ setzt sich allgemein aus einem Leitungsstrom $\vec{j}_c$ und einem Verschiebungsstrom $\vec{j}_d$ zusammen, wohingegen im weiteren Verlauf der Verschiebungsstrom vernachlässigt werden kann ($\vec{j}_d = 0$), da dieser im Plasmabulk im Vergleich zum Randschichtbereich deutlich geringer ist. Demnach ergibt sich für den Gesamtstrom:

$$\vec{j}_{tot} = \vec{j}_c = \sigma_{dc}\vec{E} = const. \tag{2.3}$$

Damit der Gesamtstrom in einem Atmosphärendruckplasma ortsunabhängig ist, muss sich beispielsweise bei einer sinkenden DC-Leitfähigkeit $\sigma_{dc}$ im Randschichtbereich ein hohes Driftfeld $\vec{E}$ einstellen, wie in Abbildung 2.2 b) zu erkennen ist [6]. Somit beobachtet man zum Beispiel ein lokales Maximum der Ionisation Im Randschichtbereich, obwohl die Randschicht zu diesem Zeitpunkt kollabiert ist.

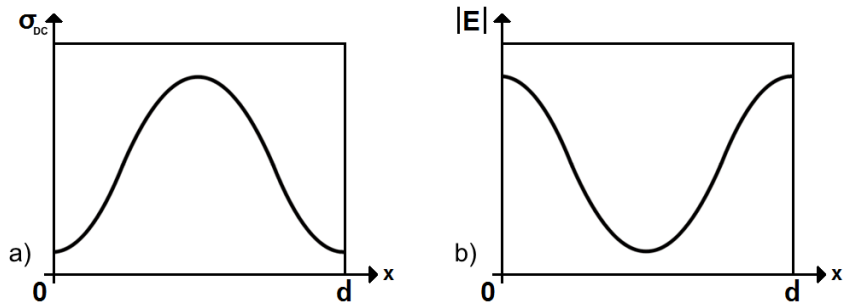

Abbildung 2.2: Schematische Darstellung der DC-Leitfähigkeit a) und des Driftfeldes b) in einem Atmosphärendruckplasma [20].

Abbildung 2.3 a) zeigt einen raum- und phasenaufgelösten Plot der Emission bei 706,5 nm eines mit Helium und Sauerstoff betriebenen COST-Jets im Ω-Modus mit einer angelegten Peak-to-Peak-Spannung von 500 $V_{pp}$. Im gesamten Zentrum der Entladung zeigt sich eine ausgeprägte Emission. Die Emissionslinie wurde bei den PROES-Messungen gewählt, da diese eine Energieschwelle für die Elektronenstoßanregung aus dem Grundzustand bei 22,7 eV besitzt und sehr sensitiv für hochenergetische Elektronen ist [6].

**Penning-Modus**
Wird beispielsweise die Eingangsleistung als externer Kontrollparameter ausreichend erhöht und die reaktive Gasbeimischung konstant gehalten, wechselt die Plasmaentladung von dem Ω-Modus in den Penning-Modus [6]. Der Penning-Modus stellt das Analogon zum γ-Modus im Niederdruck dar, da die Ionisationsmaxima ebenfalls an den Randschichtkanten zum Zeitpunkt der maximalen Randschichtspannung während einer RF-Periode beobachtet werden können. Der Übergang von dem Ω-Modus in den Penning-Modus ist in der PROES-Messung in Abbildung 2.3 b) für eine angelegte Peak-to-Peak-Spannung von 700 $V_{pp}$ zu erkennen. Hierbei ist eine schwache Emission im Plasmabulk durch den Ω-Modus zu erkennen und zusätzlich sind Ionisationsmaxima an den Randschichkanten durch den Penning-Modus zu erkennen. In Abbildung 2.3 c) ist die PROES-Messung bei 850 $V_{pp}$ gezeigt, bei der sich nun die Plasmaentladung im reinen Penning-Modus befindet und nur noch die Ionisationsmaxima an den Randschichkanten zu erkennen sind.

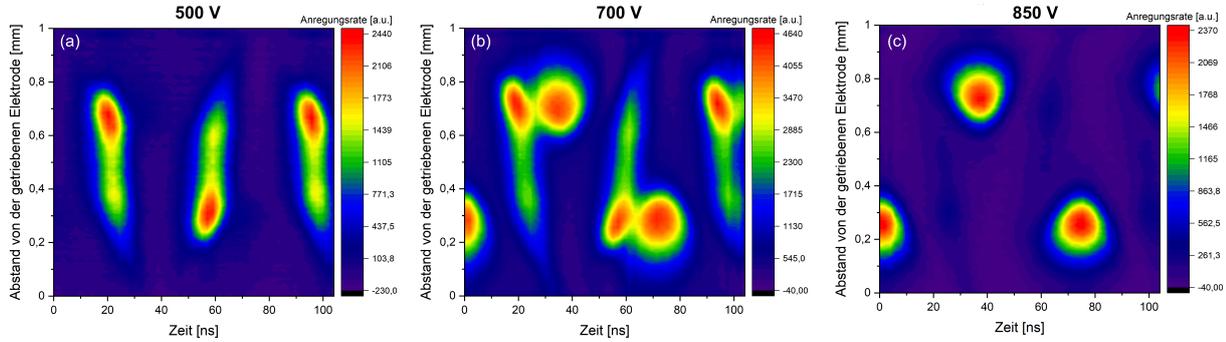

Abbildung 2.3: PROES-Messungen der Emissionslinie bei 706,5 nm des mit 1 slm He und 0,5 sccm $O_2$ betriebenen COST-Jets in Abhängigkeit der angelegten Peak-to-Peak-Spannung: a) 500 $V_{pp}$ (Ω–Modus), b) 700 $V_{pp}$ (Übergang vom Ω-Modus in den Penning-Modus) und c) 850 $V_{pp}$ (reiner Penning-Modus).

Der Heizungsmechanismus im Penning-Modus beruht jedoch im Vergleich zum $\gamma$-Modus nicht ausschließlich auf der ioneninduzierten Sekundärelektronenemission und der anschließenden Beschleunigung der Sekundärelektronen in Richtung des Plasmabulks durch das elektrische Feld in der Randschicht, sondern auf der Überlagerung der Penning-Ionisation (direkter Kanal) und der ioneninduzierten Sekundärelektronenemission (indirekter Kanal) [6, 21]. An der Penning-Ionisation sind metastabile Teilchen beteiligt, die durch eine Elektronenstoßanregung erzeugt werden können. Während der Randschichtexpansion erhalten die an der Elektronenstoßanregung beteiligten Elektronen durch das elektrische Feld die benötigte Energie im Bereich der Anregungsschwelle. Ein Atom oder Molekül kann dann anschließend mit einem metastabilen Teilchen zusammenstoßen. Bei diesem Stoß wird die innere Energie des Metastabilen auf den Stoßpartner übertragen, der dann durch eine Ionisation ein Elektron freisetzt [22]. Dieser Stoßprozess kann anhand des folgenden Beispiels gezeigt werden:

$$\text{He}^* + \text{N}_2 \longrightarrow \text{N}_2^+ + \text{He} + \text{e}^-. \tag{2.4}$$

Die erzeugten Metastabilen können dann durch Stöße in der Randschicht Atome oder Moleküle ionisieren. Ist die Randschicht expandiert, werden die durch die Penning-Ionisation erzeugten Elektronen durch das elektrische Feld in Richtung Plasmabulk beschleunigt. Der sogenannte indirekte Kanal, der auf der ioneninduzierten Sekundärelektronenemission basiert, ist ein weiterer Mechanismus im Penning-Modus. Hierbei werden die durch die Penning-Ionisation erzeugten Ionen, wie beispielsweise die in Gleichung 2.4 durch die Penning-Ionisation erzeugten $N_2^+$ Ionen, von dem elektrischen Feld in der Randschicht in Richtung der naheliegenden Elektrode beschleunigt. Durch den Ionenbeschuss auf die Elektrode können dann Sekundärelektronen herausgelöst werden. Die herausgelösten Sekundärelektronen werden dann von dem elektrischen Feld in der Randschicht in Richtung Plasmabulk beschleunigt. Somit treten im Penning-Modus Maxima der Elektronenstoßanregung an der Randschichtkante zu den Zeitpunkten der maximalen Randschichtausdehnung auf [6]. Im Penning-Modus kann wie im $\gamma$-Modus eine inhomogene Leuchtemission beobachtet werden, wie Abbildung 2.1 b) zeigt. Es ist eine hohe Leuchtintensität in den Randschichtbereichen und eine vergleichsweise schwächere Leuchtemission im Plasmabulk zu beobachten.

Um die Beiträge der jeweiligen Kanäle des Penning-Modus zu identifizieren, variieren Bischoff et al. in Abbildung 2.4 a) - d) die Oberflächenkoeffizienten in PIC/MCC-Simulationen (engl. Particle-in-cell/Monte Carlo collision, PIC/MCC) bei ansonsten konstanten Entladungsbedingungen des COST-Jets. Die Entladungsbedingungen werden so gewählt, dass sich der COST-Jet im Penning-Modus befindet. Für die Untersuchung wird die ioneninduzierte Sekundärelektronenemission (engl. Secondary Electron Emission, SEE) und die Elektronenreflektionswahrscheinlichkeit $\alpha$ an- und ausgeschaltet, um den Einfluss von Oberflächenprozessen auf den Penning-Modus zu untersuchen.

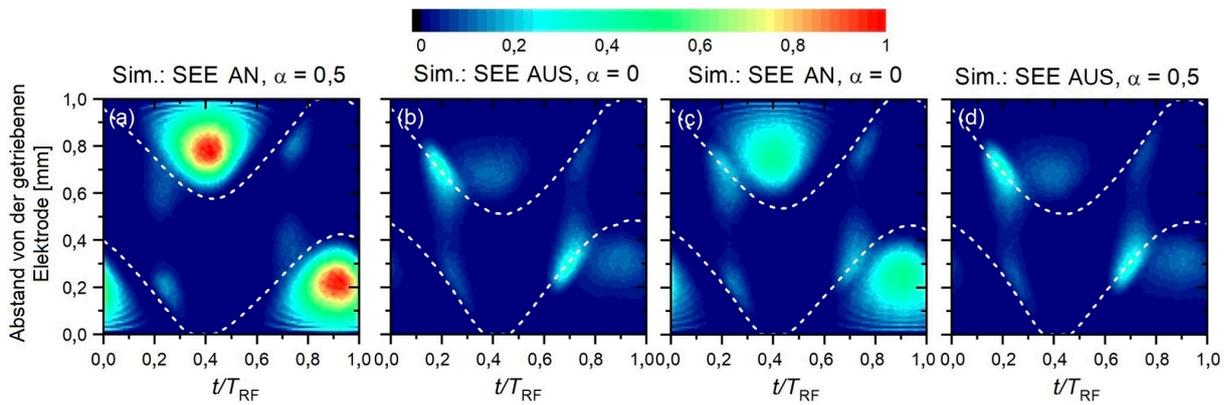

Abbildung 2.4: PIC/MCC-Simulationen der normierten räumlich-zeitlichen Elektronenstoßanregungsrate der 706,5 nm He I-Linie des COST-Jets in Abhängigkeit der ioneninduzierten Sekundärelektronenemission und der Elektronenreflektion $\alpha$. Die angelegte Peak-to-Peak-Spannung beträgt 355 $V_{pp}$ und der verwendete Gasfluss beträgt von 1 slm He und 0,5 sccm $N_2$. Der Randschichtverlauf ist als eine weiße gestrichelte Linie dargestellt [6].

In Abbildung 2.4 a) sind beide Oberflächenkoeffizienten eingeschaltet und die Plasmaentladung befindet sich im Penning-Modus. Beide Anregungsmaxima sind hingegen stark reduziert, wenn die Oberflächenkoeffizienten ausgeschaltet sind, wie in Abbildung 2.4 b) erkennbar ist. Unter diesen Bedingungen ist nur noch eine schwache Anregung zu den Zeiten der maximalen Randschichtausdehnung zu erkennen, die ausschließlich vom direkten Kanal getragen wird. Aus dem Ergebnis wird ebenfalls deutlich, dass der auf den Oberflächenprozessen basierende indirekte Kanal einen wichtigen Beitrag zum Penning-Modus leistet und somit nicht vernachlässigt werden darf [6]. Aus der Abbildung 2.4 c) - d) lässt sich ebenfalls schließen, dass die Elektronenreflektion zusätzlich zur ioneninduzierten Sekundärelektronenemission das Anregungsmaximum verstärkt. Aufgrund der kurzen mittleren freien Weglänge bei Atmosphärendruck und dem relativ geringen reduziertem elektrischen Feld an den Elektro-denoberflächen, können einige der erzeugten Sekundärelektronen durch elastische Stöße mit den Neutralteilchen zurück zur Elektrode beschleunigt werden [23]. Ist der Wert der Elektronenreflektionswahrscheinlichkeit $\alpha \neq 0$, so werden die Elektronen von der Elektrode reflektiert und anschließend wieder in Richtung des Plasmabulks vom elektrischen Feld in der Randschicht beschleunigt. Ist $\alpha = 0$, so werden die Elektronen von der Elektrode absorbiert [6].

Ein Übergang von dem Penning-Modus in den $\Omega$-Modus kann durch die Erhöhung der reaktiven Gasbeimischung bei einer konstanten Eingangsleistung induziert werden. Durch die erhöhte Zugabe eines reaktiven Gases wird die Erzeugung von Heliummetastabilen reduziert, da durch die erhöhte Stoßfrequenz die Elektronen nicht mehr genügend Energie aufnehmen können, die für die Erzeugung von metastabilen Teilchen notwendig ist. Somit werden durch die fehlenden Metastabilen auch weniger Elektronen durch die Penning-Ionisation erzeugt. Außerdem können die in der Randschicht erzeugten Elektronen aufgrund der häufigen inelastischen Stöße nicht mehr genug Energie aufnehmen, um sich durch Stöße weiter zu vervielfältigen [6]. Der Übergang von dem Penning-Modus in den $\Omega$-Modus kann außerdem durch eine erhöhte Zugabe von einem elektronegativen Gas wie zum Beispiel $O_2$ induziert werden. Dadurch befinden sich mehr negative Ionen im Plasma und die Elektronendichte sinkt. Aus der geringeren Elektronendichte resultiert eine geringere DC-Leitfähigkeit und somit stellt sich im Plasmabulk ein hohes Driftfeld ein [24].

**Kontrahierter Modus**
Die Eingangsleistung kann soweit erhöht werden, bis der COST-Jet im sogenannten kontrahierten Modus betrieben wird. Hierbei fällt die Spannung rapide ab und es kommt zu einem starken Anstieg des Entladungsstromes [25]. Dieser instabile Modus ist in Abbildung 2.1 c) durch einen Lichtbogen am Ende des Entladungskanals zu erkennen. Schließlich wird eine hohe Energieeintragung erreicht und die Gastemperatur übersteigt die Raumtemperatur. Dieser Modus eignet sich nicht für einen kontinuierlichen Plasmabetrieb, da es in der Regel schon nach wenigen Sekunden zu einer Beschädigung des COST-Jets kommt [17].

## 2.3 Möglichkeiten zur Kontrolle der Elektronenheizungsdynamik

### 2.3.1 Voltage Waveform Tailoring

Aktuelle Ergebnisse aus dem Forschungsprojekt A4 des Sonderforschungsbereiches SFB 1316 zeigen, dass mit dem sogenannten Voltage Waveform Tailoring (VWT) die Elektronenheizungsdynamik des COST-Jets und die Energieverteilungsfunktion der Elektronen verändert und optimiert werden können. Bei dem Voltage Waveform Tailoring wird eine maßgeschneiderte Spannungsform wie beispielsweise eine „Peaks"-, „Sägezahn"- oder „Täler"-Spannungsform an eine der Elektroden angelegt [26]. Die angelegte maßgeschneiderte Spannungsform wird dabei grundsätzlich als eine endliche Fourierreihe von aufeinanderfolgenden Harmonischen einer Grundfrequenz realisiert. Die Anzahl $N$ der aufeinanderfolgenden Harmonischen einer Grundfrequenz $f$, die Amplituden $\Phi_k$ und die Phasenverschiebungswinkel $\Theta_k$ der Harmonischen lassen sich dabei so einstellen, dass eine individuelle Spannungsform $\Phi(t)$ entsteht [27]:

$$\Phi(t) = \sum_{k=1}^{N} \Phi_k cos(2\pi k f t + \Theta_k) \, . \tag{2.5}$$

Eine „Peaks"-Spannungsform ($N > 1$) wird realisiert, wenn in der Gleichung 2.5 ein Phasenverschiebungswinkel $\Theta_k = 0$ gewählt wird. Die Spannungsamplituden der Harmonischen sind dabei wie folgt definiert: $\Phi_k = \Phi_0(N - k + 1)/N$, mit $\Phi_0 = \Phi_{pp} 2N/(N + 1)^2$. Wird beispielsweise $\Theta_k = 180°$ verwendet, so wird das Signal umgekehrt und es entsteht eine „Täler"-Spannungsform [28].

Die Abbildung 2.5 a) - f) zeigt die normierten raum- und zeitaufgelösten Plots der Elektronstoßanregungsrate von dem Grundzustand in den Zustand He I $^3S_1$ des COST-Jets für eine verschiedene Anzahl an aufeinanderfolgenden $N$ Harmonischen der Grundfrequenz $f$ = 13,56 MHz, die experimentell (erste Reihe) und aus PIC/MCC-Simulationen (zweite Reihe) bestimmt wurden [29]. Die dritte Reihe, g) - i), zeigt die angelegte Spannungsform in Abhängigkeit der Anzahl $N$ der aufeinanderfolgenden Harmonischen der Grundfrequenz. Die vierte Reihe, j) - l), zeigt die jeweilige gemittelte Elektronenenergiewahrscheinlichkeitsfunktion (engl. Electron Energy Probability Function, EEPF). Für den Standardfall $N = 1$, bei dem eine sinusförmige Spannung am COST-Jet anliegt, erkennt man eine typische zeitlich und räumlich symmetrische Elektronstoßanregungsrate. Unter diesen Bedingungen befindet sich der COST-Jet im $\Omega$-Modus, bei dem die zwei Maxima der Anregungsrate während der kollabierten Randschicht auftreten [29].

Wird eine „Peaks"-Spannungsform ($N > 1$) angelegt, so zeigt die Dynamik der Elektronenstoßanregung eine räumliche und zeitliche Asymmetrie, die sich mit zunehmendem $N$ vergrößert. Des Weiteren ist das Anregungsmaximum an der getriebenen Elektrode deutlich ausgeprägter als an der gegenüberliegenden geerdeten Elektrode, wie in Abbildung 2.5 e) zu erkennen ist [29]. Das Anregungsmaximum an der getriebenen Elektrode tritt genau zu dem Zeitpunkt des Peaks auf, bei dem die angelegte Spannung positiv ist. Zu den restlichen Zeiten nimmt die Spannung negative Werte an. Dies wirkt sich auf die Randschichtdynamik aus, da im Zeitraum des positiven Peaks die Randschicht an der getriebenen Elektrode kollabiert ist. Diese Randschicht ist jedoch für die angelegte „Peaks"-Spannungsform einen längeren Zeitraum expandiert als kollabiert [29]. Die Elektronen gelangen somit nur für einen kurzen Moment zu der getriebenen Elektrode. Der Fluss der Elektronen $\Gamma_e = n_e \cdot u_e$ und der Fluss der Ionen $\Gamma_i = n_i \cdot u_i$ muss sich jedoch an den Elektroden im zeitlichen Mittel einer RF-Periode kompensieren, damit der Plasmazustand aufrechterhalten bleibt [29]. Hierbei bezeichnet $u_e$ beziehungsweise $u_i$ die Geschwindigkeit der Elektronen beziehungsweise die der Ionen in der Randschicht. Somit stellt sich ein hohes elektrisches Feld ein, das in diesem kurzen Zeitraum effektiv Elektronen in Richtung der getriebenen Elektrode beschleunigt, damit die Flusserhaltung gewährleistet ist. Die beschleunigten Elektronen besitzen dann genügend Energie, um mit anderen Teilchen in diesem Zeitraum Anregungsstöße auszuführen [29]. Außerdem wird durch eine angelegte „Peaks"-Spannungsform die EEPF beeinflusst, wie in Abbildung 2.5 k) - l) gezeigt ist. So ist beispielsweise der hochenergetische Schwanz der EEPF während des Randschichtkollaps an der getriebenen Elektrode aufgrund der starken Elektronenbeschleunigung durch das hohe elektrische Feld verstärkt [29].

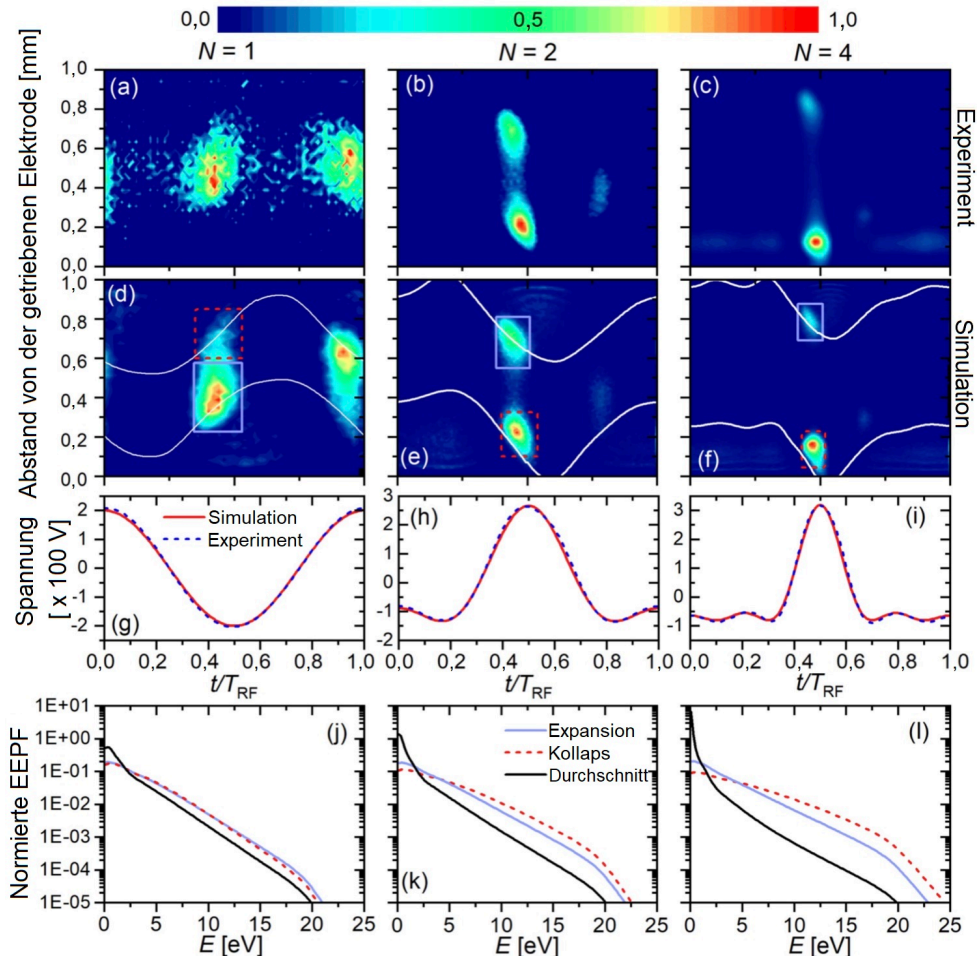

Abbildung 2.5: Normierte raum- und zeitaufgelöste Plots der Elektronstoßanregungsrate von dem Grundzustand in den Zustand He I $^3S_1$, die experimentell (erste Reihe) und aus PIC/MCC-Simulationen (zweite Reihe) bestimmt wurden, für eine verschiedene Anzahl $N$ der aufeinanderfolgenden Harmonischen der Grundfrequenz $f = 13{,}56$ MHz. Die angelegte maßgeschneiderte Spannungsform ist in der dritten Reihe abgebildet. Die Position der Randschicht ist in der zweiten Reihe als eine weiße Linie markiert. In der vierten Reihe ist die jeweilige EEPF abgebildet, die über den gesamten Elektrodenabstand und eine RF-Periode gemittelt ist. Der Gasfluss beträgt 1 slm He und 1 sccm $N_2$ [29].

### 2.3.2 Ionen­induzierte Sekundärelektronenemission

Ein wichtiger Oberflächenprozess eines Plasmas ist die Sekundärelektronenemission. Eine Oberfläche kann dabei die Elektrode oder auch das zu behandelnde Substrat sein. Sekundärelektronen können durch den Beschuss der Oberfläche durch Ionen, Elektronen, Neutralteilchen, Metastabilen oder auch Photonen herausgelöst werden [7, 30]. Der wichtigste Mechanismus für die Emission von Sekundärelektronen ist typischerweise im COST-Jet der Ionenbeschuss einer Oberfläche, der auch in dem indirekten Kanal des Penning-Modus eine wichtige Rolle spielt [6]. Die so erzeugten Elektronen führen zu einer Erhöhung der Elektronendichte im Plasma und tragen zu Ionisationsprozessen bei [31].

Im Folgenden soll der Mechanismus der Sekundärelektronenemission bei niedrigen Ionenenergien über den Auger-Effekt anhand Abbildung 2.6 näher erläutert werden.

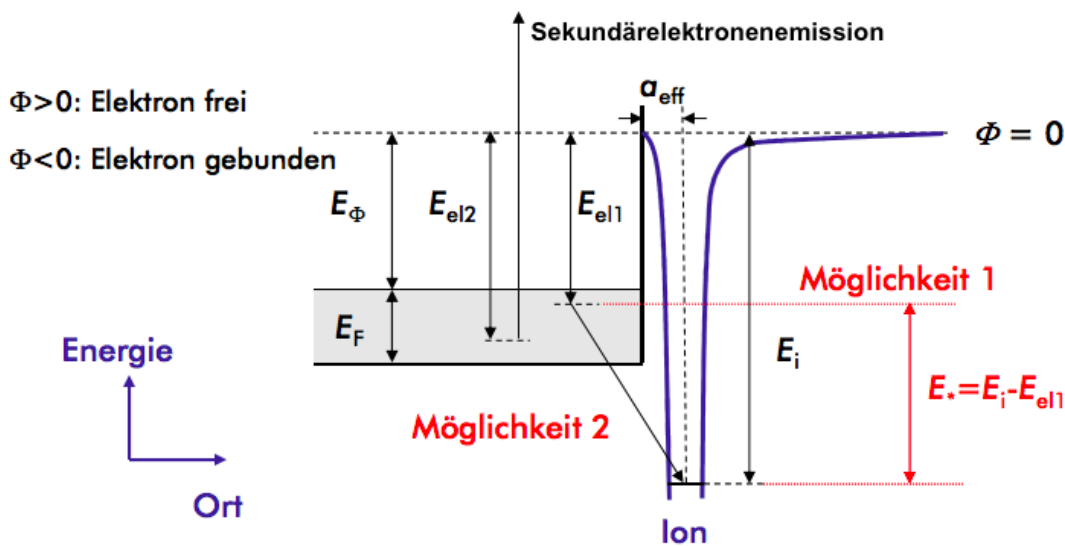

Abbildung 2.6: Potenzialmodell eines Festkörpers und die sich ergebenden Möglichkeiten der Sekundärelektronenemission über den Auger-Effekt [7, 32].

Es wird von einem Potenzialmodell ausgegangen, bei dem die Energie der Ladungsträger als Funktion des Ortes in einem Festkörper dargestellt ist. Als Festkörper kann beispielsweise ein Metall betrachtet werden. Die gebundenen Elektronen können bis zu einem bestimmten Energiemaximum verschiedene Energieniveaus im Leitungsband besetzten. Dieses Energiemaximum wird auch als Ferminiveau bezeichnet und liegt um $E_F$ unterhalb der potentiellen

Energie Φ = 0 für freie Elektronen. Liegt beispielsweise das gebundene Elektron an der Oberkante des Leitungsbandes, wird dieses im Festkörper durch eine Potentialbarriere der Höhe $E_\Phi$ eingeschlossen. Demnach muss mindestens eine Arbeit von $eE_\Phi$ verrichtet werden, um das Elektron, welches mit einem Abstand von $x = a_{\text{eff}}$ von der Oberfläche entfernt ist, aus dem Festkörper zu transportieren [7, 31]. Nähert sich nun ein positiv geladenes Ion der Oberfläche des Festkörpers und ist die Ionisierungsenergie $E_i$ größer als die Bindungsenergie eines Elektrons im Festkörper, so wird das Ion wie folgt neutralisiert:

$$A^+ + e^- \longrightarrow A. \tag{2.6}$$

Dabei wandert das Elektron aus dem Leitungsband des Festkörpers in das Ion, indem es durch die Coulomb-Barriere tunnelt. Im Allgemeinen gibt es nun zwei Möglichkeiten, wie dieser Prozess weiter abläuft [31]:

1. Das neutralisierende Elektron besetzt einen angeregten Zustand des entstandenen Neutralteilchens und regt sich durch die Emission eines Photons in den Grundzustand ab. Hierbei entsteht kein Sekundärelektron. Regt sich bei dieser Möglichkeit hingegen das Elektron strahlungslos in den Grundzustand ab, wird die frei werdende Energie von der Oberfläche absorbiert. Beträgt die absorbierte Energie mindestens die Austrittsenergie eines gebundenen Elektrons, kann durch die Energieabsorbtion nun ein Sekundärelektron aus dem Festkörper gelöst werden.

2. Das neutralisierende Elektron kann sofort den Grundzustand des entstandenen Neutralteilchens besetzen. Die frei gewordene Energie kann dann von einem im Festkörper gebundenen Elektron im Festkörper absorbiert werden, welches dann als Sekundärelektron den Festkörper verlassen kann.

Die Wahrscheinlichkeit zur ioneninduzierten Sekundärelektronenemission wird durch den ioneninduzierten Sekundärelektronenkoeffizienten $\gamma_i$ angegeben [33]. Dieser gibt die Anzahl der Sekundärelektronen an, die pro auftreffendem Ion erzeugt werden. Ein häufig verwendeter und empirischer Ausdruck für den Koeffizienten lautet [30]:

$$\gamma_i = 0{,}016(E_i - 2E_\Phi) \,. \tag{2.7}$$

Der Koeffizient hängt somit vereinfacht von der auftreffenden Ionenspezies, der Ionenenergie ($E_i$) und dem Oberflächenmaterial (beeinflusst $E_\Phi$) sowie dessen Struktur ab. Außerdem können Verunreinigungen auf dem Oberflächenmaterial und der auftreffende Winkel des Ions den Koeffizienten beeinflussen [31].

Außerdem konnten Daksha et al. durch PIC/MCC-Simulationen zeigen, dass die Wahl des Elektrodenmaterials einen Einfluss auf die Betriebsmodi einer in Argon betriebenen kapazitiven RF-Entladung bei 130 Pa hat [31]. In Abbildung 2.7 sind die raum- und zeitaufgelösten Plots der Elektronenstoßionisationsrate in Abhängigkeit des Elektrodenmaterials für eine konstante angelegte Spannungsamplitude dargestellt. Durch die verschiedenen Elektrodenmaterialien wird jeweils der ioneninduzierte Sekundärelektronenemissionskoeffizient beeinflusst. Für die theoretische Ermittlung des ioneninduzierten Sekundärelektronenemissionskoeffizienten wurde das sogenannte Hagstrum-Modell angenommen, mit dem sich für einen Oberflächenbeschuss mit $Ar^+$-Ionen und einem Oberflächenmaterial aus Molybdän ein $\gamma_i$ = 0,101, für

Kupfer ein $\gamma_i$ = 0,078 und für Platin $\gamma_i$ = 0,02 ergibt [31]. Vergleicht man die Plots der raum- und zeitaufgelösten Elektronenstoßionisationsrate in Abbildung 2.7 a) und b) für die Elektrodenmaterialen Molybdän und Kupfer, so erkennt man ein ähnliches Verhalten. Die beiden Plasmaentladungen befinden sich in dem sogenannten $\gamma$-Modus, der auf dem Heizungsmechanismus der ioneninduzierten Sekundärelektronenemission basiert [21].

Die raum- und zeitaufgelöste Elektronenstoßionisationsrate in 2.7 c) für die Plasmaentladung mit dem Elektrodenmaterial Platin zeigt hingegen ein typisches Verhalten im sogenannten $\alpha$-Modus, bei dem die stochastische Heizung als dominierender Elektronenheizungsmechanismus auftritt [18]. Aufgrund des höheren ioneninduzierten Sekundärelektronenemissionskoeffizienten für die Elektrodenmaterialien Molybdän und Kupfer im Vergleich zu Platin, werden mehr Sekundärelektronen pro auftreffendem Ar$^+$-Ion in der Randschicht erzeugt. Die Sekundärelektronen vervielfältigen sich durch Stöße in der Randschicht und es sind mehr Ladungsträger im Plasma vorhanden. Die Plasmadichte erhöht sich und somit sinkt auch die Größe der Randschicht für die Plasmaentladungen für die Elektrodenmaterialien Molybdän und Kupfer im Vergleich zu der Plasmaentladung mit dem Elektrodenmaterial Platin [31].

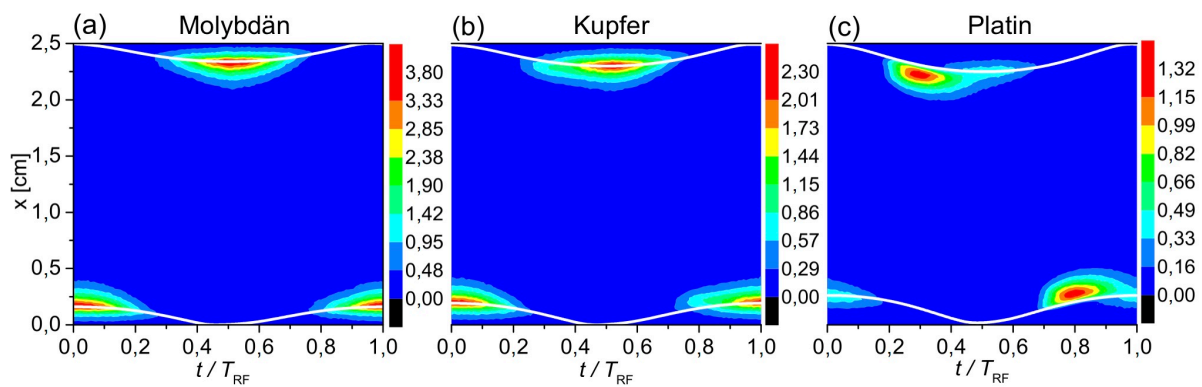

Abbildung 2.7: Raum- und zeitaufgelöste Plots der Elektronenstoßionisationsrate einer in Argon betriebenen kapazitiven RF-Entladung bei 130 Pa (13,56 MHz, 300 V$_{pp}$) in Abhängigkeit des Elektrodenmaterials, erhalten aus PIC/MCC-Simulationen: a) für Molybdän, b) für Kupfer und c) für Platin. Die Farbskala ist in der Einheit $10^{21}$ m$^{-3}$ s$^{-1}$ angegeben [31].

Die Erkenntnisse bieten ein vielversprechendes Potential, um durch die Wahl verschiedener Elektrodenmaterialen die Betriebsmodi und somit die Heizungsmechanismen bei einem Mikro-Atmosphärendruckplasma ebenfalls zu beeinflussen.

## 2.3.3 Strukturierte Elektroden

Im Vergleich zu planaren Elektroden besitzen strukturierte Elektroden rechteckige, runde oder dreieckige Gräben und beeinflussen so ebenfalls den Plasmaprozess. Sie können beispielsweise die Heizungsdynamik beeinflussen und die lokale Plasmadichte bei einem kapazitiv gekoppelten Niederdruckplasma erhöhen. Dies hat eine hohe technologische Relevanz bei plasmabasierten Beschichtungsverfahren, da so eine Erhöhung der Prozessgeschwindigkeit erreicht werden kann [34].

Schmidt et al. verdeutlichen am Beispiel einer rechteckigen Struktur in einer der Elektroden eines kapazitiv gekoppelten Niederdruckplasmas bei einem Druck von 10 Pa, dass die effiziente Heizung von Elektronen innerhalb der rechteckigen Struktur durch die oszillierenden Randschichten an den Grabenwänden verursacht wird [34]. Die Randschicht- und Elektronendynamik innerhalb einer solchen rechteckigen Struktur ist in Abbildung 2.8 a) zu der Zeit t = 10 - 15 ns während der Ausbreitungsphase der Randschicht gezeigt.

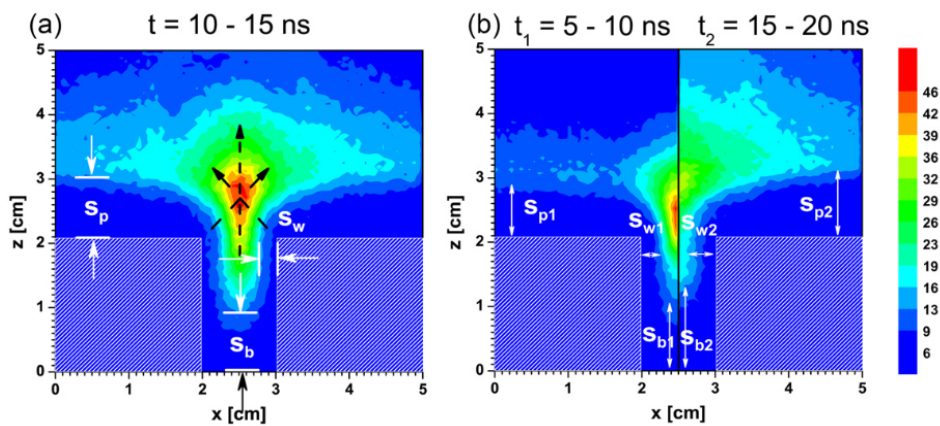

Abbildung 2.8: Randschicht- und Elektronendynamik innerhalb eines Grabens während der Expansionsphase, erhalten aus der Anregungsrate von dem Grundzustand in Ne $2p_1$, angegeben in relativen Einheiten. Der Gasdruck beträgt 10 Pa und die angelegte Spannung beträgt 264 $V_{pp}$ [34].

Die Randschichtbreite s variiert beim Vorhandensein der Struktur nicht nur mit der Zeit, sondern auch mit der Position entlang der Elektrode. Hier bezeichnet p die Randschicht über dem planaren Teil der Elektrode, w die vertikale Randschicht vor den Strukturwänden und b die Randschicht am Boden der rechteckigen Struktur.

Abbildung 2.8 b) vergleicht die Randschicht- und Elektronendynamik für zwei verschiedene Zeiträume. Der Index in dieser Abbildung beschreibt die Zeit $t_1 < t_2$. Bricht die Randschicht zusammen, gehen alle energetischen Elektronen innerhalb der Struktur an die Wände verloren. Die verbleibenden Elektronen innerhalb der Struktur sind kalt, wodurch diese Elektronen lange in der Struktur verweilen [34]. Durch eine Ausdehnung der Randschichten werden hingegen hochenergetische Elektronen erzeugt. Die Randschichtausdehnung führt zu vertikalen Elektronenstrahlen über der ebenen Elektrode ($s_{p2} > s_{p1}$). Innerhalb der Struktur werden

Elektronen mit einer horizontalen Geschwindigkeitskomponente erzeugt ($s_{w2} > s_{w1}$), die durch die gegenüberliegende Randschicht im Graben verweilen können. Dadurch entsteht innerhalb des Grabens eine höhere Plasmadichte als außerhalb ($s_w < s_p$) [34]. Am Grabenboden ist die Ionendichte und somit auch die Plasmadichte gering. Daher ist die Randschicht während der Expansionsphase am Grabenboden größer als die seitlichen Randschichten im Graben ($s_b > s_w$). Dies bewirkt, dass die vertikale Randschichtgeschwindigkeit höher als die horizontale Geschwindigkeit ist und letztendlich Elektronen durch die sich ausdehnende Randschicht von dem Grabenboden aus der Struktur heraus beschleunigt werden ($s_{b2} > s_{b1}$) [34]. Außerdem werden zusätzlich Elektronen Richtung Plasmabulk durch die Ausdehnung der Randschichten an den Strukturkanten beschleunigt. Dieser Effekt wird in der Abbildung 2.8 a) durch die gestrichelten schwarzen Pfeile verdeutlicht. Insgesamt lässt sich dadurch eine sehr starke lokale Elektronenstoßanregung über der Strukturöffnung beobachten, die mit einer Erhöhung der Ionisationsrate in diesem Bereich einhergeht [34].

Die Ergebnisse bieten somit ein aussichtsreiches Potential, um durch die Verwendung von strukturierten Elektroden die Heizungsmechanismen beim COST-Jet zu beeinflussen.

## 2.4 Phasenaufgelöste optische Emissionsspektroskopie

### 2.4.1 Aufbau und Funktionsweise einer ICCD-Kamera

Bei der phasenaufgelösten optischen Emissionsspektroskopie (PROES) wird mit Hilfe einer ICCD-Kamera (engl. intensified charge-coupled device, ICCD) die räumliche und zeitliche Beobachtung eines Plasmas ermöglicht, ohne dieses aktiv zu beeinflussen. Die Methode eignet sich für eine Untersuchung der räumlichen und zeitlichen Elektronheizungsdynamik einer Plasmaentladung innerhalb eines RF-Zykluses ($T_{RF}$ = 74 ns). Eine ICCD-Kamera kann das einfallende Licht um mehrere Größenordnungen verstärken und eignet sich daher zur Aufnahme von Bildern mit sehr kurzen Belichtungszeiten. In dieser Arbeit wird für die PROES-Messungen die ICCD-Kamera 4 Picos der Firma Stanford Computer Optics verwendet, deren innerer Aufbau in Abbildung 2.9 schematisch dargestellt ist.

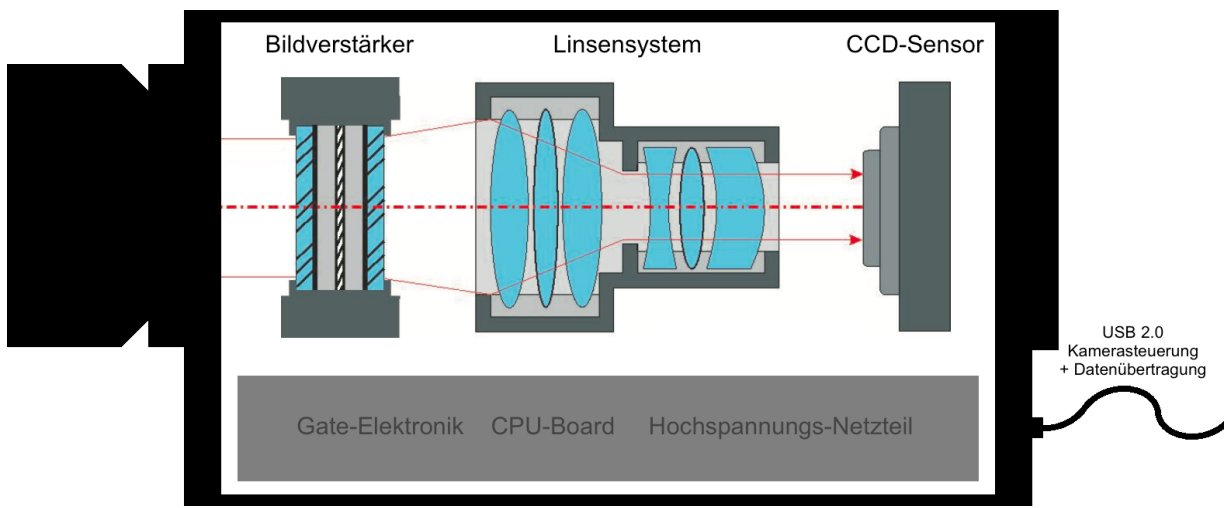

Abbildung 2.9: ICCD-Kamera 4 Picos der Firma Stanford Computer Optics [35].

Die verwendete ICCD-Kamera setzt sich grundsätzlich aus einem CCD-Sensor zusammen, vor dem ein Linsensystem und ein Bildverstärker geschaltet sind. Zu Beginn wird das auftreffende Licht von dem Bildverstärker verarbeitet. Die Photonen treffen dabei zunächst durch ein Eintrittsfenster auf eine Photokathode und lösen dort Elektronen heraus [35]. Hinter der Photokathode befindet sich eine Mikrokanalplatte (eng. Micro Channel Plate, MCP). Zwischen der Photokathode und der Mikrokanalplatte ist eine einstellbare Spannung $U_{MCP}$ angelegt [36]. Die an der Photokathode durch Photoionisation herausgelösten Elektronen werden von der angelegten Spannung in Richtung MCP beschleunigt und lösen dort durch Stöße mit der Wand des MCPs Sekundärelektronen heraus. Anschließend treffen die Elektronen auf einen Phosphorschirm und regen den Phosphor zum Leuchten an. Hinter dem Bildverstäker befindet sich ein Linsensystem, welches die Ausgangsstrahlung des Phoshorschirms in den dahinter liegenden

CCD-Detektor unter Beibehaltung der optischen Auflösung einkoppelt. Der CCD-Detektor wandelt abschließend die Lichtsignale in elektrische Signale um [35].

## 2.4.2 Erstellung der raum- und phasenaufgelösten Plots der Elektronenstoßanregung

Das Kamerasystem ist mit der angelegten RF-Spannung der zu untersuchenden Plasmaentladung synchronisiert. Das Kamerasystem erzeugt hierfür ein sogenanntes Gate-Signal, welches mit der RF-Spannung der Plasmaentladung synchronisiert ist. Das Gate-Signal steuert die Öffnung des Bildverstärkers. Die Zeitdauer des geöffneten Bildverstärkers wird als Gate-Breite bezeichnet und kann mit Hilfe der Kamerasoftware eingestellt werden. Außerdem kann eine Verzögerungszeit (Delay) gesetzt werden, die bestimmt, zu welchem Zeitpunkt die Öffnung des Bildverstärkers beginnen soll. Das Gate-Signal durchläuft den RF-Zyklus und die Verzögerung wird dabei in gleichbleibenden Schritten erhöht. Durch eine zusätzliche Einstellung der Belichtungszeit kann eingestellt werden, wie viele RF-Zyklen bei gleicher Phasenlage der Zeitfenster durchlaufen werden sollen [37]. Dies ermöglicht Aufnahmen mit einem gutem Signal-zu-Rausch-Verhältnis trotz niedriger Leuchtintensität. Die Kamerasoftware integriert dann die in dem Zeitfenster gemessene Emission über die eingestellte Anzahl der RF-Zyklen und generiert dann ein Bild für eine bestimmte Phasenlage. Die so entstehenden schwarz-weiß Bilder bieten dann mit Hilfe eines Matlab-Codes die Möglichkeit, die orts- und zeitaufgelöste Emission und anschließend die orts- und zeitaufgelöste Elektronenstoßanregungsrate in einem Konturdiagramm darzustellen, wie in Abbildung 2.10 gezeigt ist.

Für die Erstellung eines Konturdiagramms der Emission werden zunächst die Positionen der Elektroden ausgewählt, damit nur die Emission zwischen den Elektroden betrachtet wird. In Abbildung 2.10 ist zusätzlich ein sogenannter Bin-Bereich dargestellt. Dieser dient dazu, einen relevanten Bereich in der Messung zu definieren und über die horizontale Achse einen Mittelwert zu bilden. Dieser Ablauf wird für jedes der 74 Bilder automatisch wiederholt. Die Ergebnisse werden dann abschließend in einem Bild zusammengefasst. Die horizontale Achse wird dabei durch eine Zeit-Achse ersetzt, während die Auflösung senkrecht zu den Elektroden erhalten bleibt. Das zusammengefasste PROES-Bild zeigt somit die Emission zeitlich und senkrecht zu den Elektroden aufgelöst.

Zwischen der ICCD-Kamera und dem Plasma wird ein Interferenz-Filter eingesetzt, wodurch die Emission einer einzelnen Emissionslinie der Entladung betrachtet werden kann. Wird die Strahlung des Plasmas jedoch ohne einen Interferenz-Filter aufgenommen, so erhält man wellenlängenintegrierte PROES-Messungen und somit komplexe und oft unbekannte Überlagerungen von Beiträgen verschiedener Emissionslinien [6]. Bei einer geeigneten Wahl der Emissionslinie ist es möglich, aus der gemessenen Emission die Elektronenstoßanregung aus dem Grundzustand zu berechnen und diese in den PROES-Bildern darzustellen. Grundlage für die Berechnung ist die An- und Abregungsbilanz des betrachteten angeregten Zustands $i$ [37]:

$$\frac{dn_\text{i}(t)}{dt} = n_0 E_{0,\text{i}}(t) + \sum_\text{m} n_\text{m} E_{\text{m},\text{i}}(t) + \sum_\text{c} A_\text{ci} n_\text{c}(t) - A_\text{i} n_\text{i}(t). \qquad (2.8)$$

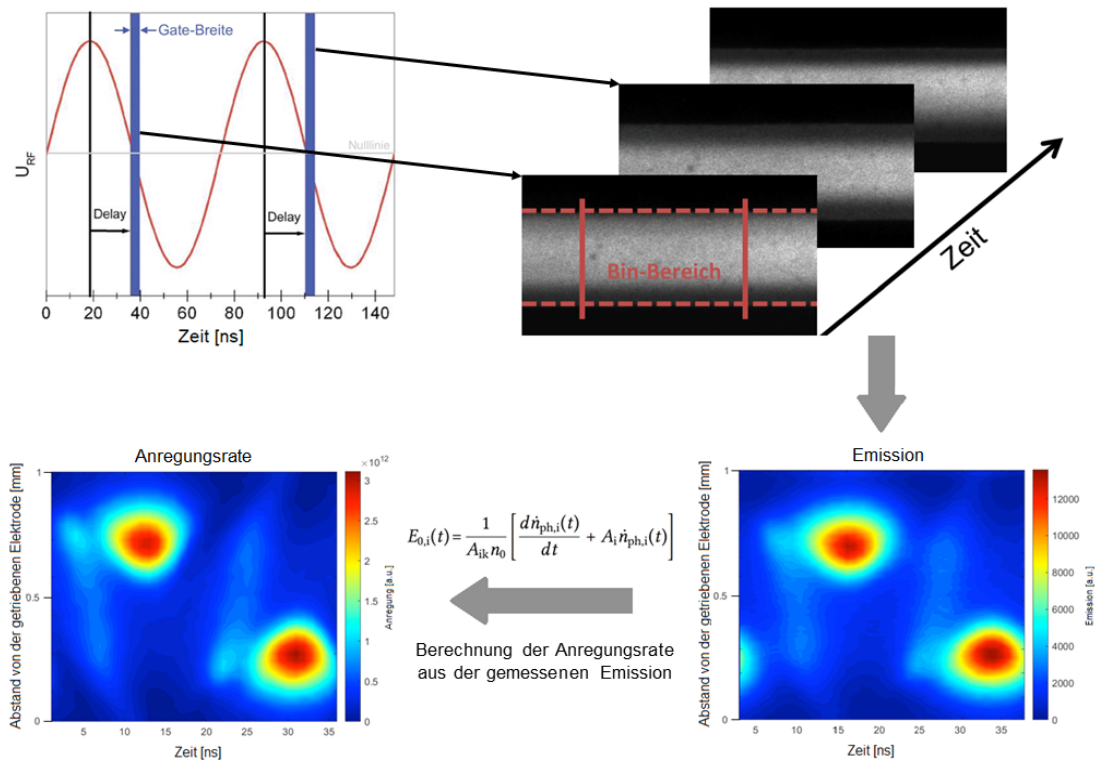

Abbildung 2.10: Illustration der Erzeugung von PROES-Bildern [37].

Der Term $n_0 E_{0,i}(t)$ beschreibt die Elektronenstoßanregung aus dem Grundzustand in den Zustand $i$. Die Besetzungsdichte des Grundzustands wird dabei mit $n_0$ und die Elektronenstoßanregungsrate aus dem Grundzustand mit $E_{0,i}(t)$ angegeben. Die Anregung aus metastabilen Zuständen wird durch den Term $\sum_m n_m E_{m,i}(t)$ beschrieben, bei dem $n_m$ die Besetzungsdichte des metastabilen Zustands bezeichnet und $E_{m,i}(t)$ die Anregungsrate des Übergangs von dem metastabilen Zustand in den betrachteten energetisch höheren Zustand $i$ bezeichnet. Der Term $\sum_c A_{ci} n_c(t)$ stellt die Anregung durch Kaskaden dar, bei dem $A_{ci}$ eine Übergangsrate aus dem Kaskadenniveau $c$ in den Zustand $i$ repräsentiert [37]. Der Term $A_i n_i(t)$ beschreibt die Abregung durch die spontane Emission und durch das Quenching, bei dem $A_i$ den Einsteinkoeffizienten des Zustands $i$ bezeichnet und folgendermaßen definiert ist:

$$A_i = 1/\tau \tag{2.9}$$

Hierbei entspricht $\tau$ der effektiven Lebenszeit des betrachteten Zustands $i$. Die Lebenszeit des betrachteten Zustands muss für die Durchführung der PROES-Messung kleiner als die betrachtete RF-Periode sein (74 ns). Um die Gleichung 2.8 zu vereinfachen, wird die Anregung durch metastabile Niveaus und durch Kaskaden vernachlässigt. Bei einem Atmosphärendruckplasma ist der Einfluss der strahlungslosen Stoßabregung (Quenching) jedoch nicht zu vernachlässigen. Unter diesen Bedingungen lässt sich die zeitliche Änderung der Besetzungsdichte des angeregten Zustands $i$ folgendermaßen definieren [38]:

$$\frac{dn_{\mathrm{i}}(t)}{dt} = n_0 E_{0,\mathrm{i}}(t) - A_{\mathrm{i}} n_{\mathrm{i}}(t). \tag{2.10}$$

Stellt man die vereinfachte Gleichung nach der Anregungsrate $E_{0,\mathrm{i}}(t)$ um, erhält man mit der von der Kamera gemessenen Anzahl an Photonen pro Raumvolumen und Zeit $\dot{n}_{\mathrm{ph,i}}(t) = A_{\mathrm{ik}} n_{\mathrm{i}}(t)$ [38]:

$$E_{0,\mathrm{i}}(t) = \frac{1}{A_{\mathrm{ik}} n_0} \left[ \frac{d\dot{n}_{\mathrm{ph,i}}(t)}{dt} + A_{\mathrm{i}} \dot{n}_{\mathrm{ph,i}}(t) \right]. \tag{2.11}$$

In dieser Arbeit wird ein Interferenz-Filter eingesetzt, der eine zentrale Wellenlänge von 700 nm und eine Halbwertsbreite von 15 nm besitzt. Somit wird eine Heliumlinie bei 706,5 nm beobachtet, die eine Energieschwelle für die Elektronenstoßanregung aus dem Grundzustand bei 22,7 eV besitzt [6]. Die Linie besitzt eine mittlere Lebenszeit von $\tau$ = 5 ns und ist somit deutlich kleiner als die betrachtete RF-Periode.

## 2.5 Zweiphotonen laserinduzierte Fluoreszenzspektroskopie

### 2.5.1 Erzeugungsprozesse von atomarem Sauerstoff

In einem mit Helium und einer Beimischung von Sauerstoff betriebenen COST-Jet finden zahlreiche Stöße zwischen den einzelnen Teilchenspezies statt, die so zur Aufrechterhaltung der Plasmachemie beitragen. Die Untersuchung der atomaren Sauerstoffproduktion in einem Mikro-Atmosphärendruckplasma ist von großer Bedeutung, da das Vorhandensein der im Plasma erzeugten reaktiven Spezies eine wichtige Rolle bei der Behandlung von lebendem Gewebe spielt. Die Produktion von atomaren Sauerstoff wird durch folgende dominierende Stoßprozesse im COST-Jet erreicht [39]:

$$O^* + He \longrightarrow O + He \tag{2.12}$$

$$e^- + O_2 \longrightarrow O + O^* + e^- \tag{2.13}$$

$$e^- + O_2 \longrightarrow 2\,O + e^- \tag{2.14}$$

$$O^* + O_3 \longrightarrow O_2 + 2\,O. \tag{2.15}$$

Es handelt sich somit um Prozesse, die durch Elektronen und Schwerteilchen aufrechterhalten werden. Zu den wichtigsten Prozessen, die atomaren Sauerstoff abbauen, zählen [39]:

$$O + O_2 + He \longrightarrow O_3 + He \tag{2.16}$$

$$2\,O + He \longrightarrow O_2 + He \tag{2.17}$$

$$O + O_3 \longrightarrow 2\,O_2. \tag{2.18}$$

### 2.5.2 Die Zweiphotonenanregung

Die Zweiphotonen laserinduzierte Fluoreszenz (engl. two photon absorption laser induced fluoreszenz, TALIF-) Spektroskopie ermöglicht die räumlich und zeitlich aufgelöste Analyse der Grundzustandsdichte des atomaren Sauerstoffs im COST-Jet. Der Prozess und der experimentelle Aufbau der Zweiphotonen laserinduzierten Fluoreszenzspektroskopie ist in Abbildung 2.11 a) und b) vereinfacht dargestellt. Der Prozess basiert auf der Anregung eines Sauerstoffatoms aus dem Grundzustand $|1\rangle$ in den angeregten Zustand $|3\rangle$ durch die Absorption zweier Photonen mit einer jeweiligen Wellenlänge von 225,65 nm [40]. Dabei regt das erste Photon das Atom in einen virtuellen Zwischenzustand an, der eine sehr geringe Lebensdauer hat und wodurch das zweite Photon schon nach einer sehr kurzen Zeit das Atom anschließend in den höheren Zustand anregen muss. Die Photonen stammen aus einer abstimmbaren schmalbandigen Laserstrahlung im UV-Bereich, die in das Beobachtungsvolumen mit Hilfe einer Linse fokussiert wird. Beim Nachweis von atomaren Sauerstoffspezies im Grundzustand werden zur Anregung zwei Photonen anstatt einem Photon benötigt, da zur Überbrückung der großen Energiedifferenz von circa 11 eV zwischen dem Grundzustand und dem angeregten Zustand ein Photon mit einer Wellenlänge im VUV-Bereich benötigt würde [40]. Eine Wellenlänge im VUV-Bereich ist im Vergleich zu einer Wellenlänge im UV-Bereich experimentell schwer zugänglich. Die Abregung in den Zustand $|2\rangle$ bei einer Wellenlänge von 844,87 nm, die durch einen Einsteinkoeffizienten $A_{32}$ für spontane Emission beschrieben werden kann, wird durch einen Photomultiplier detektiert. Durch die Fokussierung der Laserstrahlung auf ein bestimmtes räumliches Gebiet, bei gleichzeitiger Abbildung des Volumens der emittierenden Fluoreszenzstrahlung auf den Photomultiplier, wird die räumliche Auflösung erreicht. Die zeitliche Auflösung wird durch den zeitlich begrenzten Anregungsmechanismus der gepulsten Laserstrahlung bestimmt [40].

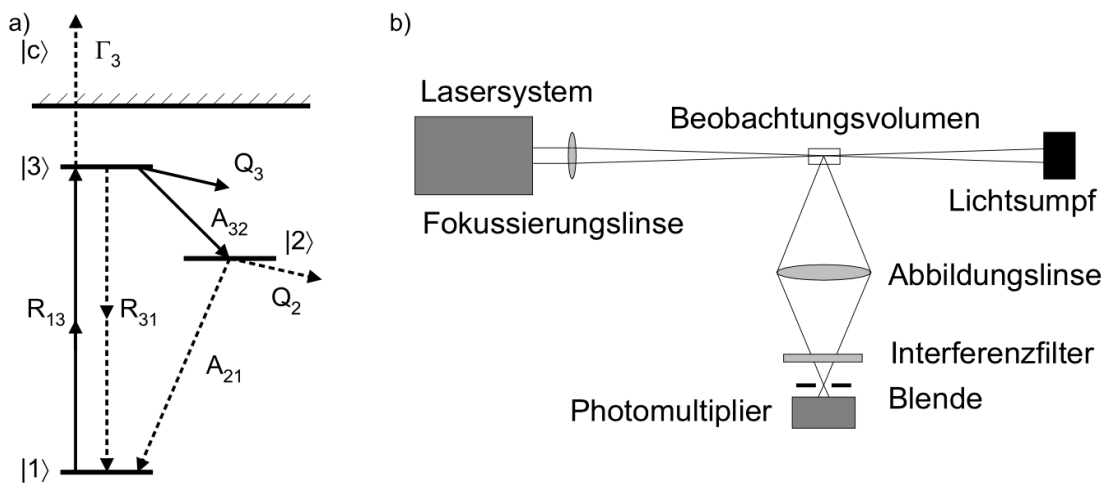

Abbildung 2.11: a) Schema der Sauerstoff An- und Abregung und b) schematischer Aufbau eines Fluoreszenzexperiments zur Bestimmung der atomaren Sauerstoffdichte [40].

Bei niedrigen Laserenergien können Verlustprozesse, wie beispielsweise die Entvölkerung des Zustandes $|3\rangle$ durch die Photoionisation $\Gamma_i$ oder die Zweiphotonenemission $R_{31}$, welche durch den Laser induziert werden, vernachlässigt werden. Nur das Quenching $Q_3$ muss hier als Konkurrenzprozess zur spontanen Emission berücksichtigt werden. Das Quenching, welches einen strahlungslosen Stoßabregungsprozess bezeichnet, ist vor allem in einem Plasma bei Atmosphärendruck stark ausgeprägt [40].

Demnach lässt sich die zeitliche Änderung der Besetzungsdichte des angeregten Zustandes $|3\rangle$ durch folgende reduzierte Ratengleichung beschreiben [40]:

$$\frac{dn_3}{dt} = R_{31} n_1 - (A_{32} + Q_3) n_3. \tag{2.19}$$

Als Lösung dieser Gleichung erhält man die Besetzungsdichte $n_3$, die man dann über die Zeit und das Anregungsvolumen integrieren kann. Schließlich erhält man die Zahl der Fluoreszenzphotonen, die linear von der Grundzustandsdichte abhängig ist [40].

### 2.5.3 TALIF-Kalibrierung mit Xenon

Das Ziel der Kalibrierung einer TALIF-Messung ist es, einen Zusammenhang zwischen dem Fluoreszenzsignal und der absoluten Grundzustandsdichte des atomaren Sauerstoffes zu bekommen. Die genaue Bestimmung der Zahl der Fluoreszenzphotonen ist experimentell nur eingeschränkt möglich, da zum Beispiel hierfür der Einfluss aller optischen Komponenten im Laserstrahlengang explizit bekannt sein muss [40].

Für die Kalibrierung wird eine TALIF-Vergleichsmessung mit einem Edelgas durchgeführt. Das Edelgas wird dabei in einer definierten Menge in eine Vakuumkammer eingelassen. Für die Untersuchung wird Xenon als Referenzgas verwendet, da dieses eine Zweiphotonenresonanz besitzt, die in der unmittelbaren Nähe der spektralen Zweiphotonenresonanz des atomaren Sauerstoffs liegt [41]. Die Grundzustandsdichte des Xenons $n_{Xe}$ ist hier für verschiedene Drücke bekannt. Ein Vergleich der vereinfachten Energieniveauschemata von atomarem Sauerstoff und Xenon ist in Abbildung 2.12 abgebildet. Außerdem muss eine ähnliche Erzeugung und Fokussierung der UV-Anregungsstrahlung ohne Modifikation des Lasersystems gewährleistet sein, damit eine ähnliche räumliche, spektrale und zeitliche Laserintensitätsverteilung im Fokus gegeben ist [40].

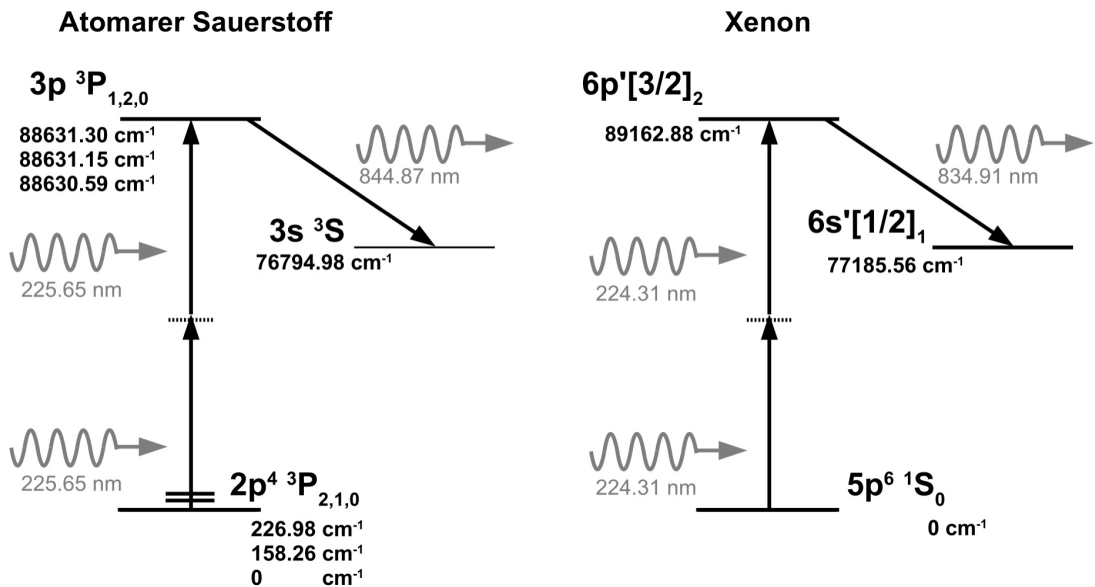

Abbildung 2.12: Zweiphotonenanregung und Fluoreszenzschema für atomaren Sauerstoff und Xenon. [42].

Aus der bekannten Xenondichte $n_{Xe}$ im Grundzustand, die aus dem Fluoreszenzsignal $S_{Xe}$ berechnet wird, kann dann die absolute Sauerstoffdichte im Grundzustand über folgenden Zusammenhang berechnet werden [41]:

$$n_O = \frac{T_{Xe}\eta_{Xe}}{T_O\eta_O} \frac{a_{Xe}}{a_O} \frac{E_O^2}{E_{Xe}^2} \frac{\sigma_{Xe}^{(2)}}{\sum_{J'}^{1,2,0} \sigma_{O,J\to J'}^{(2)}} \frac{S_O}{S_{Xe}} \beta_T n_{Xe} \qquad (2.20)$$

$$n_\mathrm{O} = \chi S_\mathrm{O} \tag{2.21}$$

Die Fluoreszenzsignale des atomaren Sauerstoffes $S_\mathrm{O}$ und des Xenons $S_\mathrm{Xe}$ sind die auf die Laserintensität normierten Spannungen des Photomultipliers. $T_\mathrm{Xe}$ und $T_\mathrm{O}$ bezeichnen die Transmission des verwendeten Filters für Xenon und Sauerstoff. Die jeweilige Photonenausbeute des Photomultipliers bei gegebener Wellenlänge wird mit $\eta_\mathrm{Xe}$ und $\eta_\mathrm{O}$ angegeben. Der Term $\frac{\sigma^{(2)}_\mathrm{Xe}}{\sum_{J'}^{1,2,0} \sigma^{(2)}_{\mathrm{O},J \to J'}}$ beschreibt das Verhältnis des Zweiphotonenabsorptionsquerschnitts und berücksichtigt mit $J'$ alle möglichen Ebenen des Grundzustands [41]. Für $J' = 2$ ergibt sich zum Beispiel ein Übergang aus dem unteren Zustand $(2p^4\,{}^3P_2)$ in den angeregten Zustand $(3p^3P_{1,2,0})$.

Der Quotient der Photonenenergien $\frac{E_\mathrm{O}^2}{E_\mathrm{Xe}^2}$ der beiden Fluoreszenzsignale kann hier vernachlässigt werden, da die beiden Photonenenergien identisch sind. Der Korrekturfaktor $\beta_\mathrm{T}$ berücksichtigt die Verteilung der thermischen Besetzung der drei Unterebenen des Grundzustands. Bei der Kalibriermessung wird $\chi$ bestimmt [41].

# 3 Experimenteller Aufbau

## 3.1 Der COST-Jet

### 3.1.1 Konzept und Aufbau

In dieser Arbeit wird der sogenannte COST-Jet als Plasmaquelle eingesetzt, der in Abbildung 3.1 schematisch dargestellt ist. Es handelt sich dabei um einen kapazitiv gekoppelten Mikro-Atmosphärendruckplasmajet ($\mu$-APPJ), welcher von Schulz-von der Gathen et al. entwickelt wurde und auf dem Konzept des von Selwyn et al. entwickelten Atmosphärendruckplasmajets (APPJ) beruht [43]. Der COST-Jet wurde durch eine Zusammenarbeit von Wissenschaftlern (engl. Cooperation in Science and Technology, COST) entwickelt und als eine Referenzquelle für die Erforschung nichtthermischer Atmosphärendruckplasmen und deren biomedizinische Anwendung konzipiert. Die Idee war es, ein kostengünstiges, einfaches und robustes Design zu erstellen, welches in verschiedenen Labors anwendbar ist und für die Forschungsgemeinschaft offen zur Verfügung steht. Als Referenzquelle ermöglicht der COST-Jet die Vergleichbarkeit der Ergebnisse verschiedener Forschungsgruppen und stärkt somit die Zusammenarbeit auf dem Gebiet der Plasmamedizin [43].

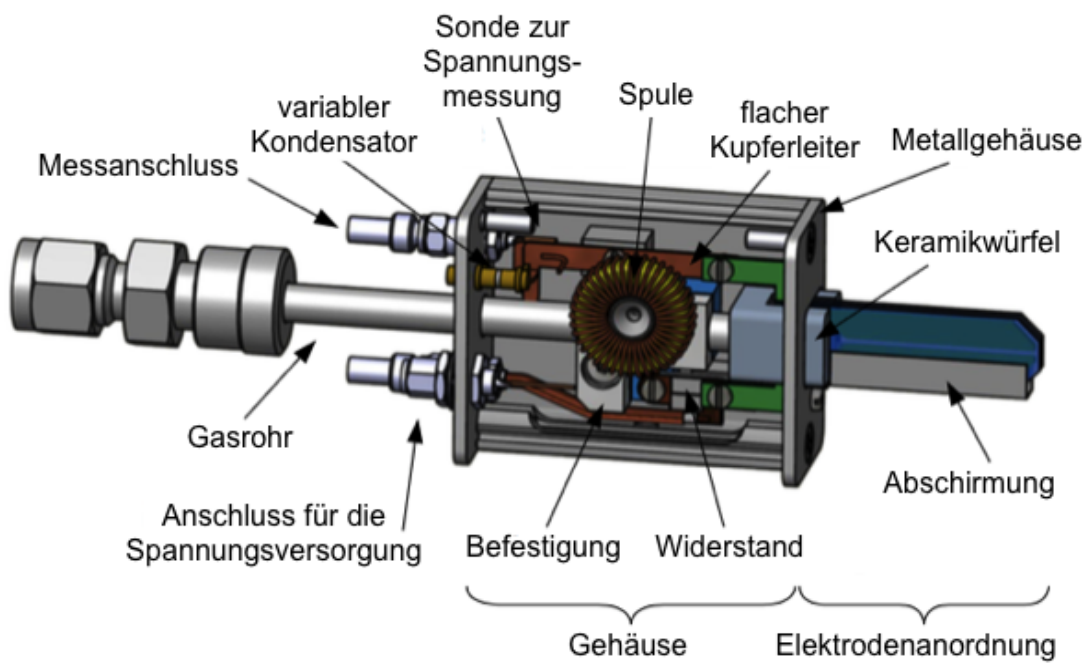

Abbildung 3.1: Schematische Darstellung des COST-Jets [43].

Zur Anregung des Plasmas wird typischerweise eine sinusförmige 13,56 MHz-Wechselspannung angelegt, wodurch der zuvor beschriebene Streamer-zu-Bogen-Übergang (Kapitel 2.1.2) verhindert werden kann. Der µ-APPJ kann mit reinem Helium oder mit einer Beimischung von molekularen Gasen wie zum Beispiel Stickstoff oder Sauerstoff betrieben werden. Die Beimischung von molekularen Gasen führt zur Bildung hochreaktiver Spezies im Plasma wie beispielsweise atomarem Sauerstoff [43].

Abbildung 3.2 zeigt die Elektrodenanordnung, die mit der Gasleitung über einen Keramikwürfel verbunden ist. Diese ist mit Schraubenverbindungen in dem Metallgehäuse des COST-Jets fixiert, wodurch ein einfacher Austausch verschiedener Elektrodenanordnungen für diese Arbeit ermöglicht wird [43].

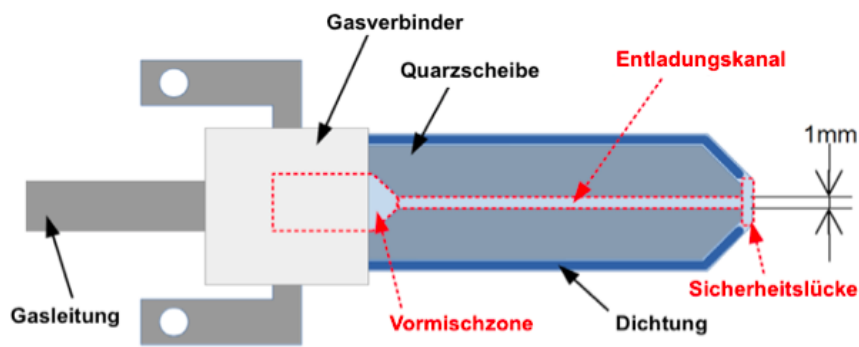

Abbildung 3.2: Skizze des Kopfes und der Gasleitung des COST-Jets [43].

Der Entladungskanal, der den Kopf des COST-Jets bildet, besteht aus zwei planparallelen Edelstahlelektroden, die von zwei Quarzscheiben umschlossen sind. Die Quarzscheiben ermöglichen einen breitbandigen optischen Diagnostikzugriff auf den Entladungskanal in einem Bereich von 170 nm - 3500 nm. Das Entladungsvolumen beträgt 1 mm (Elektrodenabstand) × 1 mm (Elektrodenbreite) × 30 mm (Entladungskanallänge). Dieses Entladungsvolumen wurde gewählt, da frühere Messungen gezeigt haben, dass bei dieser Größe eine homogene Plasmaentladung erreicht wird. Vor dem Entladungskanal befindet sich eine Vormischzone mit einem Elektrodenabstand von 5 mm, in der kein Plasma gezündet wird und in der sich das Gasgemisch lediglich zunächst umverteilen kann. Mit Hilfe eines Zweikomponentenklebstoffes (TorrSeal®) wird zwischen den Elektroden und den Quarzscheiben die Dichtigkeit des Entladungskanals gewährleistet und die Elektrodenanordnung in einem Keramikwürfel befestigt. Der Keramikwürfel stellt eine elektrisch isolierende Verbindung zur dahinter liegenden Gasleitung dar. Ein Gasrohradapter ermöglicht einen flexiblen Anschluss an standardisierten Gasleitungen [43].

In dem elektrisch isolierten Metallgehäuse des COST-Jets, welches einen zufälligen elektrischen Kontakt des Benutzers verhindert, befindet sich ein internes Anpassnetzwerk aus einem LC-Schwingkreis. Der LC-Schwingkreis besteht hierbei aus einer Spule (L = 9.6 µH, 41 Windungen) und einem abstimmbaren Kondensator (C = 0.8 pF bis 8 pF). Der abstimmbare

Kondensator kann für die Resonanz bei der Radiofrequenz eingestellt werden, womit die Generatorleistung effektiv in das Plasma eingekoppelt werden kann. Somit ermöglicht es eine Impedanzanpassung der komplexen Impedanz der Plasmaentladung an die reelle Ausgangsimpedanz des Generators [43].

Eine SMA-Verbindung an dem Metallgehäuse ermöglicht eine Verbindung des Generators an den COST-Jet. Des Weiteren sind in dem Gehäuse des COST-Jets zwei interne Sonden für jeweils eine Strom- und Spannungsmessung integriert, um eine kontinuierliche Überwachung der Kontrollparameter zu gewährleisten. Die internen Sonden können über SMC-Verbindungen an einem Oszilloskop angeschlossen werden. Vor einer Messung muss die interne Spannungssonde kalibriert werden. Dafür wird eine parallel ablaufende Spannungsmessung am COST-Jet mit einer externen Spannungssonde benötigt [43].

### 3.1.2 Maßgeschneiderte Elektroden

Im Folgenden sollen die in dieser Arbeit verwendeten Elektroden vorgestellt werden, deren grundsätzlichen Strukturentypen in Abbildung 3.3 a) - d) schematisch dargestellt sind. Der Kopf des COST-Jets, der aus zwei gegenüberliegenden Elektroden besteht, wird mit Hilfe der Bauanleitung von Golda et al. angefertigt [44].

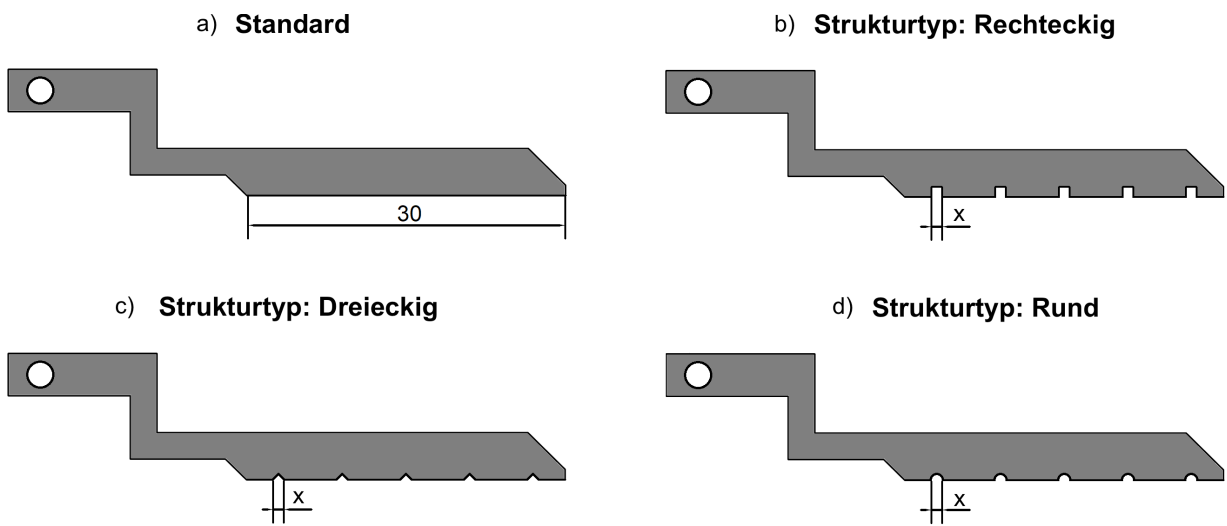

Abbildung 3.3: Auswahl der maßgeschneiderten Elektroden. Der Parameter x bezeichnet die Breite einer Struktur und kann die Werte 0,5 mm, 1 mm und 3 mm annehmen. Die Tiefe der rechteckigen Strukturen beträgt konstant 1 mm und die der dreieckigen und runden Strukturen 0,5x.

Wie in Kapitel 3.1 gezeigt, ist der COST-Jet standardmäßig mit zwei planparallelen Edelstahlelektroden und einer an der geerdeten Elektrode befestigten Abschirmung ausgestattet. Auf die Abschirmung wird in dieser Arbeit verzichtet, um die Qualität der Ergebnisse zu verbessern, da sich so ein symmetrisches Plasma einstellt [6]. Zunächst werden PROES-Messungen des COST-Jets mit zwei planparallelen Elektroden aus verschiedenen Elektrodenmaterialien durchgeführt, deren Form in Abbildung 3.3 a) dargestellt ist. Es handelt sich hierbei um die standardmäßige Elektrodenform des COST-Jets. Als Elektrodenmaterialien werden Edelstahl, Kupfer und Aluminium verwendet. Die verschiedenen Elektrodenmaterialien werden verwendet, da frühere Arbeiten gezeigt haben, dass diese Materialien jeweils ein unterschiedliches Verhalten auf die Elektronenreflektion und die Sekundärelektronenemission eines Niederdruckplasmas haben [31].

Außerdem werden PROES-Messungen und TALIF-Messungen mit maßgeschneiderten Elektroden durchgeführt, die unterschiedliche Strukturen aufweisen. In dieser Arbeit werden rechteckige, dreieckige (rechtwinklig und gleichschenklig) sowie runde Strukturformen verwendet, von denen sich jeweils fünf gleichartige in einer Elektrode befinden, wie in Abbildung 3.3 b) - d) zu erkennen ist. Die in der Abbildung 3.3 b) - d) angegeben Breite x der Strukturen kann zu 0,5 mm, 1 mm und 3 mm gewählt werden. Die Breite x wurde so gewählt, dass sich die Struktur auf das Plasma auswirken kann. Das Plasma kann nämlich genau dann in eine Struktur eindringen, wenn die Strukturenbreite größer als das Doppelte der maximalen Randschichtausdehnung (0,1 mm) beträgt [34]. Für die Herstellung der Elektroden wurde ein Laserschneidverfahren mit einem $CO_2$-Laser mit einem Durchmesser im Fokuspunkt von 0,25 mm verwendet. Die Tiefe der rechteckigen Strukturen beträgt konstant 1 mm und die der dreieckigen und runden Strukturen 0,5x. Das Material der strukturierten Elektroden kann ebenfalls aus Edelstahl, Kupfer und Aluminium bestehen.

## 3.2 Messaufbau der phasenaufgelösten optischen Emissionsspektroskopie

Im Rahmen dieser Arbeit wird zunächst die Elektronenheizungsdynamik des modifizierten COST-Jets in Abhängigkeit der Elektrodenkonfiguration, der Reaktivgasbeimischung und der Spannungsamplitude mit der phasenaufgelösten optischen Emissionsspektroskopie innerhalb eines RF-Zykluses (74 ns) untersucht. In Abbildung 3.4 ist der Versuchsaufbau schematisch dargestellt.

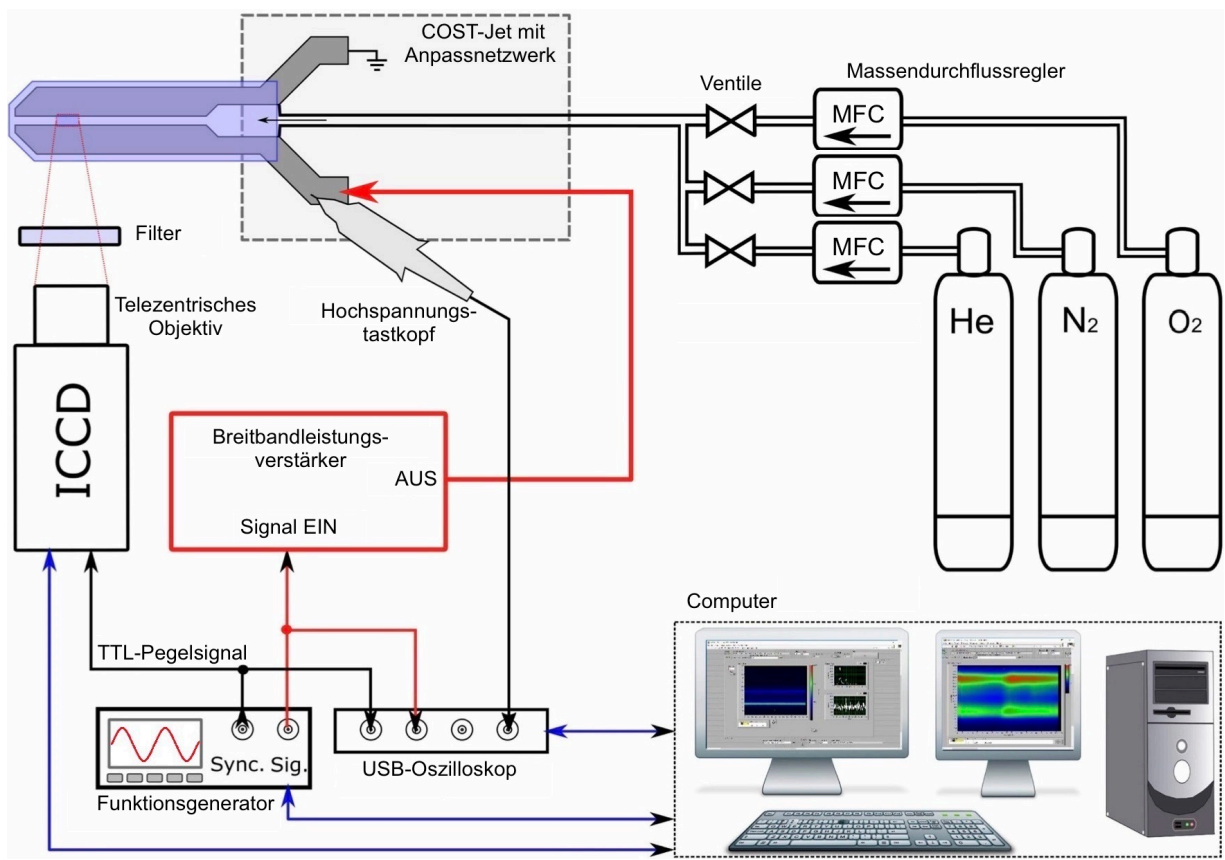

Abbildung 3.4: Schematische Darstellung des PROES-Messaufbaus [29].

Die PROES-Messungen werden mit einem konstanten Heliumfluss (99,999 % Reinheit) von 1 slm und einem Sauerstofffluss von 0,5 - 5 sccm $O_2$ durchgeführt. Außerdem werden die Messungen auch mit einem konstanten Heliumfluss von 1000 sccm und einer Stickstoffbeimischung von 0,5 - 5 sccm $N_2$ durchgeführt, um den Effekt auf die Heizungsdynamik des COST-Jets durch die Verwendung verschiedener Reaktivgase zu untersuchen. Der jeweilge Gasfluss kann über einen Massendurchflussregler (engl. Mass Flow Controller, MFC) eingestellt werden. Um die Verunreinigungen in den Gasleitungen nach einem Wechsel des Reaktivgases zu minimieren, wird mit einem an dem Versuchsaufbau benachbarten Pumpsystem, bestehend aus einer Drehschieberpumpe, einer mechanischen Vakuumpumpe und einem Hochvakuumschieber, das restliche Gasgemisch aus den Gasleitungen abgepumpt. Anschließend wird vor einem Messdurchgang der Verunreinigungsgrad und die Stabilität des Systems durch die zeitintegrierte optische Emissionsspektroskopie unter Verwendung eines USB-Spektrometers (Ocean Optics QE65000, Spektralbereich: 200 - 980 nm, spektrale Auflösung: 0,8 nm) kontrolliert.

Die PROES Messungen werden mit einer ICCD-Kamera (4 Picos von Stanford Computer Optics) durchgeführt, vor der sich ein telezentrisches Objektiv (Edmund Optics, TECHSPEC Compact TL) mit einem Arbeitsabstand von 40 mm und einer 4-fachen optischen Vergrößerung befindet. Das telezentrische Objektiv bietet den Vorteil, dass nur achsparallele Lichtstrahlenbündel vom COST-Jet in das Objektiv fallen und somit eine gleichmäßige, vom Objektabstand unabhängige Vergrößerung ohne eine perspektivische Verzerrung ermöglicht wird. An dem telezentrischen Objekt ist ein Interferenzfilter (Thorlabs) mit einer zentralen Wellenlänge von 700 nm und einer Halbwertsbreite von 15 nm angebracht, um die Emission der Heliumlinie bei 706,5 nm aufzunehmen.

Über einen Funktionsgenerator (Keysight 33600A) wird zunächst die gewünschte Frequenz und Amplitude des sinusförmigen Spannungssignals generiert. In dieser Arbeit wird der COST-Jet ausschließlich mit einer sinusförmigen Wechselspannung mit einer Frequenz von 13,56 MHz betrieben. Das ausgewählte Spannungssignal wird dann von einem Breitbandleistungsverstärker (AR RF/Microwave Instrumentation, Leistung: 500 W) verstärkt und anschließend an die getriebene Elektrode des COST-Jets angelegt. Die interne Spannungssonde des COST-Jets ist direkt mit einem USB-Oszilloskop (Picoscope 6402c) verbunden, welches mit einer LabVIEW-Software ausgelesen werden kann, die von Dr. Ihor Korolov entwickelt wurde. Die LabVIEW-Software ermöglicht die kontinuierliche Überwachung der an dem COST-Jet angelegten Spannung. Vor einer Messreihe muss die interne Spannungssonde des COST-Jets kalibriert werden. Hierfür wird ein externer Hochspannungstastkopf an die getriebene Elektrode angeschlossen, die damit gemessenen Spannungswerte mit denen der internen Spannungssonde verglichen und ein Kalibrierungsfaktor bestimmt. Die ICCD-Kamera ist ebenfalls über den Funktionsgenerator mit der angelegten Wechselspannung synchronisiert. Der Funktionsgenerator generiert einen Trigger-Impuls in Form eines TTL-Pegelsignals, welches hinsichtlich der Frequenz und der Phase des erzeugten Spannungssignals identisch ist. Der Triggerimpuls wird an den internen Delay-Generator der ICCD-Kamera übertragen und initialisiert dann das Gate-Signal.

Vor einem Messdurchgang wird zunächst die ICCD-Kamera im sogenannten freilaufenden Modus betrieben, bei dem das aufgenommene Bild des CCD-Detektors kontinuierlich dargestellt wird. In diesem Modus kann die Beobachtungsposition der Plasmaentladung justiert werden und eine korrekte Fokussierung des verwendeten telezentrischen Objektivs gefunden werden. Danach werden die PROES-Messungen gestartet. In dieser Arbeit wird am Bildverstärker konstant eine Spannung von $U_{MCP}$ = 1470 V eingestellt. Außerdem wird eine Gate-Breite von 2 ns eingestellt. Die Belichtungszeit wird für jede PROES-Messungen individuell bestimmt.

## 3.3 TALIF-Messaufbau zur Bestimmung der atomaren Sauerstoffdichte

Der in dieser Arbeit verwendete TALIF-Messaufbau ist in Abbildung 3.5 schematisch dargestellt und besteht grundsätzlich aus einem Lasersystem (Kapitel 3.3.1), einem Detektionsaufbau (Kapitel 3.3.2) und einem Vakuumsystem (Kapitel 3.3.3).

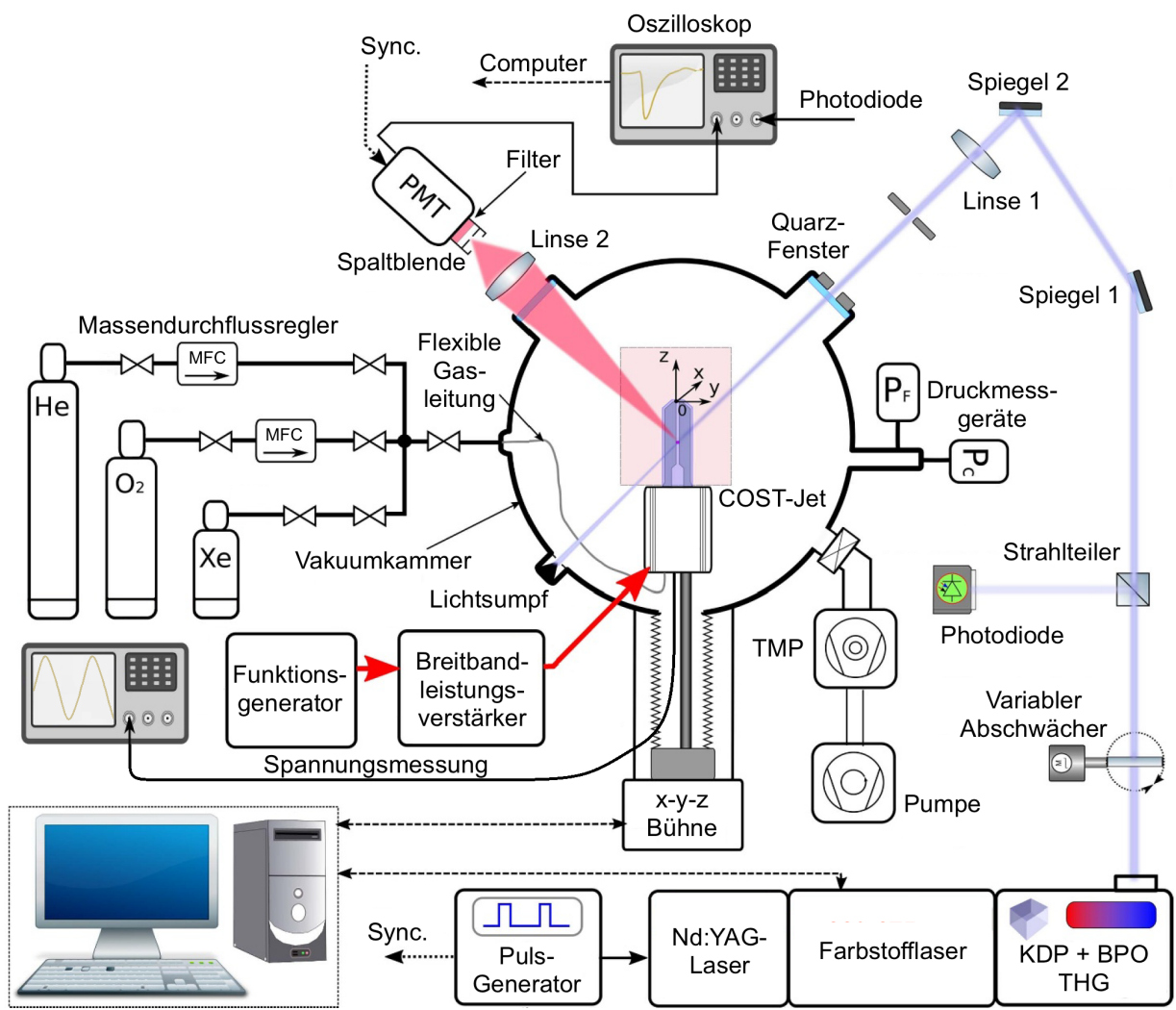

Abbildung 3.5: Schematische Darstellung des TALIF-Messaufbaus [41].

Die atomare Sauerstoffdichte im Grundzustand wird in Abhängigkeit der ausgewählten Elektrodenkonfigurationen und der Spannungsamplitude an verschiedenen Positionen innerhalb der Plasmaentladung des COST-Jets gemessen. Der COST-Jet wird während der TALIF-Messungen mit einer konstanten Gasmischung von 1000 sccm He und 5 sccm $O_2$ betrieben. Außerdem wird der COST-Jet mit einer sinusförmigen Wechselspannung mit einer Frequenz von 13,56 MHz betrieben, die von einem Funktionsgenerator (Keysight 33600A) und einem Breitbandleistungsverstärker (AR RF/Microwave Instrumentation, Leistung: 500 W) bereitgestellt wird. Die interne Spannungssonde des COST-Jets ist mit einem Oszilloskop (Tektronix TDS 2014B) verbunden, um das angelegte Spannungssignal an der getriebenen Elektrode kontinuierlich zu überwachen. Vor den verschiedenen TALIF-Messungen wird die interne Spannungssonde kalibriert und ein Kalibrierungsfaktor bestimmt.

### 3.3.1 Das Lasersystem

Das Lasersystem besteht aus einem Farbstofflaser (Continuum ND6000), der optisch von einem Nd:YAG-Laser (Continuum Powerlite 8000) gepumpt wird. Der Pumplaser emittiert eine Wellenlänge von 1064 nm und wird mit einer Frequenz von 10 Hz von einem Puls-Generator (Stanford Reasearch Systems DG535) getriggert. Die Länge eines Laserpulses beträgt 6 - 7 ns und die Energie 70 mJ. Mit Hilfe frequenzverdoppelnder Kristalle wird der Laserstrahl des Pumplasers auf eine Wellenlänge von 532 nm gebracht und in den Farbstofflaser geleitet.

Das aktive Medium des Farbstofflasers ist eine Farbstofflösung, die eine Mischung aus Pyridin I, DCM (roter Farbstoff) und Methanol enthält. Die Emissionswellenlänge des Farbstofflasers kann mit Hilfe eines Gitters (Dual 2400 1/mm) in einem Bereich von 650 - 715 nm in 1 pm-Schritten variiert werden. Das Gitter kann schrittweise mit einem Motor gedreht werden, der über einen Computer gesteuert wird. Anschließend wird die Wellenlänge des Farbstofflasers mittels einer Kristallanordnung aus einem Kaliumdihydrogenphosphat-Kristall (KDP) und einem Beta-Bariumborat-Kristall (BBO) verkürzt, da eine Wellenlänge von 225,6 nm für die Anregung von atomarem Sauerstoff benötigt wird. Anschließend wird der Laserstrahl durch ein Dispersionsprisma geleitet, sodass nur die gewünschte Wellenlänge selektiert wird. Danach wird der Laserstrahl, der nun einen Durchmesser von 3 mm und eine Pulsenergie von 0,1 mJ aufweist, über einen Spiegel in einen variablen Abschwächer geleitet. Der variable Abschwächer besteht aus einer dielektrisch beschichteten Quarzplatte, deren Neigungswinkel schrittweise mit einem Motor variiert werden kann. Somit ist es möglich, die Energie des Laserstrahls von 0% - 95% zu reduzieren.

Der Laserstrahl wird anschließend mit einem Strahlteiler in zwei Teile aufgeteilt. Ein Teil des Laserstrahls (8% - 10%) wird in eine Photodiode geleitet, um die Energie des Referenzsignals zu messen, mit dem das Fluoreszenzsignal normiert wird. Der andere Teil des Laserstrahls wird über eine Spiegelanordnung zu einer Linse gelenkt, wodurch der Laserstrahl fokussiert wird und in die Mitte der Vakuumkammer gelangt. Der COST-Jet ist in einer x-y-z-Verschiebung in der Vakuumkammer montiert, welche über computergesteuerte Schrittmotoren bewegt werden kann. Der Laserstrahl durchquert die Plasmaentladung und erreicht an der gegenüberliegenden Kammerwand einen Lichtsumpf, in dem der Laserstrahl absorbiert wird. Somit werden Reflexionen des Laserstrahls innerhalb der Vakuumkammer vermieden.

### 3.3.2 Der Detektionsaufbau

Durch die Fokussierung des Laserstrahls in die Plasmaentladung des COST-Jets wird der im Plasma erzeugte atomare Sauerstoff angeregt. Durch eine Abregung aus dem höheren Zustand $(3p^3P_{1,2,0})$ in den unteren Zustand $(3s^3S)$ mittels spontaner Emission ergibt sich ein Fluoreszenzlicht mit einer Wellenlänge von 844,87 nm. Für die Kalibrierung wird Xenongas verwen-det. Das Fluoreszenzlicht des Xenons entsteht aus dem Übergang von dem laserangeregten Zustands (6p'[3/2]$_2$) in den unteren Zustand (6s'[1/2]$_1$) mit einer Wellenlänge von 834,91 nm. Das Fluoreszenzlicht wird senkrecht zu dem Laserstrahl beobachtet und mit Hilfe einer BK7-Linse (f = 300 mm) durch einen Filter und eine Spaltblende (0,5 mm) auf einen Photomultiplier (Burle C31034A) geleitet. Die Spaltblende schirmt das Licht aus anderen Bereichen ab, womit die Auflösung verbessert wird. Der Photomultiplier ist ebenfalls mit dem Puls-Generator synchronisiert, sodass nur das laserinduzierte Signal gemessen wird. Das Fluoreszenzsignal und das Referenzsignal der Photodiode wird mit einem Oszilloskop (HP54510A) abgebildet und mit Hilfe einer LabVIEW-Software am Computer ausgelesen, die von Dr. Ihor Korolov erstellt wurde.

### 3.3.3 Das Vakuumsystem

Der COST-Jet ist in einer Vakuumkammer platziert, um die Kalibrierung der atomaren Sauerstoffdichte im Grundzustand mit einer bekannten Xenondichte durchzuführen. Beim Betrieb des COST-Jets bleibt die Vakuumkammer geöffnet. Für die Kalibrierung wird hingegen die Vakummkammer geschlossen und zunächst mit einem Pumpsystem auf einen Druck von circa $10^{-3}$ Pa abgepumpt. Das Pumpsystem besteht aus einer Turbomolekularpumpe (Pfeiffer DCU 100) und einer Vorpumpe (ABM-4EKF63CX). Der Druck in der Vakuumkammer kann mit einem kapazitiven Druckmessgerät (Edwards 600AB), in Abbildung 3.5 als „$P_\mathrm{C}$" gekennzeichnet, kontrolliert werden. Anschließend wird das Xenongas über ein Nadelventil in die Vakuumkammer geleitet bis ein gewünschter Druck im Bereich von 1 - 40 Pa erreicht ist. Die Xenondichte lässt sich ebenfalls über ein Druckmessgerät (Edwards W60022811) messen, welches in Abbildung 3.5 als „$P_\mathrm{F}$" bezeichnet wird.

# 4 Ergebnisse und Diskussion

## 4.1 Auswertung der phasenaufgelösten optischen Emissionsspektroskopie

Mit Hilfe der phasenaufgelösten optischen Emissionsspektroskopie wird in diesem Kapitel die Elektronenheizungsdynamik des COST-Jets in Abhängigkeit der maßgeschneiderten Elektroden, der Gasmischung und der angelegten Spannungsamplitude untersucht.

### 4.1.1 Planparallele Elektrodenanordnung aus Edelstahl, Kupfer und Aluminium

Aus den raum- und phasenaufgelösten Plots der Elektronenstoßanregungsrate, die durch die gemessene raum- und zeitliche Emission unter Verwendung eines Filters mit einer zentralen Wellenlänge von 700 nm und einer Halbwertsbreite von 15 nm erzeugt werden, wird die Elektronenheizungsdynamik des COST-Jets analysiert. Mit Hilfe eines USB-Spektrometers konnte zuvor gezeigt werden, dass in dem ausgewählten Wellenlängenbereich ausschließlich die Emissionslinie bei 706,5 nm beobachtet wird und keine weiteren Emissionslinien dort auftreten. Somit basieren die Ergebnisse allein auf einer Elektronenstoßanregungsrate, die eine Energieschwelle aus dem Grundzustand bei 22,7 eV besitzt. In jedem raum- und phasenaufgelösten Plot der Elektronenstoßanregung entspricht die vertikale Achse der räumlichen Auflösung des Elektrodenabstandes und die horizontale Achse zeigt die zeitliche Auflösung von 0 - 100 ns. Die zeitliche Auflösung ist hier größer als eine RF-Periode (74 ns) gewählt, da so die Stabilität des Messsystems besser überprüft werden kann. Anhand der Farbskala, die in relativen Einheiten angegeben ist, lässt sich erkennen, wie stark die Elektronenstoßanregung zu einem bestimmten Zeitpunkt zwischen den Elektroden ist. Für alle Ergebnisse hat sich eine DC-Self Bias-Spannung von $V_{bias} \approx 0$ V eingestellt. Der DC-Self Bias ist ein wichtiger Symmetrieparameter, der die Elektronenheizungsdynamik in den Randschichten beeinflussen kann. Dieser stellt sich beispielsweise bei einer geometrisch stark asymmetrischen Entladung oder bei der Verwendung des Voltage Waveform Tailorings ein. Bei einem Atmosphärendruckplasma ist dieser jedoch deutlich geringer als bei einem Niederdruckplasma [29].

Zunächst werden PROES-Messungen des COST-Jets mit zwei planparallelen Elektroden aus verschiedenen Elektrodenmaterialien durchgeführt, um den Einfluss des Elektrodenmaterials auf das Plasma zu untersuchen. Die Absolutwerte der Elektronenstoßanregung in den verschiedenen Plots können untereinander verglichen werden. In Abbildung 4.1 a) - f) sind die raum- und phasenaufgelösten Plots der Elektronenstoßanregung für eine planparallele Elektrodenanordnung aus den Materialien Edelstahl, Aluminium und Kupfer in Abhängigkeit der Peak-to-Peak-Spannungs-amplitude für einen Gasfluss von 1 slm He und 0,5 sccm $O_2$ gezeigt.

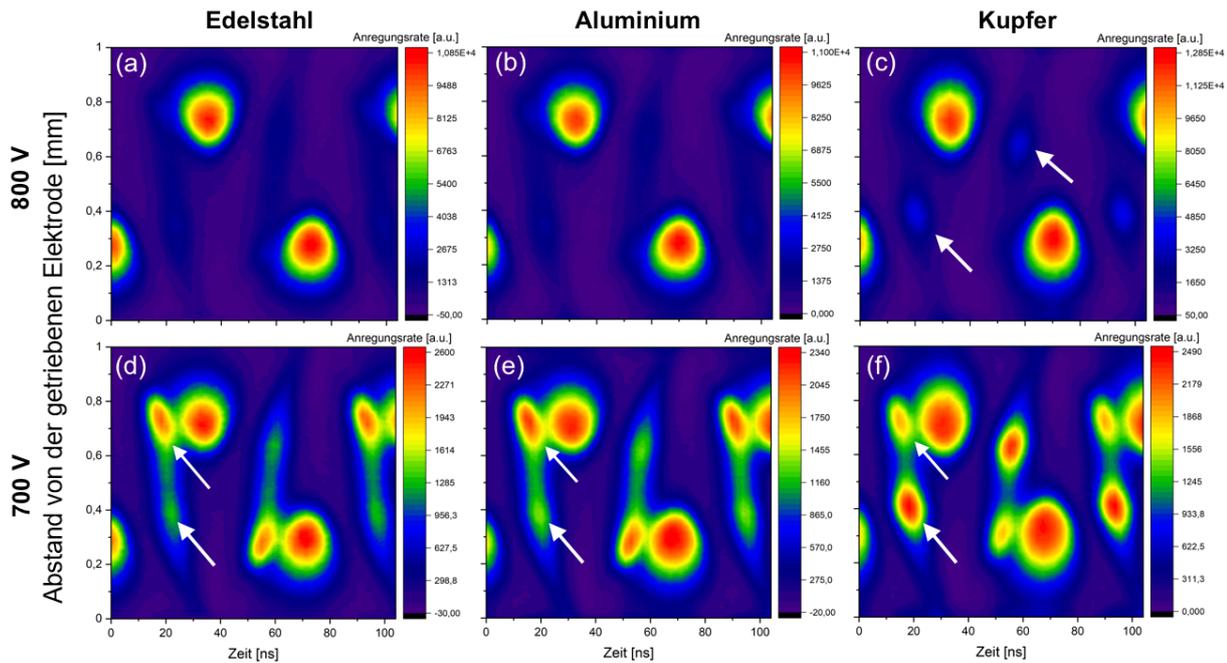

Abbildung 4.1: Raum- und phasenaufgelöste Plots der Elektronenstoßanregung der 706,5 nm He I-Linie des COST-Jet mit einer planparallelen Elektrodenanordnung in Abhängigkeit des Elektrodenmaterials und der Peak-to-Peak-Spannung. Der Gasfluss beträgt 1 slm He und 0,5 sccm $O_2$.

Als Elektrodenmaterialien werden Edelstahl, Kupfer und Aluminium verwendet, die jeweils ein unterschiedliches Verhalten auf die Sekundärelektronenemission eines Plasmas aufweisen. Die geerdete und die getriebene Elektrode bestehen aus dem gleichen Material. Abbildung 4.1 a) - c) zeigt die Elektronenstoßanregung des COST-Jets mit einer angelegten Peak-to-Peak-Spannung von 800 $V_{pp}$ für die verschiedenen Elektrodenmaterialien. Die gezeigten Plasmaentladungen des COST-Jets befinden sich für diese Kontrollparameter im Penning-Modus, wie an den beiden Anregungsmaxima in den Randschichtbereichen vor den Elektroden (während der Zeiträume 20 ns - 40 ns und 60 ns - 80 ns) zu erkennen ist. Für diese Zeiträume ist die Randschicht an der jeweiligen Elektrode expandiert. Die Intensität der Anregungsmaxima im Penning-Modus sind für eine Elektrodenanordnung aus Aluminium circa 10% höher als für eine Elektrodenanordnung aus Edelstahl. Für eine Elektrodenanordnung aus Kupfer beträgt die Erhöhung der Intensität circa 20%. Bei einer Elektrodenanordnung aus Kupfer ist innerhalb des Plasmabulks zusätzlich eine schwache Anregung des Ω-Modus (während der Zeiträume 10 ns - 20 ns und 50 ns - 60 ns) zu beobachten, wie in Abbildung 4.1 c) dargestellt ist und mit Pfeilen markiert ist. Wird die angelegte Peak-to-Peak-Spannung auf 700 $V_{pp}$ verringert, findet ein Übergang von dem Penning-Modus in den Ω-Modus statt, wie Abbildung 4.1 d) - f) zeigt. Die Maxima des Ω-Modus treten genau dann auf, wenn sich die Randschicht an der geerdeten Elektrode ausdehnt und die Randschicht an der getriebenen Elektrode kollabiert. Zu diesen Zeiten ergibt sich ein hohes Driftfeld. Das hohe Driftfeld wird hier einerseits durch eine verminderte Leitfähigkeit aufgrund der hohen Stoßfrequenz bei atmosphärischem Druck erzeugt, wie bereits in Kapitel 2.2 erklärt wurde. Andererseits wird das Driftfeld durch die Elektronegativität verstärkt, die durch die Beimischung von Sauerstoff entsteht und somit die Elektronen-dichte im Plasma verringert. Dieser Effekt lässt sich mit dem Drift-Ambipolar-Modus in einem elektronegativen Niederdruckplasma vergleichen. Vergleicht man die Anregungsmaxima im Plasmabulk (markiert durch Pfeile), so erkennt man auch hier einen geringen Unterschied bei der Verwendung des Elektrodenmaterials Kupfer im Vergleich zu den anderen Elektrodenmaterialien. So erkennt man in Abbildung 4.1 f) ein zusätzliches Anregungsmaxima vor der getriebenen Elektrode in dem Zeitraum von 10 ns - 30 ns. Hier kann somit angenommen werden, dass die Kupferelektroden die Elektronegativität des Plasmas stärker beeinflussen als die anderen Elektrodenmaterialien. So kann beispielsweise durch die Kupferelektroden die Elektronendichte im Randschichtbereich erhöht sein, weil hier mehr Sekundärelektronen herausgelöst werden können. Diese Elektronen können dann von dem Driftfeld beschleunigt werden und Anregungsstöße ausführen. Die PROES-Messungen werden auch unter den selben Bedingungen mit einer Stickstoffbeimischung wiederholt, jedoch zeigen diese keinen nennenswerten Unterschied und werden aus diesem Grund nicht gezeigt.

Da die Unterschiede, die sich durch die Variation des Elektrodenmaterials ergebenen, sehr gering sind, können diese zum Beispiel auch durch Veränderungen am Quarzglas, Verunreinigungen in den Gasleitungen, Temperaturveränderungen im Labor oder durch Fehler bei der Spannungskalibrierung verursacht werden. Aus diesem Grund werden zusätzlich Elektrodenanordnungen entworfen, die einen direkten Vergleich zwischen den Elektrodenmaterialien und deren Einfluss auf die Heizungsdynamik unter identischen Entladungsbedingungen ermöglichen. In Abbildung 4.2 a) - b) sind die verwendeten Elektrodenanordnungen abgebildet. Die getriebene Elektrode besteht hierbei aus Kupfer oder Aluminium und die geerdete Elektrode jeweils aus Edelstahl.

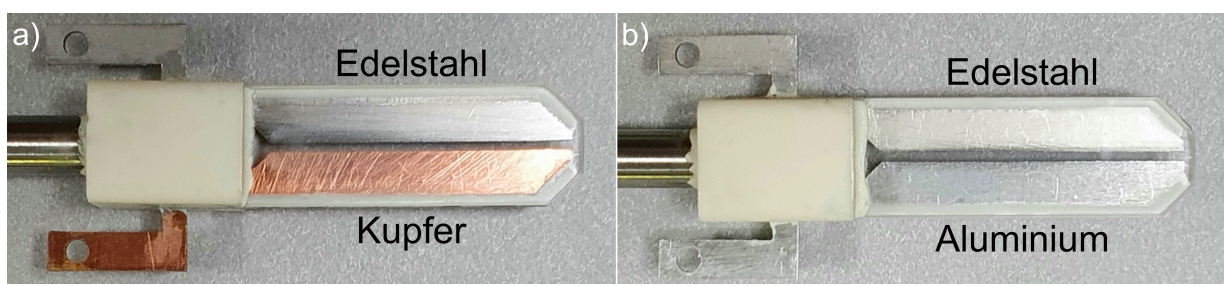

Abbildung 4.2: Fotos der planparallelen Elektrodenanordnungen aus verschiedenen Materialien: a) Kupfer und Edelstahl und b) Aluminium und Edelstahl.

Die PROES-Messungen mit den gezeigten Elektrodenanordnungen werden mit einem konstanten Heliumfluss von 1 slm und einem Sauerstofffluss von 0,5 - 5 sccm durchgeführt. Außerdem werden die Messungen mit einem konstanten Heliumfluss von 1 slm und einer Stickstoffbeimischung von 0,5 - 5 sccm wiederholt. Die angelegte Peak-to-Peak-Spannung wird jeweils zu 800 $V_{pp}$, 700 $V_{pp}$ und 600 $V_{pp}$ gewählt.

Im Folgenden werden zunächst die Ergebnisse der PROES-Messungen mit einem konstanten Heliumfluss von 1 slm und einem Sauerstofffluss von 0,5 - 5 sccm vorgestellt. Die Abbildung 4.3 a) - i) zeigt die raum- und phasenaufgelösten Plots der Elektronenstoßanregungsrate für einen COST-Jet mit einer planparallelen Elektrodenanordnung aus den Materialien Aluminium (getriebene Elektrode) und Edelstahl (geerdete Elektrode). Die Zeilen zeigen die Ergebnisse in Abhängigkeit der Peak-to-Peak-Spannung und die Spalten zeigen die Ergebnisse in Abhängigkeit der Beimischung des Sauerstoffflusses.

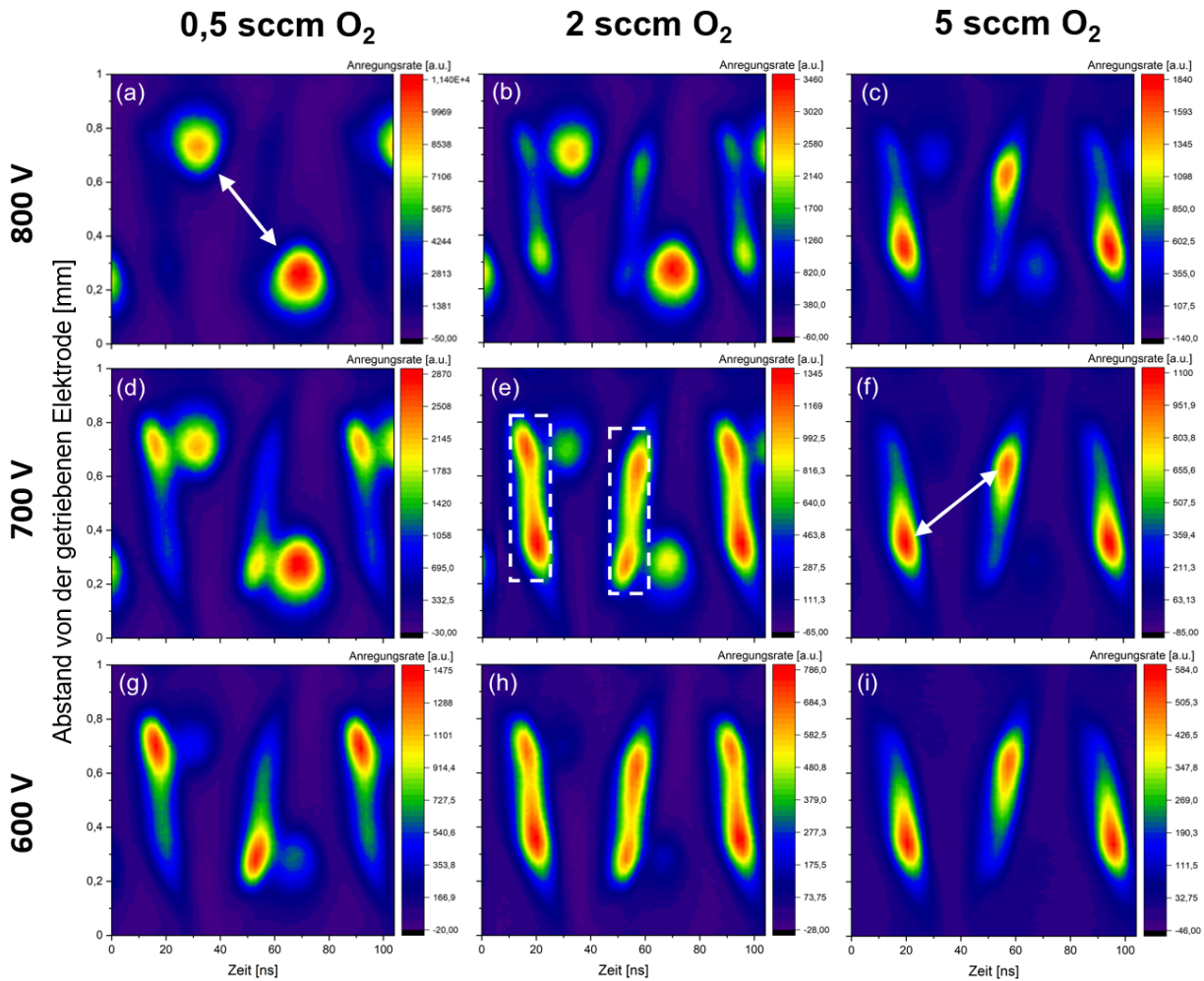

Abbildung 4.3: Raum- und phasenaufgelöste Plots der Elektronenstoßanregung der 706,5 nm He I-Linie eines COST-Jet mit einer planparallelen Elektrodenanordnung aus den Materialien Aluminium (getriebene Elektrode) und Edelstahl (geerdete Elektrode) in Abhängigkeit der Peak-to-Peak-Spannung und der Beimischung des Sauerstoffflusses. Der Heliumgasfluss beträgt 1 slm.

Die erste Spalte zeigt die Ergebnisse der Elektronenstoßanregungsrate für eine Sauerstoffbeimischung von 0,5 sccm unter Variation der angelegten Peak-to-Peak-Spannung. Abbildung 4.3 a) zeigt die Elektronenstoßanregung des COST-Jets mit einer angelegten Peak-to-Peak-Spannung von 800 $V_{pp}$. Die Plasmaentladung des COST-Jets befindet sich hierfür im Penning-Modus. Die beiden Anregungsmaxima in den Randschichtbereichen treten in den Zeiträumen 20 ns - 40 ns und 60 ns - 80 ns (markiert durch Pfeile) auf. Jedoch erkennt man auch eine deutli-che Asymmetrie in der Intensität der Anregungsmaxima. So ist die Intensität des Anregungsmaximums vor der getriebenen Elektrode, die hier aus Aluminium besteht, deutlich größer im Vergleich zum Anregungsmaximum vor der geerdeten Elektrode. Dieser Effekt zeigt, dass die Elektrodenmaterialien den indirekten Kanal des Penning-Modus unterschiedlich beeinflussen.

Wie bereits in Kapitel 2.2 näher erläutert, basiert der indirekte Kanal des Penning-Modus auf Oberflächenprozessen wie beispielsweise der ioneninduzierten Sekundärelektronenemission. Hierbei werden die durch die Penning-Ionisation erzeugten Ionen von dem elektrischen Feld in der Randschicht in Richtung der naheliegenden Elektrode beschleunigt und können dort bei einer ausreichenden Ionenenergie Sekundärelektronen herauslösen. Die herausgelösten Sekundärelektronen werden dann von dem elektrischen Feld in der Randschicht in Richtung Plasmabulk beschleunigt. Außerdem ist der ioneninduzierte Sekundärelektronenkoeffizient $\gamma_i$ abhängig von dem Elektrodenmaterial. Da das Anregungsmaximum vor der getriebenen Elektrode ausgeprägter ist, kann die Annahme getroffen werden, dass Aluminium im Vergleich zu Edelstahl einen höheren ioneninduzierten Sekundärelektronenemissionskoeffizienten besitzt. Somit werden im Vergleich zu der geerdeten Elektrode pro auftreffendem Ion mehr Sekundärelektronen in der Randschicht vor der getriebenen Elektrode erzeugt. Die Sekundärelektronen vervielfältigen sich durch Stöße in der Randschicht und es sind mehr Ladungsträger im Plasma vorhanden, die zu dem Anregungsmaximum beitragen können.

Wird die angelegte Peak-to-Peak-Spannung auf 700 $V_{pp}$ beziehungsweise 600 $V_{pp}$ verringert, findet ein Übergang von dem Penning-Modus in den $\Omega$-Modus statt, wie in Abbildung 4.3 d) und g) beobachtet werden kann. Bei einer ausreichenden Reduzierung der angelegten Spannung können die Elektronen nicht mehr genügend Energie aufnehmen, um Anregungsstöße in der Randschicht auszuführen, da sich das elektrische Feld in der Randschicht ebenfalls verringert. Folglich zeigt sich eine Verringerung der Anregungsmaxima des Penning-Modus in den Randschichtbereichen. Schließlich nimmt der Heizungsmechanismus des $\Omega$-Modus zu, der durch ein hohes Driftfeld im Plasmabulk hervorgerufen wird und durch ein Anregungsmaximum während des Randschichtkollaps gekennzeichnet ist . Die Mehrzahl der energetischen Elektronen (über 22,7 eV) resultieren dann aus dem hohen Driftfeld. In Abbildung 4.3 g) ist zudem zu erkennen, dass die Intensität der beiden Elektronenstoßanregungsmaxima (während der Zeiträume 10 ns - 20 ns und 50 ns - 60 ns), die durch den Heizungsmechanismus des $\Omega$-Modus entstehen, identisch sind. Dieser Effekt zeigt, dass der $\Omega$-Modus nicht von Oberflächenprozessen beeinflusst wird.

Die zweite Spalte in Abbildung 4.3 zeigt die Ergebnisse für eine Erhöhung des Sauerstoffflusses auf 2 ccm für verschiedene angelegte Peak-to-Peak-Spannungen. Vergleicht man diese Ergebnisse mit den Ergebnissen für einen Sauerstofffluss von 0,5 sccm, so erkennt man beispielsweise in Abbildung 4.3 e) für 700 $V_{pp}$, dass die Anregungsmaxima im Randschichtbereich durch den Heizungsmechanismus im Penning-Modus deutlich geringer sind und eine stärkere Anregung im Plasmabulk durch den $\Omega$-Modus (markierte Bereiche) zu beobachten ist. Die schwachen Anregungsmaxima des Penning-Modus zeigen jedoch auch hier eine Asymmetrie in der Intensität. In der dritten Spalte in Abbildung 4.3 sind die Ergebnisse für einen Sauerstofffluss von 5 sccm für verschiedene angelegte Peak-to-Peak-Spannungen abgebildet und der $\Omega$-Modus ist im Vergleich zu einem Sauerstofffluss von 2 sccm noch deutlicher zu erkennen.

Vergleicht man zum Beispiel Abbildung 4.3 e) und f), so befindet sich die Plasmaentladung bei einem Sauerstofffluss von 5 sccm und einer Peak-to-Peak-Spannung von 700 $V_{pp}$ im reinen $\Omega$-Modus, wohingegen die Plasmaentladung bei einem Sauerstofffluss von 2 sccm zusätzlich eine schwache Anregung in den Randschichten durch den Heizungsmechanismus des Penning-Modus aufweist. Wird ein reaktives Gas in einer größeren Menge wie zum Beispiel 5 sccm $O_2$ hinzugefügt, so ist die Anzahl der Sauerstoffmoleküle im Plasma im Vergleich zu einer Plasmaentladung mit einer Sauerstoffzugabe von 0,5 sccm deutlich höher und es finden mehr Stöße statt. Die Teilchen können nicht mehr genügend Energie aufnehmen, die für die Erzeugung von Heliummetastabilen benötigt wird. Somit werden durch die geringere Anzahl an Heliummetastabilen auch weniger Elektronen durch die Penning-Ionisation erzeugt. Außerdem können die in der Randschicht erzeugten Elektronen durch die zunehmenden inelastischen Stöße nicht mehr genug Energie aufnehmen, um sich durch Stöße weiter zu vervielfältigen [6, 24]. Außerdem befinden sich durch die erhöhte Zugabe von $O_2$ mehr negative Ionen im Plasma und die Elektronendichte im Plasmabulk sinkt. Aus der geringeren Elektronendichte resultiert eine geringere DC-Leitfähigkeit und somit stellt sich im Plasmabulk ein hohes Driftfeld ein [24].

Außerdem erkennt man eine Asymmetrie in der Intensität der Anregungsmaxima in Abbildung 4.3 f), die dort mit Pfeilen markiert ist. Dieser Effekt kann beispielsweise bei einem Sauerstofffluss von 0,5 sccm nicht beobachtet werden. Durch eine größere Beimischung von Sauerstoff, wie zum Beispiel 5 sccm, können viel mehr reaktive Sauerstoffspezies im Plasmabulk erzeugt werden, die dann die Elektrodenoberfläche erreichen können. Durch Oberflächenprozesse, die abhängig von dem Elektrodenmaterial sind, kann es zu einem unterschiedlich starken Verlust der reaktiven Sauerstoffspezies an den beiden Elektroden kommen. Außerdem kann diese Asymmetrie auch durch die erhöhte Elektronegativität bei einem Sauerstofffluss von 5 sccm begründet werden. Aus dem Elektrodenmaterial Aluminium (getriebene Elektrode) können mehr Sekundärelektronen herausgelöst werden im Vergleich zu einer Elektrode aus Edelstahl. Das verstärkte Driftfeld (Drift-Ambipolar-Modus) beschleunigt dann die Elektronen und das Anregungsmaximum vor der getriebenen Elektrode ist somit stärker als vor der geerdeten Elektrode. Die experimentellen Ergebnisse können für diesen Effekt keine endgültige Antwort finden, somit sind für eine umfangreiche Untersuchung PIC/MCC-Simulationen nötig.

Die Abbildung 4.4 a) - i) zeigt die Ergebnisse der raum- und phasenaufgelösten Plots der Elektronenstoßanregungsrate für einen COST-Jet mit einer getriebenen Elektrode aus Kupfer in Abhängigkeit der Peak-to-Peak-Spannung (Zeilen) und der Beimischung des Sauerstoffflusses (Spalten). Die Elektronenstoßanregung zeigt ein ähnliches Verhalten im Vergleich zu den vorherigen Ergebnissen für eine getriebene Elektrode aus Aluminium. Auch hier zeigt sich im Penning-Modus eine Asymmetrie der Anregungsmaxima in den Randschichtbereichen, wie beispielsweise in Abbildung 4.4 a) zu erkennen ist. In Abbildung 4.4 f) beobachtet man für einen hohen Sauerstofffluss ebenfalls eine Asymmetrie der Anregungsmaxima im $\Omega$-Modus.

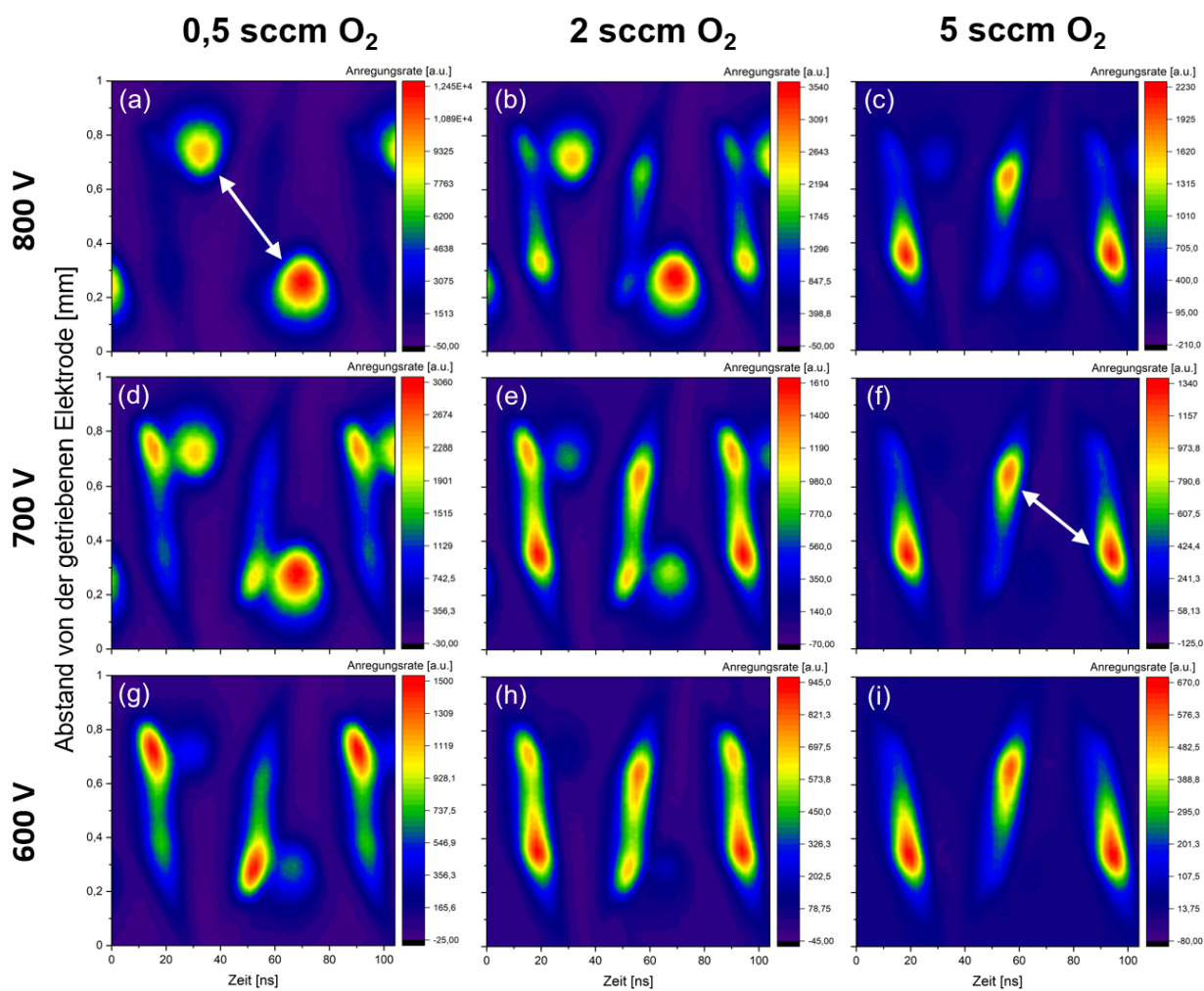

Abbildung 4.4: Raum- und phasenaufgelöste Plots der Elektronenstoßanregung der 706,5 nm He I-Linie für einen COST-Jet mit einer planparallelen Elektrodenanordnung aus den Materialien Kupfer (getriebene Elektrode) und Edelstahl (geerdete Elektrode) in Abhängigkeit der Peak-to-Peak-Spannung und der Beimischung des Sauerstoffflusses. Der Heliumgasfluss beträgt 1 slm.

Außerdem werden die PROES-Messungen mit einem konstanten Heliumfluss von 1 slm und einer Stickstoffbeimischung von 0,5 - 5 sccm wiederholt. Die Abbildung 4.5 a) - i) zeigt die raum- und phasenaufgelösten Plots der Elektronenstoßanregungsrate für einen COST-Jet mit einer planparallelen Elektrodenanordnung aus den Materialien Aluminium und Edelstahl. Die Zeilen zeigen die Ergebnisse in Abhängigkeit der Peak-to-Peak-Spannung und die Spalten in Abhängigkeit der Beimischung des Stickstoffflusses. Wie in den vorherigen Ergebnissen der Elektronenstoßanregungsrate für eine Sauerstoffbeimischung zeigt sich auch bei einer Stickstoffbeimischung ein Einfluss des Elektrodenmaterials auf den Heizungsmechanismus im Penning-Modus. So ist hier ebenfalls eine Asymmetrie in der Intensität der Maxima der Elektronenanregungsrate im Penning-Modus zu erkennen, wie beispielsweise Abbildung 4.5

a) zeigt. So ist die Intensität des Anregungsmaximums vor der getriebenen Elektrode ebenfalls deutlich größer im Vergleich zum Anregungsmaximum vor der geerdeten Elektrode.

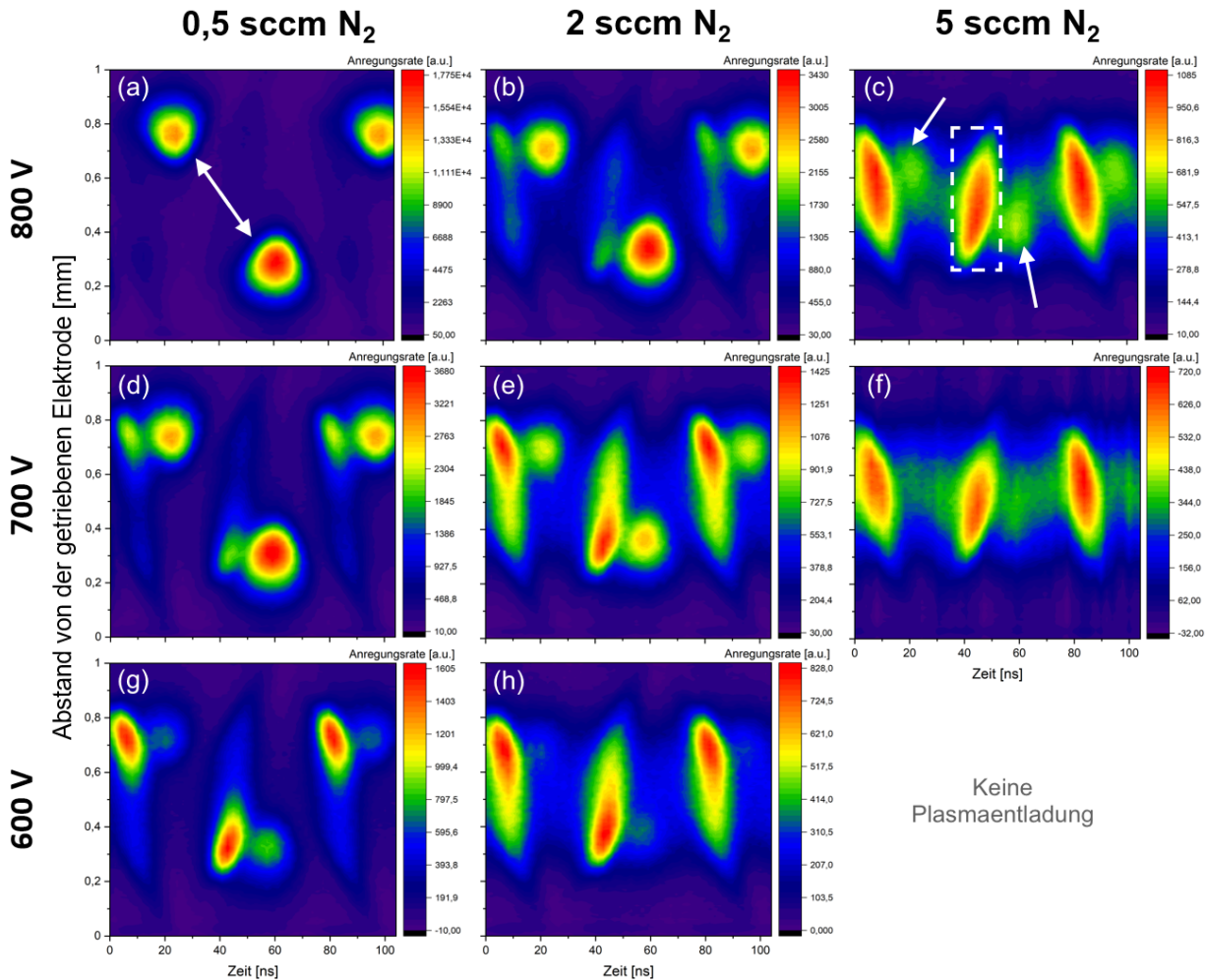

Abbildung 4.5: Raum- und phasenaufgelöste Plots der Elektronenstoßanregung der 706,5 nm He I-Linie für einen COST-Jet mit einer planparallelen Elektrodenanordnung aus den Materialien Aluminium (getriebene Elektrode) und Edelstahl (geerdete Elektrode) in Abhängigkeit der Peak-to-Peak-Spannung und der Beimischung des Stickstoffflusses. Der Heliumgasfluss beträgt 1 slm.

Die Abbildung 4.6 a) - i) zeigt die raum- und phasenaufgelösten Plots der Elektronenstoßanregungsrate für eine planparallele Elektrodenanordnung aus den Materialien Kupfer und Edelstahl. Auch hier zeigt sich ein sehr ähnliches Verhalten im Vergleich zu einer Elektrodenanordnung mit einer getriebenen Elektrode aus Aluminium. Wird die Stickstoffbeimischung erhöht, findet ebenso ein Übergang von dem Penning-Modus in den Ω-Modus statt. In der Abbildung 4.5 c) beziehungsweise Abbildung 4.6 c) ist die Elektronenstoßanregungsrate für einen Stickstofffluss von 5 sccm und Peak-to-Peak-Spannungs-amplituden von 800 V$_{pp}$ gezeigt. Aus den Abbildungen ist zu erkennen, dass die Elektronenstoßanregung im Plasmabulk während der Zeiträume 30 ns und 50 ns dominiert (markiert durch einen in weiß gestrichelten Bereich). Zusätzlich ist eine schwache Anregung des Penning-Modus zu erkennen (mit Pfeilen markiert). Die Elektronenstoßanregung zeigt hier im Ω-Modus ein symmetrisches Muster und ist somit nicht abhängig von den Elektrodenmaterialien beziehungsweise den Oberflächenprozessen.

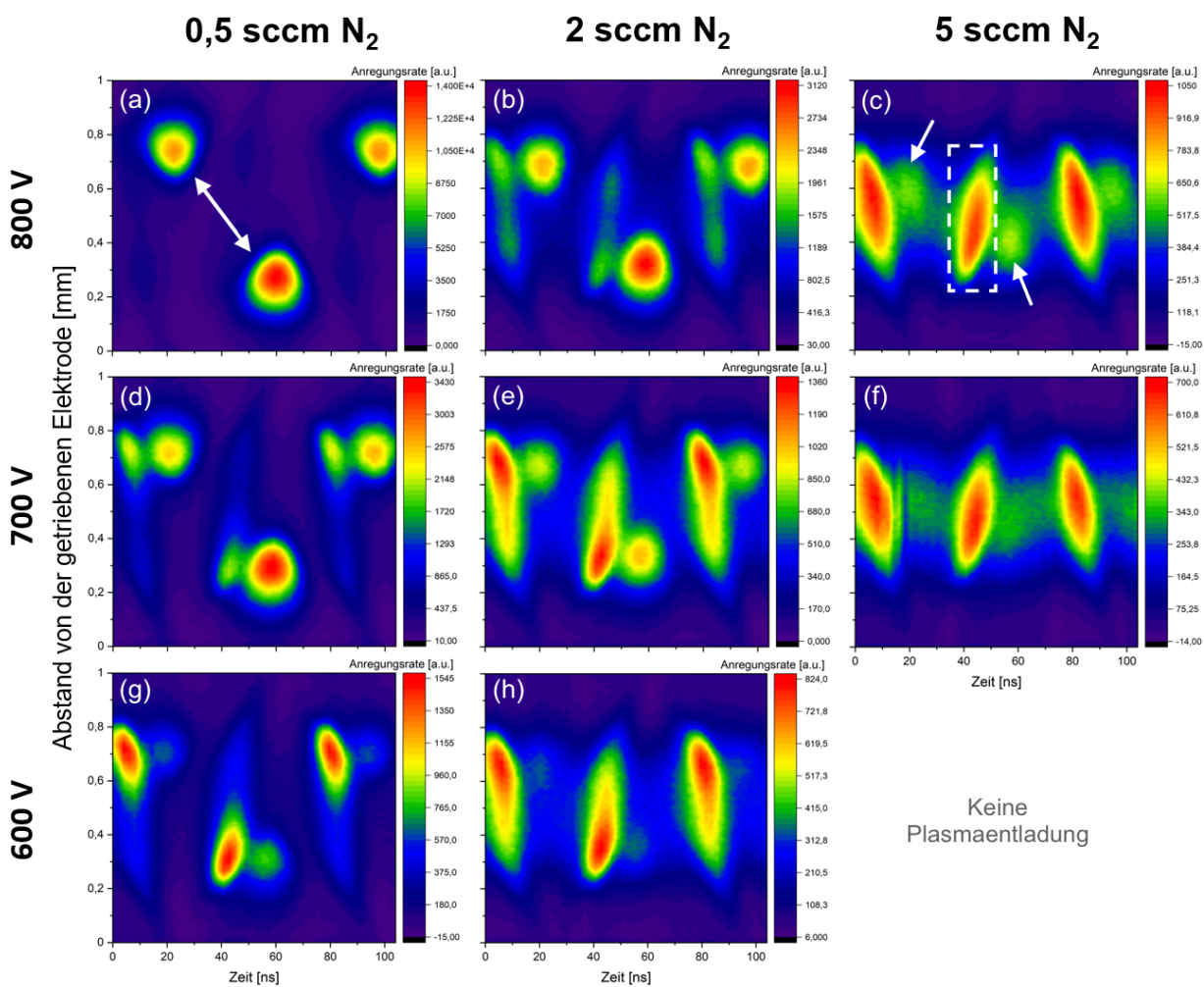

Abbildung 4.6: Raum- und phasenaufgelöste Plots der Elektronenstoßanregung der 706,5 nm He I-Linie für einen COST-Jet mit einer planparallelen Elektrodenanordnung aus den Materialien Kupfer (getriebene Elektrode) und Edelstahl (geerdete Elektrode) in Abhängigkeit der Peak-to-Peak-Spannung und der Beimischung des Stickstoffflusses. Der Heliumgasfluss beträgt 1 slm.

Außerdem erkennt man einen Unterschied in dem Elektronenstoßanregungsmuster, wenn man die Ergebnisse für die Sauerstoffbeimischung mit denen der Stickstoffbeimischung vergleicht. Der Unterschied wird anhand der Abbildung 4.6 f) und der Abbildung 4.4 f) (5 sccm $N_2/O_2$, 700 $V_{pp}$) erklärt. Die Anregungsmaxima der Plasmaentladung sind für die Stickstoffbeimischung deutlich ausgeprägter im Plasmabulk zu beobachten. Zudem ist für die Stick-stoffbeimischung eine schwache Anregung des Penning-Modus zu erkennen, wohin gegen sich die Plasmaentladung für die Sauerstoffbeimischung im reinen Ω-Modus befindet. Dieser Unterschied lässt sich durch die verschiedenen Elektronenstoßquerschnitte erklären die abhängig von der Gaszusammensetzung sind. So werden beispielsweise für inelastische Stöße (dazu zählen zum Beispiel Anregungs- und Ionisationsstöße) in einem reinen Helium-Plasma Elektronenenergien von mindestens 20 eV benötigt. Bei einem inelastischen Stoß verliert ein Elektron seine Energie. Bei einem Plasma mit einer Sauerstoff- beziehungsweise Stickstoffbeimischung ergeben sich jedoch andere Elektronenstoßquerschnitte. Die Abbildung 4.7 a) und b) zeigt eine Zusammenfassung der Elektronenstoßquerschnitte mit Sauerstoff- und Stickstoffmolekülen [45, 46].

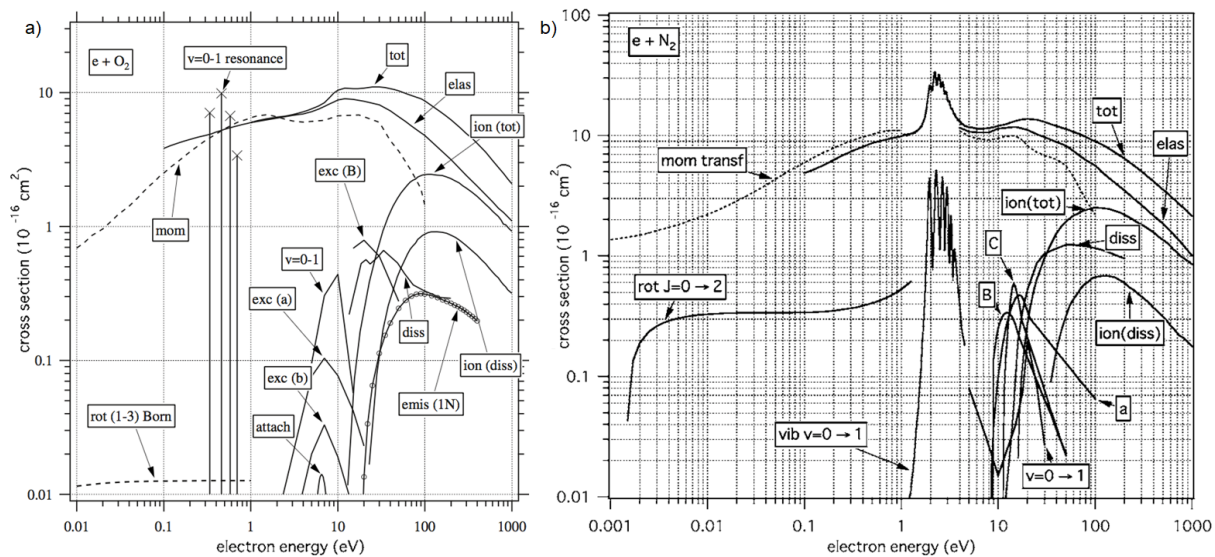

Abbildung 4.7: Zusammenfassung der Elektronenstoßquerschnitte mit a) Sauerstoffmolekülen und b) Stickstoffmolekülen [45, 46].

So ist in Abbildung 4.7 a) zu erkennen, dass Elektronen schon mit einer Energie von circa 2,5 eV ein Sauerstoffteilchen anregen können. Für die Anregung eines Stickstoffmoleküls benötigt ein Elektron hingegen circa 9 eV, wie in Abbildung 4.7 b) gezeigt ist. Somit werden in einem Plasma mit einer Beimischung eines molekularen Gases im Vergleich zu einem reinen Helium-Plasma Anregungsstöße mit einer geringeren Elektronenenergie ausgeführt. Da die Energieschwelle für einen Anregungsstoß mit einem Sauerstoffmolekül geringer ist im Vergleich zu einem Anregungsstoß mit einem Stickstoffmolekül, verlieren bei einer Sauerstoffbeimischung viel mehr Elektronen ihre Energie. Diese Elektronen fehlen schließlich im Plasma, um Anregungsstöße auszuführen. Außerdem können Elektronen mit einer Energie von circa 6 eV mit Sauerstoffmolekülen Anlagerungsstöße ausführen, wodurch negative Ionen entstehen und somit Elektronen verloren gehen.

Zusammenfassend lässt sich durch die Asymmetrie der Anregungsmaxima im Penning-Modus zeigen, dass durch die verschiedenen Elektrodenmaterialien die Heizungsdynamik des COST-Jets beeinflusst werden kann. Die Heizungsdynamik im Ω-Modus ist für eine Stickstoffbeimischung unabhängig von Oberflächenparametern. Nur für eine größere Beimischung eines

Sauerstoffflusses zeigt sich im Ω-Modus eine leichte Asymmetrie in der Intensität der Anregungsmaxima im Plasmabulk. Für eine genauere Untersuchung sollten jedoch zusätzlich PIC/MCC-Simulationen durchgeführt werden.

### 4.1.2 Einseitig rechteckig strukturierte Elektrodenanordnung aus Edelstahl

In diesem Kapitel werden zunächst die Ergebnisse der PROES-Messungen des COST-Jets mit verschiedenen rechteckig strukturierten Elektrodenkonfigurationen aus Edelstahl vorgestellt, um den Einfluss der verschiedenen Strukturbreiten auf die Plasmaentladung des COST-Jets zu untersuchen. Die untersuchten Elektrodenkonfigurationen sind in Abbildung 4.8 a) - c) schematisch dargestellt. Die Breite der rechteckigen Strukturen wird zu 3 mm, 1 mm und 0,5 mm gewählt. Die Tiefe eines rechteckigen Strukturgrabens beträgt konstant 1 mm. Die Elektrodenanordnung ist so gewählt, dass die getriebene Elektrode eine planare Elektrode aus Edelstahl ist und die geerdete Elektrode rechteckige Strukturen besitzt. Die roten Markierungen in Abbildung 4.8 a) - c) zeigen die ausgewählten Aufnahmebereiche der ICCD-Kamera.

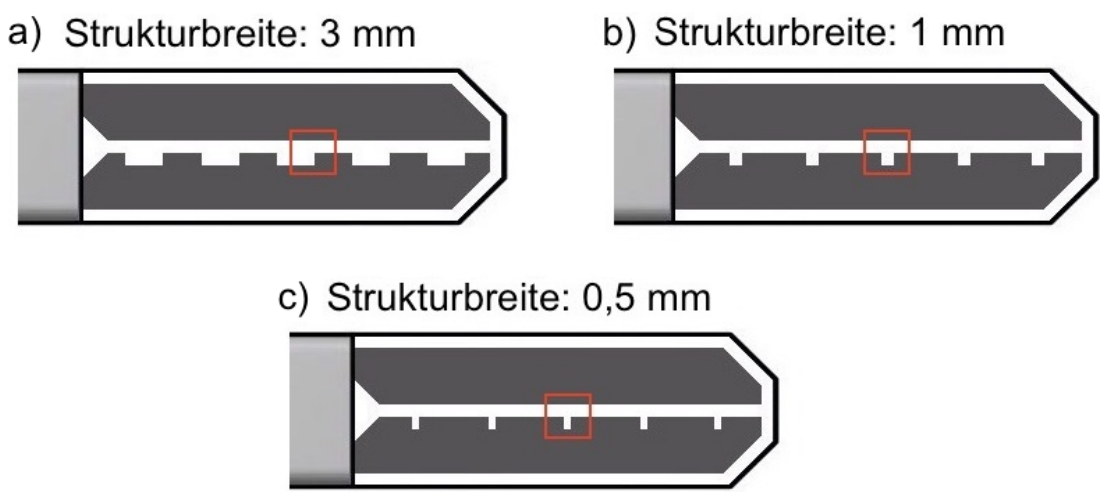

Abbildung 4.8: Überblick der rechteckig strukturierten Elektrodenanordnungen aus Edelstahl in Abhängigkeit der Strukturbreite. Die roten Markierungen zeigen die Aufnahmebereiche der ICCD-Kamera.

Die Ergebnisse werden zu Beginn für einen konstanten Gasfluss von 1 slm He und 0,5 sccm $O_2$ vorgestellt. Es wird die maximal erreichbare Peak-to-Peak-Spannung im Penning-Modus gewählt, die für die verschiedenen Elektrodenanordnungen bei circa 780 $V_{pp}$ liegt. Der Betrieb im Penning-Modus ermöglicht die Untersuchung von Oberflächeneffekten, wie bereits in Kapitel 4.1.1 gezeigt wurde. Die Abbildung 4.9 a) - c) zeigt die Fotos der Plasmaentladungen für die verschiedenen Strukturbreiten. Auf den Fotos erkennt man für die Plasmaentladungen eine typische inhomogene Leuchtemission im Penning-Modus, das heißt, in den Randschichtbereichen ist die Leuchtemission ausgeprägter als im Plasmabulk.

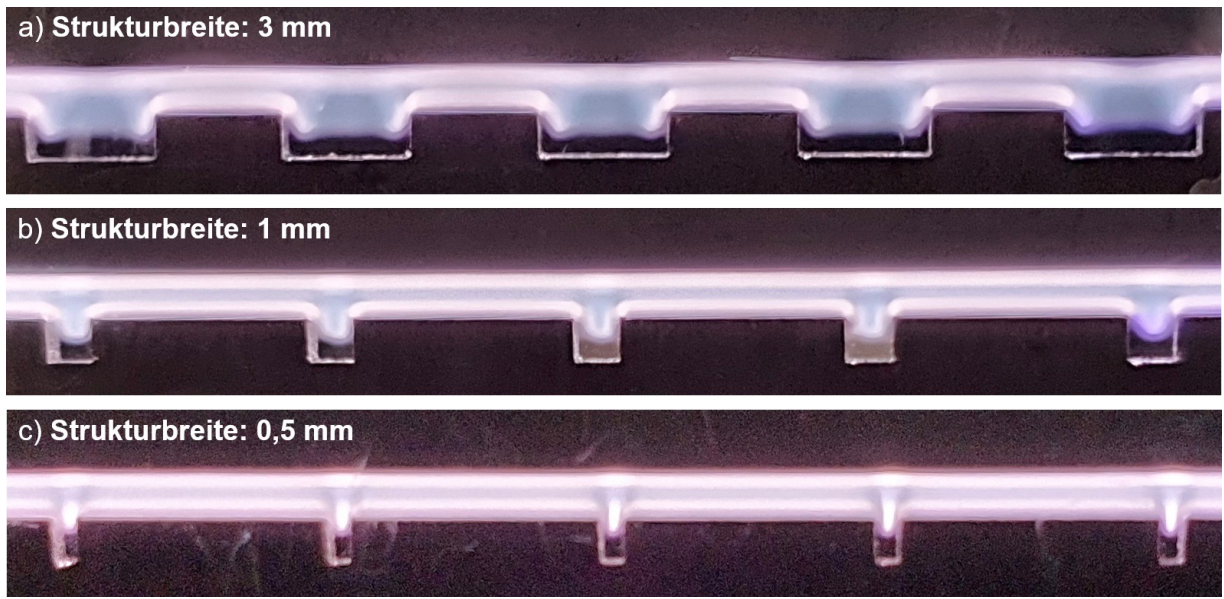

Abbildung 4.9: Fotos der Plasmaentladungen für verschiedene Strukturgrößen bei einem Gasfluss von 1 slm He und 0,5 sccm $O_2$ und einer Peak-to-Peak-Spannung von 780 $V_{pp}$.

Das Plasma kann für die gewählten Kontrollparameter grundsätzlich in jede der Strukturen eindringen, jedoch nur circa bis zur Hälfte der maximalen Tiefe einer Struktur. Die Wahl der Strukturbreite hat ebenfalls einen Einfluss auf die Leuchtemission der Plasmaentladung. Bei einer Strukturbreite von 3 mm, wie sie in Abbildung 4.9 a) verwendet wird, ist die Leuchtemission innerhalb der Strukturen schwächer im Vergleich zu den außerhalb liegenden Bereichen, wie zum Beispiel vor der gegenüberliegenden planaren Elektrode. Der Abstand von dem Strukturboden zu der gegenüberliegenden planaren Elektrode beträgt 2 mm. Wird die Strukturbreite zu 1 mm reduziert, wie in Abbildung 4.9 b) gezeigt, so erkennt man eine Zunahme der Leuchtemission innerhalb der Struktur. Abbildung 4.9 c) zeigt die Plasmaentladung für eine Elektrodenanordnung mit einer minimalen Strukturbreite von 0,5 mm. Hier kann eine erhöhte lokale Leuchtemission innerhalb der Struktur und an der gegenüberliegenden planaren Elektrode beobachtet werden. Des Weiteren erkennt man für die Strukturbreiten von 3 mm und 1 mm eine erhöhte Leuchtemission an den Ecken der Strukturen.

Die Abbildung 4.10 a) - i) zeigt die raumaufgelösten Plots der Emission der 706,5 nm He I-Linie im Bereich einer Struktur für ausgewählte Zeitpunkte innerhalb einer RF-Periode (74 ns), die durch das PROES-Verfahren erzeugt werden. Das erzeugte Bild wird auf dem CCD-Sensor auf Pixel abgebildet und ist dabei um 180° verdreht. Der Gasfluss beträgt ebenfalls konstant 1 slm He und 0,5 sccm $O_2$ und es wird die maximal erreichbare Peak-to-Peak-Spannung im Penning-Modus gewählt (780 $V_{pp}$). So ist in Abbildung 4.10 a) - c) die raumaufgelöste Emission für eine Struktur mit einer Breite von 3 mm für ausgewählte Zeitpunkte abgebildet. Für die Zeitpunkte t = 11 ns und t = 21 ns ist die Randschicht an der geerdeten Elektrode expandiert und man erkennt an der Kante des Struktureinganges ein lokales Emissionsmaximum. Das Auftreten eines Emissionsmaximums in diesem Bereich kann durch den sogenannten Kanteneffekt erklärt werden, der beispielsweise auch an den Rändern einer parallelen Kondensatorplattenanordnung beobachtet werden kann. Bei diesem Effekt tritt an dem Rand einer Elektrode ein inhomogener Verlauf der elektrischen Feldlinien auf, der zu einer erhöhten elektrischen Feldstärke in diesem Bereich führt.

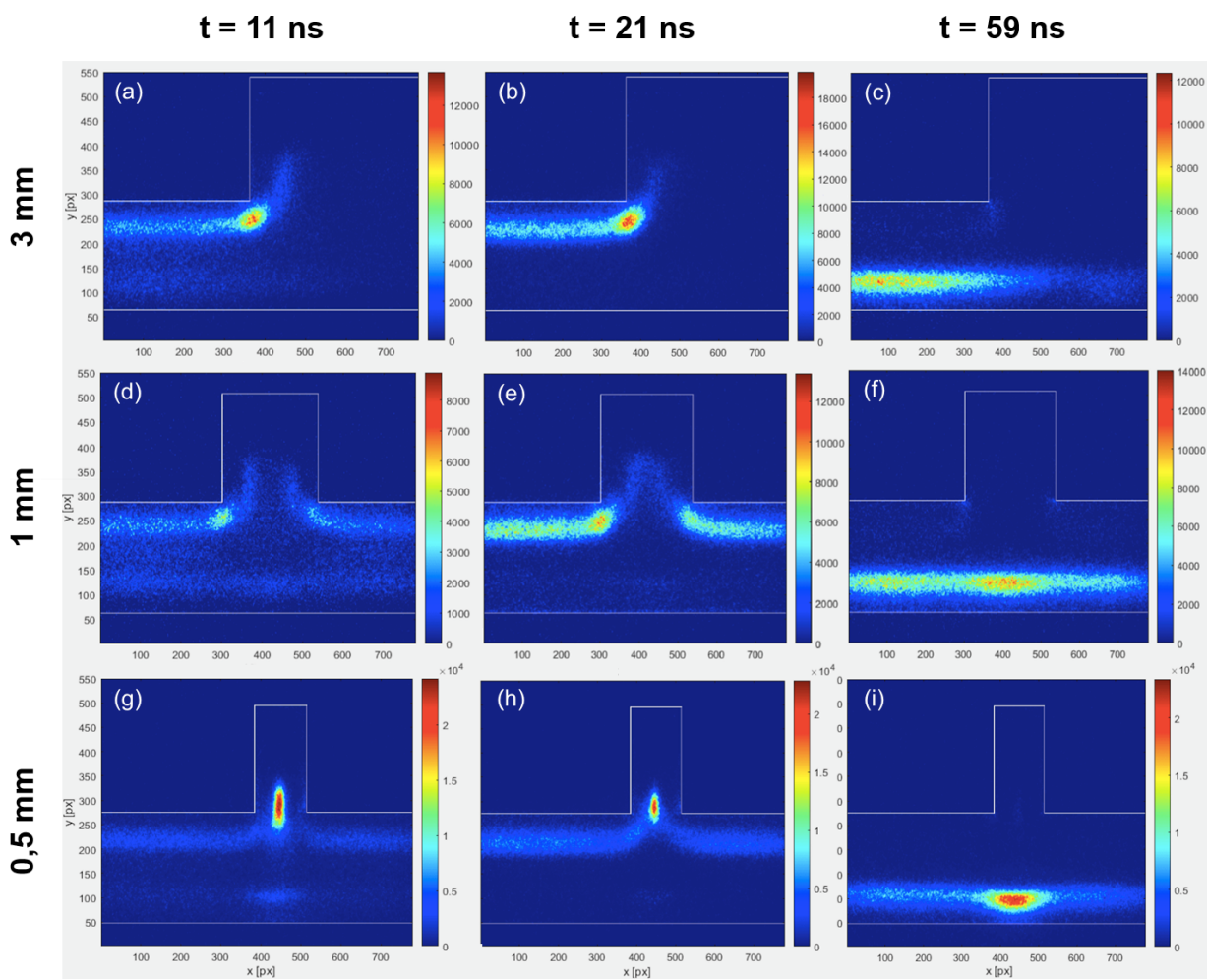

Abbildung 4.10: Raumaufgelöste Plots der Emission der 706,5 nm He I-Linie einer Struktur mit einer Breite von 3 mm, 1 mm und 0,5 mm für ausgewählte Zeitpunkte innerhalb einer RF-Periode (74 ns). Der Gasfluss beträgt 1 slm He und 0,5 sccm $O_2$ und die Peak-to-Peak-Spannung beträgt 780 $V_{pp}$.

Überträgt man den Kanteneffekt auf die gezeigte Plasmaentladung des COST-Jets, so führt die Strukturkante zu einem lokal erhöhten elektrischen Feld in der expandierten Randschicht vor der geerdeten Elektrode, wodurch die Elektronen in diesem Bereich beschleunigt werden. Diese Elektronen können dann im Randschichtbereich andere Teilchen anregen und ionisieren. Somit haben die Teilchen in diesem Bereich genug Energie, um zum Beispiel auch durch die Penning-Ionisation Sekundärelektronen zu erzeugen. Innerhalb der Struktur kann nur eine schwache Emission direkt neben dem lokalen Emissionsmaximum beobachten werden. Für den Zeitpunkt t = 59 ns ist nun die Randschicht vor der geerdeten Elektrode kollabiert und vor der getriebenen Elektrode (planare Elektrode) expandiert. Man erkennt vor der planaren Elek-trode im Randschichtbereich eine Emission, jedoch nur in dem Bereich vor dem Struktureingang. In diesem Bereich beträgt der Abstand zwischen den Elektroden 1 mm. Somit werden innerhalb der Struktur eine nicht ausreichende Anzahl an angeregten Teilchen erzeugt, da hier der Plattenabstand nach dem Paschen-Kriterium für eine Plasmazündung zu groß ist.

In Abbildung 4.10 d) - f) ist die raumaufgelöste Emission der 706,5 nm He I-Linie für eine Struktur mit einer Breite von 1 mm für ausgewählte Zeitpunkte abgebildet. Für die Zeitpunkte t = 11 ns und t = 21 ns erkennt man an beiden Kanten der Struktur ein Emissionsmaximum im Randschichtbereich, welches durch das lokal erhöhte elektrische Feld an den Kanten verursacht wird. Durch die nun verkleinerte Strukturbreite liegen die beiden gegenüberliegenden Kanten so nahe beieinander, dass sich die beiden lokal erhöhten elektrischen Felder überlagern können. Dadurch werden die Elektronen effektiver geheizt und haben genug Energie, um nun auch Teilchen innerhalb der Struktur anzuregen und zu ionisieren. Für den Zeitpunkt t = 59 ns ist die Randschicht vor der geerdeten Elektrode wieder kollabiert und vor der getriebenen Elektrode expandiert. Nun befindet sich ein Emissionsmaximum auf der gegenüberliegenden Seite der Strukturöffnung und somit vor der planaren Elektrode. Eine Erklärung für das Emissionsmaximum kann mit der Flusserhaltung gefunden werden, die für die Aufrechterhaltung eines Plasmas sorgt. Durch die angelegte RF-Spannung herrscht in der gesamten Plasmaentladung eine ausgeprägte Randschichtdynamik und somit auch innerhalb der Struktur. Dort können während der kollabierten Randschicht vor der geerdeten Elektrode mehr Teilchen abfließen, da das Oberflächenverhältnis größer im Vergleich zu der gegenüberliegenden planaren Elektrode ist. Der Fluss der Elektronen und der Ionen muss sich jedoch an den Elektroden im zeitlichen Mittel einer RF-Periode kompensieren ($<\Gamma_e> = <\Gamma_i>$), damit der Plasmazustand aufrechterhalten bleibt. Somit stellt sich ein elektrisches Feld zur Flusserhaltung vor der planaren Elektrode ein. Die so beschleunigten Elektronen können dann Anregungs- und Ionisationsstöße ausführen.

In Abbildung 4.10 g) - i) ist die raumaufgelöste Emission der 706,5 nm He I-Linie für eine Struktur mit einer Breite von 0,5 mm für ausgewählte Zeitpunkte innerhalb einer RF-Periode gezeigt. Für die Zeitpunkte t = 11 ns und t = 21 ns erkennt man ein Emissionsmaximum im Randschichtbereich der expandierten Randschicht, welches innerhalb der Struktur fokussiert ist. Die Strukturbreite ist nun so gering, dass sich die lokal erhöhten elektrischen Felder an den Kanten noch effektiver überlagern können. Durch die Überlagerung der elektrischen Felder werden die Elektronen für diese Strukturbreite noch stärker beschleunigt als bei einer Strukturbreite von 1 mm und somit kann noch effektiver angeregt und ionisiert werden. Außerdem führen die elektrischen Felder bei einer kleineren Strukturbreite zu einer Fokussierung der Anregung innerhalb der Struktur. Für den Zeitpunkt t = 59 ns ist nun die Randschicht vor der geerdeten Elektrode kollabiert und es befindet sich hier nun ebenfalls ein Emissionsmaximum auf der gegenüberliegenden Seite der Strukturöffnung, welches durch die Flusserhaltung entsteht.

Um den Einfluss der rechteckig strukturierten Elektroden auf die Heizungsdynamik der Plasmaentladung noch genauer zu untersuchen, werden im Folgenden die raum- und phasenaufgelösten Plots der Elektronenstoßanregung für die zuvor genannten Kontrollparameter für verschiedene Bereiche der rechteckigen Strukturen vorgestellt. Vergleicht man den Heizungs-mechanismus innerhalb einer Struktur mit dem der Niederdruckplasmen, so erkennt man demzufolge einen signifikanten Unterschied. Wie bereits in Kapitel 2.3.3 erwähnt, werden bei einem Niederdruckplasma durch die oszillierenden Randschichten an den Strukturwänden die Elektronen stoßfrei aus der Struktur beschleunigt [34]. Da jedoch bei einem Atmosphärendruckplasma die Stoßfrequenz deutlich höher ist, kann dieser Mechanismus hier nicht zu einer effektiven Heizung der Elektronen führen.

Die Abbildung 4.11 a) zeigt die schematische Aufnahme von der rechteckigen Struktur mit einer Breite von 3 mm. Für die ausgewählten Bereiche 1 - 3 werden die raum- und phasenaufgelösten Plots der Elektronenstoßanregung, die in Abbildung 4.11 b) - d) dargestellt sind, analysiert.

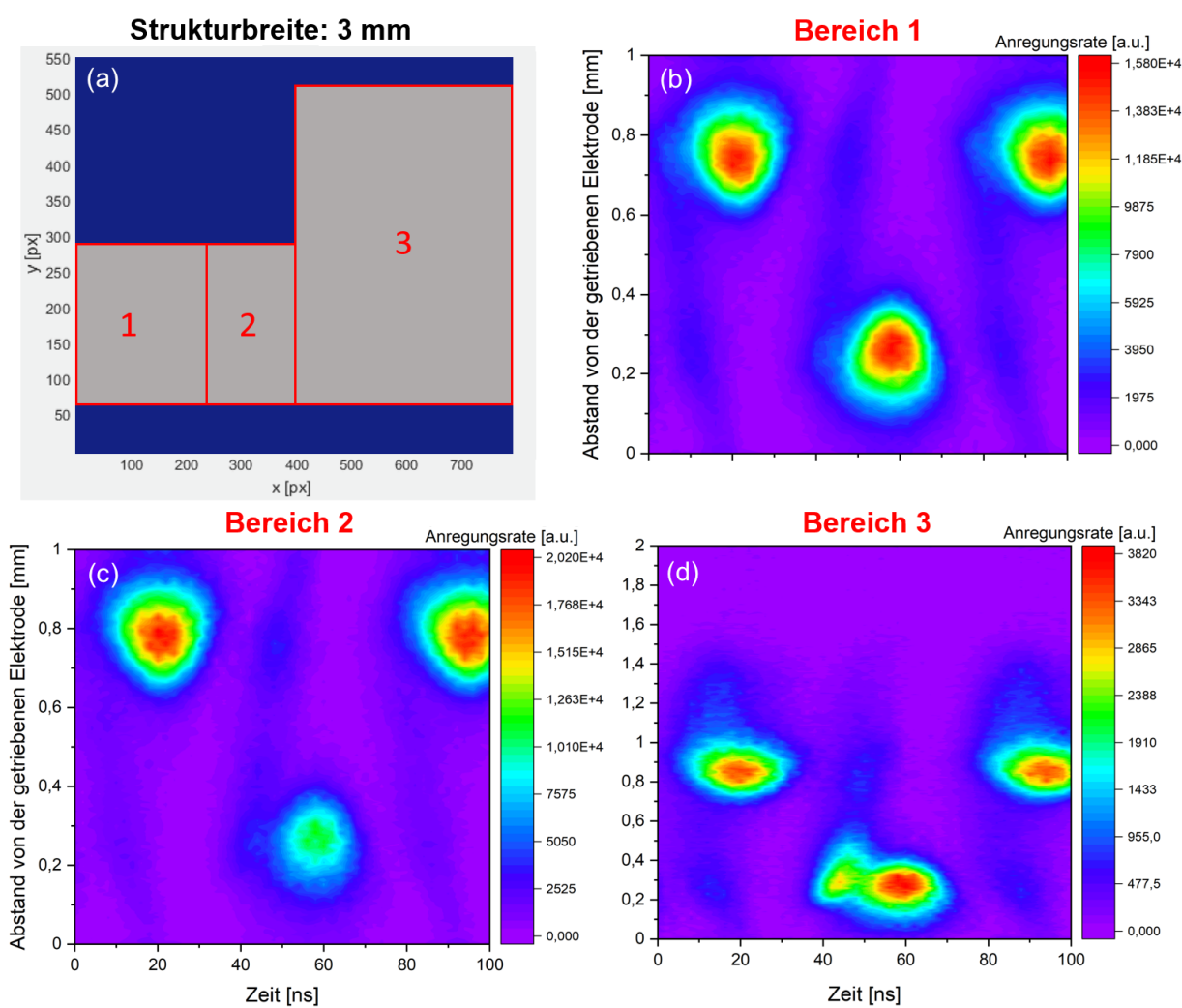

Abbildung 4.11: Raum- und phasenaufgelöste Plots der Elektronenstoßanregung für verschiedene Bereiche in der Plasmaentladung einer rechteckig strukturierten Elektrodenanordnung mit einer Strukturbreite von 3 mm. Die Peak-to-Peak-Spannung beträgt 780 $V_{pp}$ und der Gasfluss beträgt 1 slm He und 0,5 sccm $O_2$.

In Abbildung 4.11 b) ist der raum- und phasenaufgelöste Plot der Elektronenstoßanregung für den Bereich 1 gezeigt, der sich vor der Struktur befindet. Es ist zu erkennen, dass sich die Anregungsmaxima in den Randschichtbereichen befinden und die Plasmaentladung somit im Penning-Modus betrieben wird. Außerdem lässt sich im Bereich 2, der direkt vor dem Struktureingang liegt und für den die Anregungsrate in Abbildung 4.11 c) abgebildet ist, eine Asymmetrie in der Intensität der beiden Anregungsmaxima beobachten. Das heißt, die Intensität des Anregungsmaximums vor der geerdeten Elektrode ist höher als vor der getriebenen Elektrode. Dieser Effekt lässt sich durch das erhöhte elektrische Feld an der Kante der Strukturöffnung im Bereich 2 begründen. Durch dieses können die Elektronen in der Randschicht noch stärker beschleunigt werden, sodass mehr Anregungsstöße und somit mehr Metastabile erzeugt werden. Die Metastabilen können dann anschließend durch Stöße in der Randschicht Atome oder Moleküle ionisieren, welche dann wiederum Sekundärelektronen aus der Elektrode herauslösen können und somit zum Anreungsmaximum beitragen. Der raum- und phasenaufgelöste Plot der Elektronenstoßanregung für den Bereich 3 ist in Abbildung 4.11 d) gezeigt. Der Bereich 3 zeigt die Elektronenstoßanregung innerhalb der Struktur. Abbildung 4.11 c) zeigt den raum- und phasenaufgelösten Plot der Elektronenstoßanregung für den Bereich 2, der außerhalb der Struktur liegt. Der Abstand von der getriebenen Elektrode zur geerdeten Elektrode beträgt hier somit 2 mm. In diesem Bereich erkennt man die Anregungsmaxima des Penning-Modus und zusätzlich in dem Zeitraum von 38 ns - 50 ns ein Anregungsmaximum des Ω-Modus. Somit findet innerhalb der Struktur ein Modenübergang statt, da die Strukturbreite so groß gewählt ist, dass weniger Oberflächenprozesse im Penning-Modus stattfinden können.

In der Abbildung 4.12 a) - d) werden die raum- und phasenaufgelösten Plots der Elektronenstoßanregung für eine Strukturbreite von 1 mm gezeigt. In Abbildung 4.12 b) und d) sind die raum- und phasenaufgelöste Plots der Elektronenstoßanregung für den Bereich 1 und 3 gezeigt, die sich jeweils vor der Struktur befindet. In diesen Bereichen treten die Anregungsmaxima in den Randschichtbereichen auf und die Plasmaentladung befindet sich somit dort im Penning-Modus. In Abbildung 4.12 c) ist der raum- und phasenaufgelöste Plot der Elektronenstoßanregung für den Bereich 2 gezeigt, der innerhalb der Struktur liegt. Hierfür befindet sich die Plasmaentladung im reinen Penning-Modus, in dem jedoch das Anregungsmaximum vor der getriebenen Elektrode deutlich ausgeprägter ist. Das Anregungsmaximum vor der getriebenen Elektrode ergibt sich aus dem lokal erhöhten elektrischen Feld, welches durch die Flusserhaltung erzeugt wird und einen dominierenden Beitrag zu der Elektronenheizung innerhalb der Struktur liefert. Die schwächeren Anregungsmaxima befinden sich circa 0,9 mm von der getriebenen Elektrode entfernt.

In der Abbildung 4.13 a) - d) werden die raum- und phasenaufgelösten Plots der Elektronenstoßanregung für eine Strukturbreite von 0,5 mm gezeigt. In Abbildung 4.13 b) und d) sind die raum- und phasenaufgelösten Plots der Elektronenstoßanregung für den Bereich 1 und 3 gezeigt, die sich jeweils vor der Struktur befinden und für die sich die Plasmaentladung ebenfalls im Penning-Modus befindet. In Abbildung 4.13 c) ist der raum- und phasenaufgelöste Plot der Elektronenstoßanregung für den Bereich 2 gezeigt, der innerhalb der Struktur liegt.

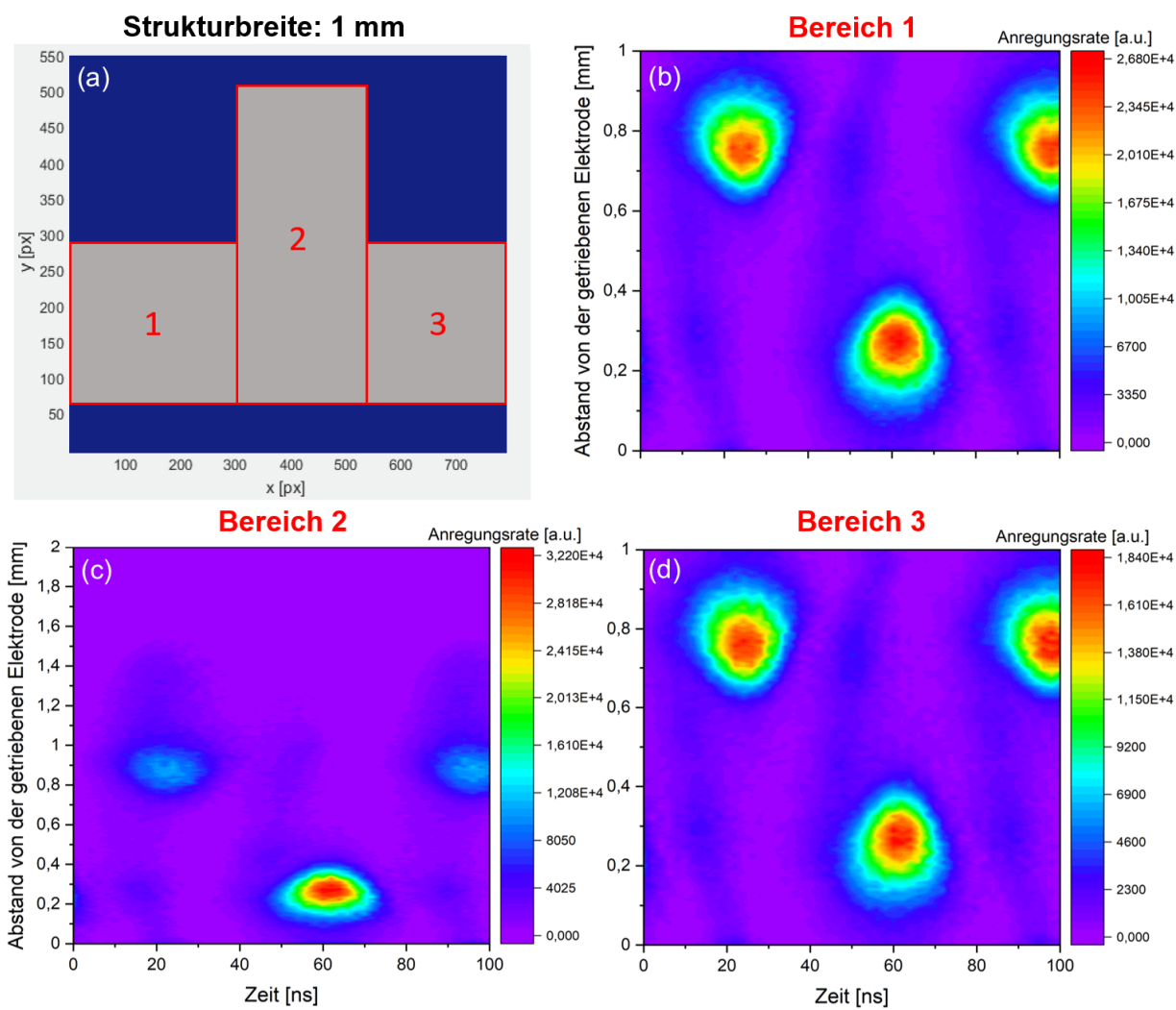

Abbildung 4.12: Raum- und phasenaufgelöste Plots der Elektronenstoßanregung für verschiedene Bereiche in der Plasmaentladung einer rechteckig strukturierten Elektrodenanordnung mit einer Strukturbreite von 1 mm für eine Peak-to-Peak-Spannung von 780 $V_{pp}$. Der Gasfluss beträgt 1000 sccm He und 0,5 sccm $O_2$.

In diesem Bereich befindet sich die Plasmaentladung im reinen Penning-Modus. Das Anregungsmaximum vor der getriebenen Elektrode ist wieder deutlich ausgeprägter, was auf das erhöhte elektrische Feld vor der getriebenen Elektrode zurückzuführen ist. Die schwächeren Anregungsmaxima befinden sich nun 1,1 mm von der getriebenen Elektrode entfernt und somit kann auch eine Anregung innerhalb der Struktur auftreten. Da sich das Anregungsmaximum im Penning-Modus nur innerhalb der Randschicht befindet, kann davon ausgegangen werden, dass die Randschichtdicke vor der getriebenen Elektrode circa 0,3 mm beträgt. Bei einer Strukturgröße von 1 mm ist die Randschicht vor der getriebenen Elektrode 0,4 mm dick. Aus diesen Ergebnissen kann man schließen, dass bei einer Strukturbreite von 0,5 mm die Plas-madichte im Bereich der getriebenen Elektrode größer ist als für eine Strukturbreite von 1 mm in diesem Bereich, da die Randschichtdicke mit der Plasmadichte abnimmt.

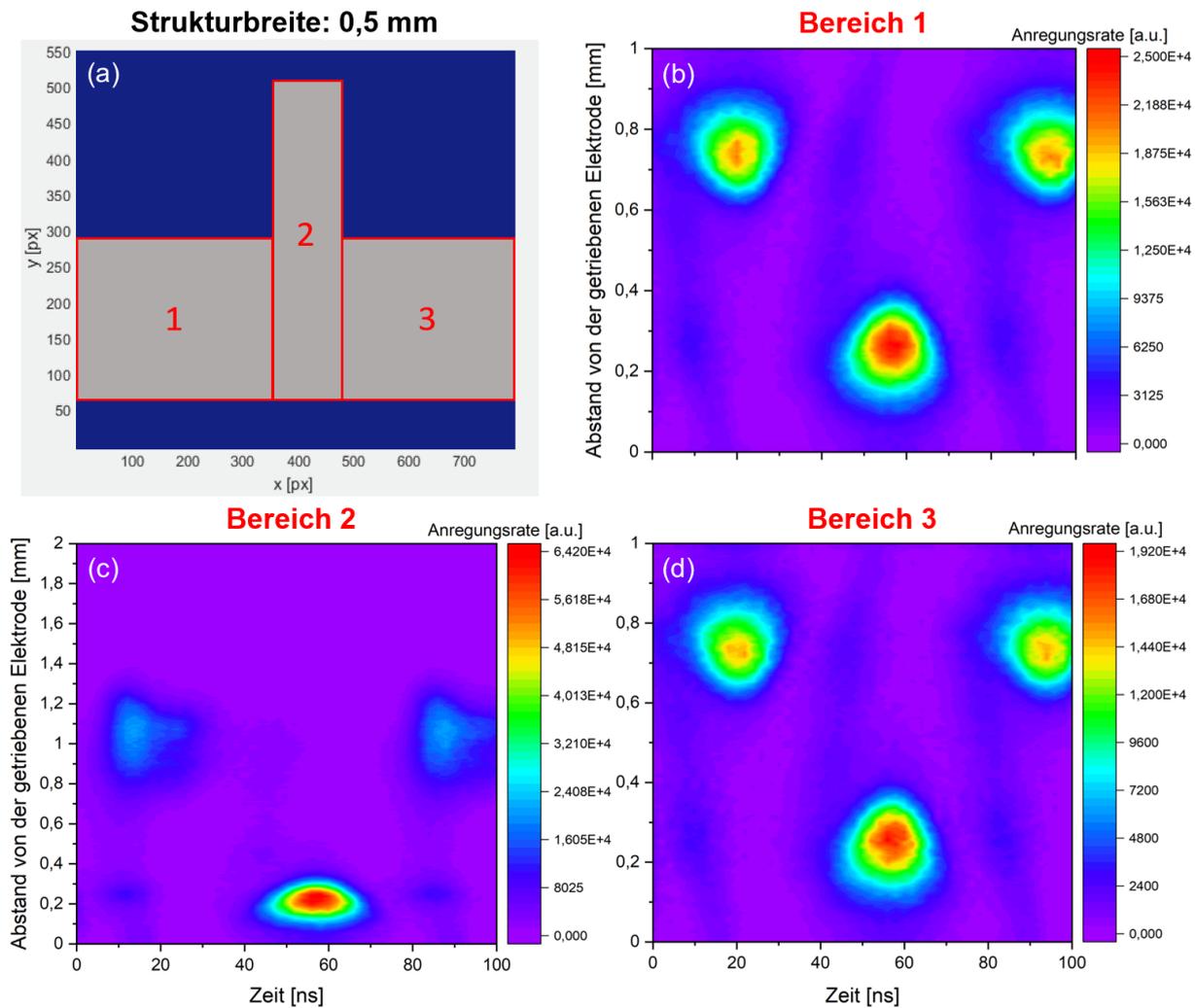

Abbildung 4.13: Raum- und phasenaufgelöste Plots der Elektronenstoßanregung für verschiedene Bereiche in der Plasmaentladung der rechteckig strukturierte Elektrodenanordnung mit einer Strukturenbreite von 0,5 mm für eine Peak-to-Peak-Spannung von 780 $V_{pp}$. Der Gasfluss beträgt 1000 sccm He und 0,5 sccm $O_2$.

Angesichts dieser Ergebnisse liegt die Schlussfolgerung nahe, dass bei einem Gasfluss von 1 slm He und 0,5 sccm $O_2$ und einer angelegten Peak-to-Peak-Spannung von 780 $V_{pp}$ die effektivste Anregung der Teilchen innerhalb der Struktur bei einer Strukturbreite von 0,5 mm beobachtet werden kann. Die Überlagerung der lokalen elektrischen Felder führt zu einer Fokussierung der Anregung innerhalb der Struktur, wodurch mehr Teilchen in diesem Bereich ionisiert werden können und beispielsweise auch mehr Sekundärelektronen aus den Oberflächen herausgelöst werden können. Um die Flusserhaltung und ein Gleichgewicht im Plasma aufrechtzuerhalten, bildet sich auch bei einer Strukturbreite von 0,5 mm das stärkste Anregungsmaximum vor der planaren Elektrode aus.

An dieser Stelle soll die Heizungsdynamik und der Übergang der Betriebsmodi für einen Gasfluss von 1 slm He und 5 sccm $N_2$ in Abhängigkeit der angelegten Peak-to-Peak-Spannungsamplitude untersucht werden. Abbildung 4.14 a) - b) zeigt die Fotos der Plasmaentladungen für eine Strukturbreite von 3 mm für eine angelegte Peak-to-Peak-Spannung von 700 $V_{pp}$ und 900 $V_{pp}$.

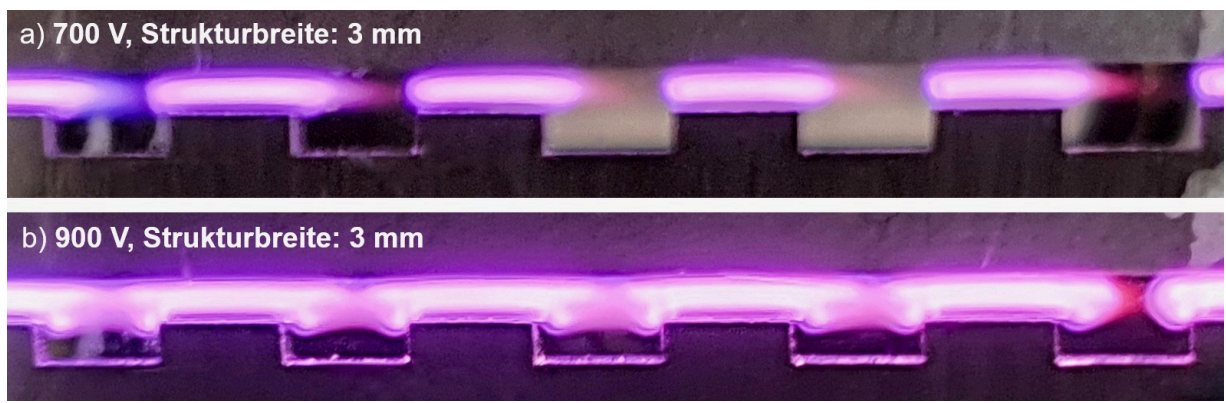

Abbildung 4.14: Fotos der Plasmaentladungen einer Elektrodenanordnung mit einer Strukturbreite von 3 mm bei 1 slm He und 5 sccm $N_2$ und einer Peak-to-Peak-Spannung von a) 700 $V_{pp}$ und b) 900 $V_{pp}$.

Zusätzlich soll anhand Abbildung 4.15 a) - d) das Verhalten der Anregungsrate für die ausgewählten Kontrollparameter für verschiedene Bereiche innerhalb der Plasmaentladung analysiert werden. In Abbildung 4.14 a) ist die Plasmaentladung für 700 $V_{pp}$ gezeigt. Es bildet sich nur in den Bereichen mit einem Elektrodenabstand von 1 mm ein Plasma aus. Innerhalb der Struktur kann keine Leuchtemission festgestellt werden. Außerdem erkennt man, dass die Leuchtemission immer nur an einer Seite über der Strukturöffnung auftritt. Somit zeigt sich hier eine Abhängigkeit der Richtung des Gasflusses. Der Gasfluss fließt auf dem Foto von rechts nach links innerhalb des abgebildeten Entladungskanals.

In Abbildung 4.15 b) ist der raum- und phasenaufgelöste Plot der Elektronenstoßanregung außerhalb der Struktur gezeigt. In diesem Bereich befindet sich die Plasmaentladung für eine angelegte Peak-to-Peak-Spannung von 700 $V_{pp}$ im Ω-Modus. Die Anregungsmaxima befinden sich ausschließlich im Plasmabulk und es können keine Oberflächeneffekte stattfinden. Für diesen Fall konnte keine Elektronenstoßanregung innerhalb der Struktur beobachtet werden und es kann sich kein Plasma in diesem Bereich ausbilden. Wird nun die angelegte Spannung auf 900 $V_{pp}$ erhöht, so erkennt man in Abbildung 4.14 b), dass nun das Plasma in die Struktur eindringen kann und sich auch über der gesamten Struktur ein Plasma ausbildet. Außerdem findet nun ein Übergang von dem Ω-Modus in den Penning-Modus außerhalb der Struktur statt, wie in dem raum- und phasenaufgelösten Plot der Elektronenstoßanregung in Abbildung 4.15 c) gezeigt ist. So erkennt man zusätzliche Anregungsmaxima innerhalb der Randschicht während des Zeitraumes von 10 ns - 30 ns und 48 ns - 68 ns. Im Penning-Modus beeinflussen nun auch wieder Oberflächeneffekte die Heizungsdynamik der Plasmaentladung wie zum Beispiel die Sekundärelektronenemission. Es werden nun ausreichend viele angeregte Teilchen erzeugt, sodass auch innerhalb der Struktur eine Elektronenstoßanregung beobachtet werden kann, wie in Abbildung 4.15 c) zu erkennen ist.

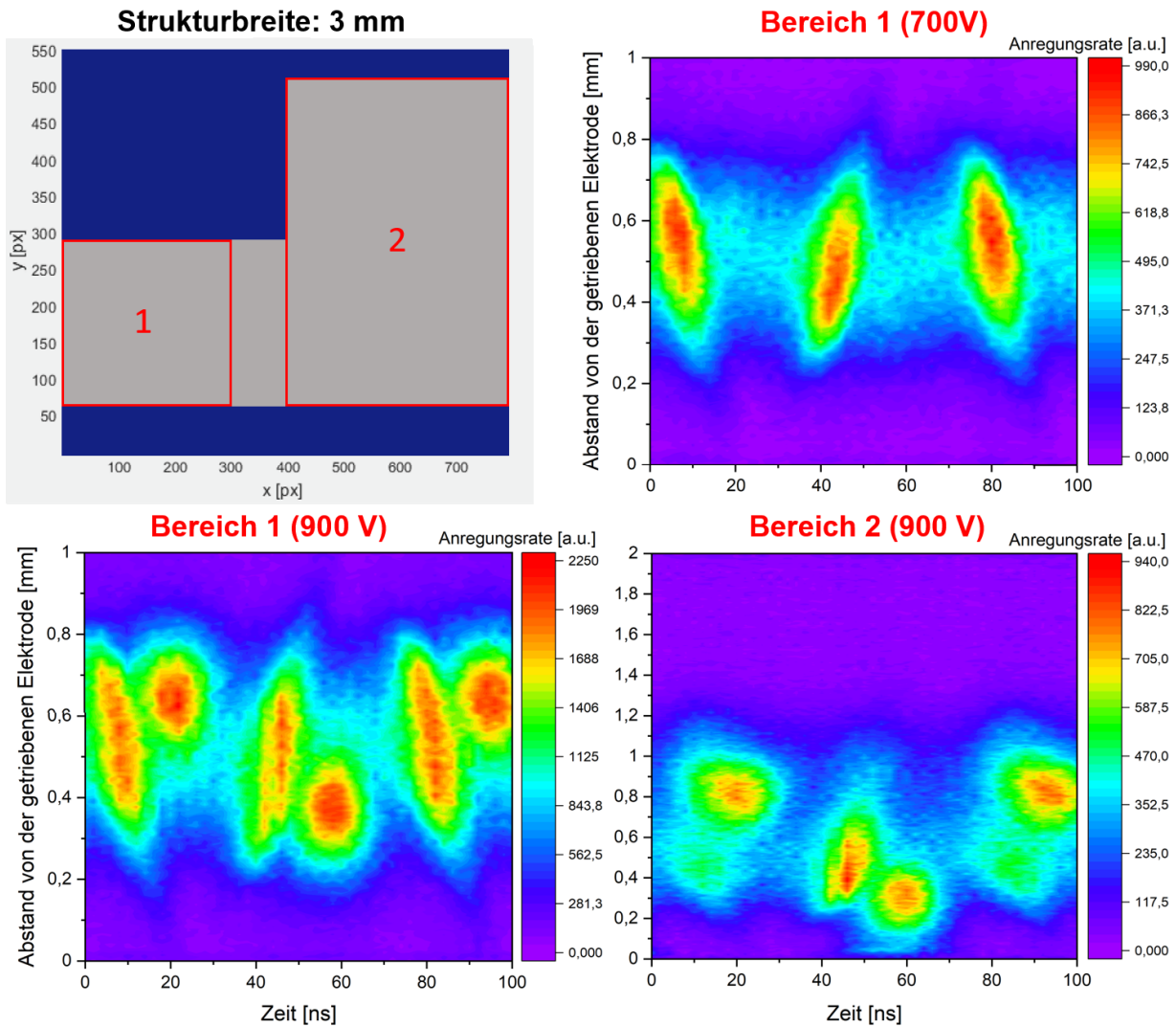

Abbildung 4.15: Raum- und phasenaufgelöste Plots der Elektronenstoßanregung für verschiedene Bereiche in der Plasmaentladung einer rechteckig strukturierten Elektrodenanordnung mit einer Strukturbreite von 3 mm in Abhängigkeit der Peak-to-Peak-Spannung. Der Gasfluss beträgt 1 slm He und 5 sccm $N_2$.

## 4.1.3 Einseitig rechteckig strukturierte Elektrodenanordnung aus Kupfer und Aluminium

In diesem Kapitel werden die PROES-Messungen des COST-Jets mit einer einseitig rechteckig strukturierten Elektrodenanordnung aus verschiedenen Elektrodenmaterialien durchgeführt, wie in Abbildung 4.16 a) - c) gezeigt. ist.

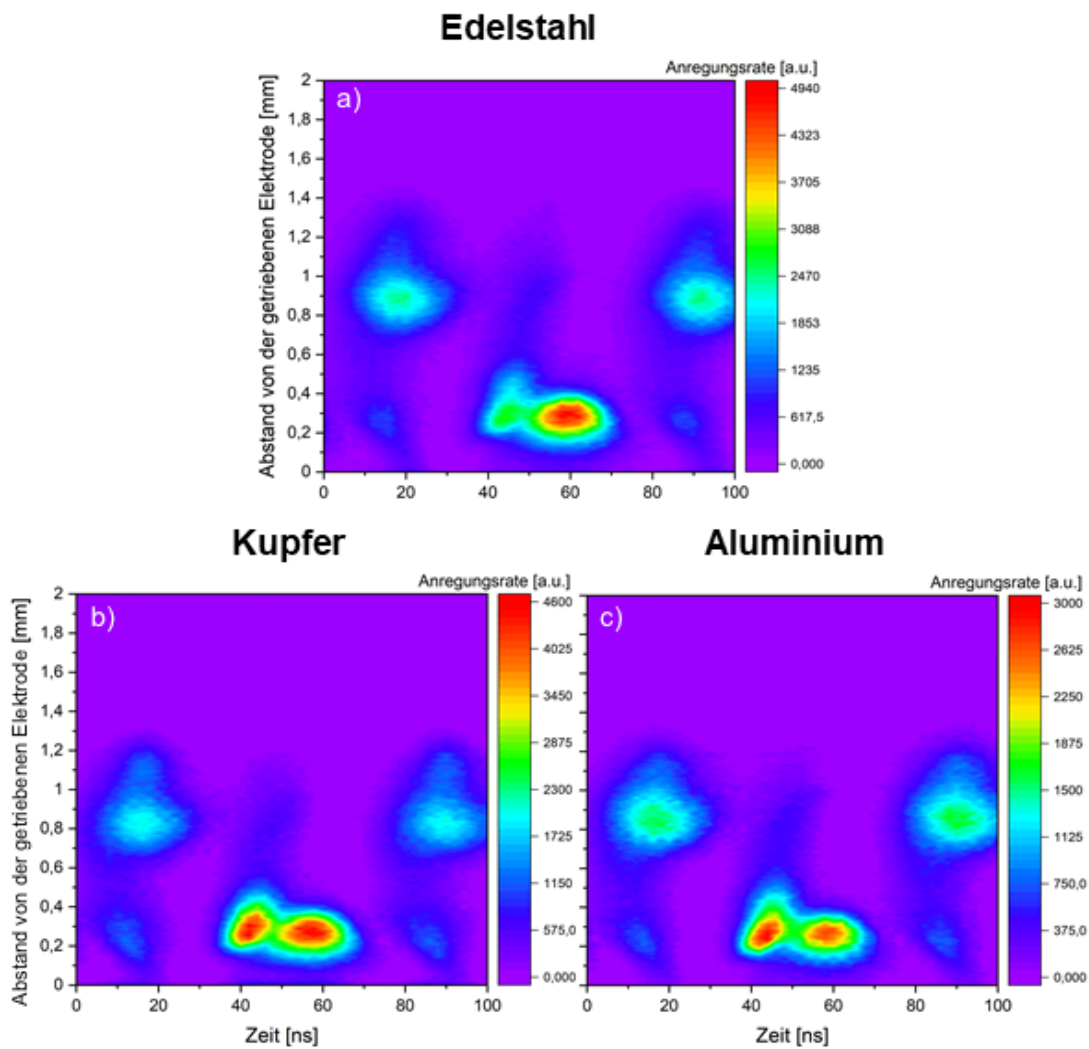

Abbildung 4.16: Raum- und phasenaufgelöste Plots der Elektronenstoßanregung einer rechteckig strukturierten Elektrodenanordnung mit einer Strukturbreite von 1 mm aus den Materialien a) Edelstahl, b) Kupfer und c) Aluminium für eine Peak-to-Peak-Spannung von 700 $V_{pp}$. Der Gasfluss beträgt 1 slm He und 0,5 sccm $O_2$.

Hiermit soll nun der Einfluss des Elektrodenmaterials auf die Plasmaentladung untersucht werden. Die Elektrodenanordnung ist so gewählt, dass die getriebene Elektrode eine planare Elektrode ist und die geerdete Elektrode rechteckige Strukturen mit einer Breite von 1 mm besitzt. Beide Elektroden bestehen aus dem selben Elektrodenmaterial. Der Gasfluss beträgt konstant 1 slm He und 0,5 sccm $O_2$ und die angelegte Peak-to-Peak-Spannung ist zu 700 $V_{pp}$ gewählt. In Abbildung 4.16 a) - c) sind die raum- und phasenaufgelösten Plots der Elektronenstoßanregung innerhalb einer Struktur in Abhängigkeit des Elektrodenmaterials gezeigt. Der hier ausgewählte Beobachtungsbereich entspricht dem Bereich 2 in der Abbildung 4.12 a) in Kapitel 4.1.2. Für alle Elektrodenmaterialien erkennt man grundsätzlich ein ähnliches Anregungsmuster. Für die gewählten Kontrollparameter zeigt sich ein dominierendes Anregungsmaximum in der Randschicht vor der getriebenen Elektrode in dem Zeitraum von 50 ns - 70 ns, welches durch den Heizungsmechanismus im Penning-Modus entsteht. Für das Elektrodenmaterial Aluminium ist dieses Anregungsmaximum etwas geringer ausgeprägt. In dem Zeitraum von 38 ns - 50 ns erkennt man ein zusätzliches Anregungsmaximum, welches auf dem Heizungsmechanismus im $\Omega$-Modus basiert. Für die Elektrodenmaterialien Kupfer und Aluminium ist dieses Anregungsmaximum am stärksten ausgeprägt. Die genannten Unterschiede in dem Anregungsmuster für die verschiedenen Elektrodenmaterialien sind so gering, sodass hier keine genaue Aussage über den Einfluss der Elektrodenmaterialien auf die Heizungsdynamik innerhalb einer Struktur getroffen werden kann. Für zukünftige Untersuchungen empfiehlt sich die Verwendung von Elektrodenmaterialien, die zum Beispiel unterschiedlichere Oberflächenparameter besitzen. Außerdem können auch dielektrische Elektrodenmaterialien getestet werden, bei denen weniger Ladungsträger an den Oberflächen während einer kollabierten Randschicht abfließen können und somit länger in der Struktur verweilen können.

### 4.1.4 Beidseitig rechteckig strukturierte Elektrodenanordnung aus Edelstahl

In diesem Kapitel werden die Ergebnisse der PROES-Messungen mit einer Elektrodenanordnung analysiert, die aus zwei rechteckig strukturierten Edelstahlelektroden mit einer Strukturbreite von 0,5 mm besteht. Die Ergebnisse in Kapitel 4.1.2 haben für die gewählte Gasmischung von 1 slm He und 0,5 sccm $O_2$ und einer maximal möglichen Peak-to-Peak-Spannung im Penning-Modus gezeigt, dass die effektivste Anregung der Teilchen innerhalb der Struktur bei einer Strukturbreite von 0,5 mm beobachtet werden kann. Hier werden jedoch die Ergebnisse für einen Gasfluss 1 slm He und 5 sccm $O_2$ vorgestellt, da diese Gasmischung auch später für die TALIF-Messungen verwendet wird. Der Gasfluss wird für die TALIF-Messungen verwendet, da frühere Arbeiten zeigen, dass bei einer Beimischung von 5 sccm $O_2$ die höchste atomare Sauerstoffdichte im COST-Jet erzeugt werden kann [41, 48]. Die Abbildung 4.17 a) - b) zeigt die Fotos der Plasmaentladungen für einen Gasfluss von 1 slm He und 5 sccm $O_2$ in Abhängigkeit der angelegten Peak-to-Peak-Spannung.

So ist in Abbildung 4.17 a) die Plasmaentladung für eine Spannung von 700 $V_{pp}$ gezeigt. Hier erkennt man in dem Bereich, bei dem der Elektrodenabstand 1 mm beträgt, dass die Leuchtemission im Plasmabulk homogen ist. Liegen zwei Strukturen gegenüber, so erhöht sich in diesem Bereich die Leuchtemission und diese dringt auch in die Strukturen ein. In Abbildung 4.17 b) wird die maximal erreichbare Peak-to-Peak-Spannung im Penning-Modus angelegt, die hier 1040 $V_{pp}$ beträgt. Für eine Peak-to-Peak-Spannung von 1040 $V_{pp}$ erkennt man in dem planparallelen Bereich eine inhomogene Leuchtemission, die typischerweise im Penning-Modus auftritt. Liegen zwei Strukturen gegenüber, so ist in diesem Bereich die Leuchtemission noch heller im Vergleich zu der Leuchtemission bei 700 $V_{pp}$. Außerdem erkennt man, dass nun die Leuchtemission noch weiter in die Strukturen eindringen kann.

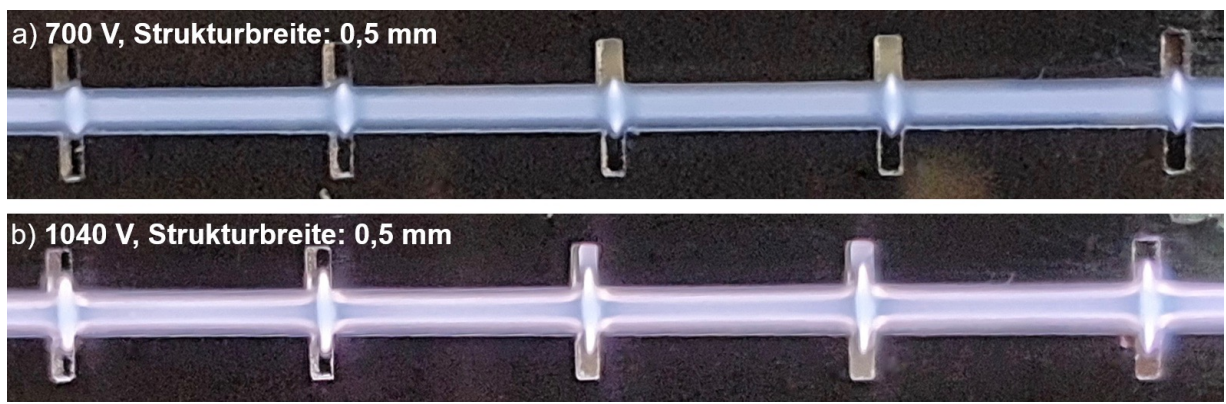

Abbildung 4.17: Fotos der Plasmaentladungen einer beidseitig rechteckig strukturierten Elektrodenanordnung mit einer Strukturbreite von 0,5 mm bei einem Gasfluss von 1 slm He und 5 sccm $O_2$ und einer Peak-to-Peak-Spannung von a) 700 $V_{pp}$ und b) 1040 $V_{pp}$.

Die Abbildung 4.18 a) - f) zeigt die raumaufgelöste Emission der 706,5 nm He I-Linie für ausgewählte Zeitpunkte innerhalb einer RF-Periode in der Struktur für die verschiedenen Peak-to-Peak-Spannungen. In Abbildung 4.18 a) - c) ist die Emission für eine angelegte Spannung von 700 $V_{pp}$ für ausgewählte Zeitpunkte gezeigt. So ist für den Zeitpunkt t = 3 ns die Randschicht vor der geerdeten Elektrode expandiert und es bildet sich vor dem Struktureingang ein Emissionsmaximum aus, welches durch das lokal erhöhte elektrische Feld in diesem Bereich entsteht. Den selben Effekt erkennt man vor der getriebenen Elektrode für den Zeitpunkt t = 39 ns in Abbildung 4.18 c). In Abbildung 4.18 b) erkennt man den Zeitpunkt t = 27 ns, bei dem die Randschicht vor beiden Elektroden kollabiert ist.

In Abbildung 4.18 d) - f) ist die Emission für eine angelegte Spannung von 1040 $V_{pp}$ zu erkennen. Für die ausgewählten Zeitpunkte, dringt das Emissionsmaximum in die jeweilige Strukturöffnung deutlich tiefer ein im Vergleich zu einer angelegten Spannung von 700 $V_{pp}$. Außerdem kann mit dieser Elektrodenanordnung eine deutlich größere maximale Peak-to-Peak-Spannung angelegt werden und die Emission kann tiefer in die Struktur eindringen im Vergleich zu einer einseitig rechteckig strukturierten Elektrodenanordnung mit der gleichen Strukturbreite.

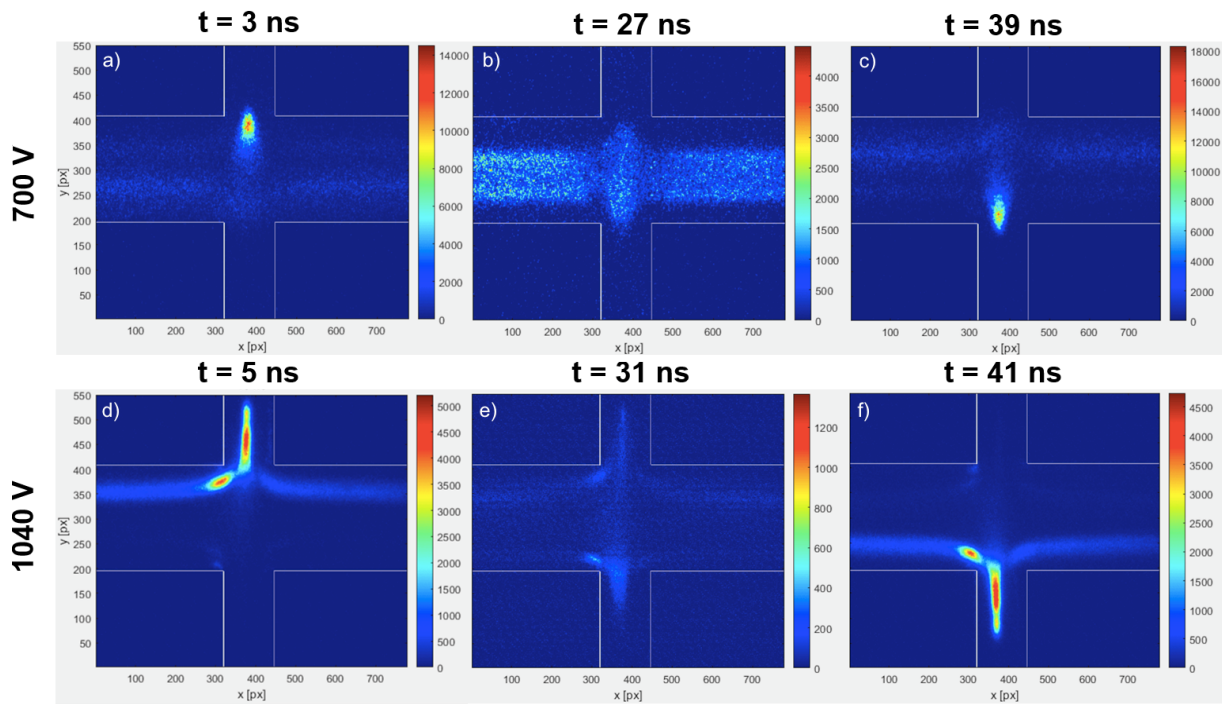

Abbildung 4.18: Raumaufgelöste Plots der Emission der 706,5 nm He I-Linie einer beidseitig rechteckigen Struktur mit der Breite 0,5 mm für ausgewählte Zeitpunkte innerhalb einer RF-Periode (74 ns). Der Gasfluss beträgt 1 slm He und 5 sccm $O_2$ und die Peak-to-Peak-Spannung beträgt a) 1040 $V_{pp}$ und b) 700 $V_{pp}$.

Die Abbildung 4.19 zeigt den Aufnahmebereich der PROES-Auswertung und die raum- und phasenaufgelösten Plots der Elektronenstoßanregung in Abhängigkeit der angelegten Peak-to-Peak-Spannung. Hier erkennt man für eine angelegte Peak-to-Peak-Spannungs-amplitude von 700 $V_{pp}$, dass sich die Anregungsmaxima im Plasmabulk befinden und somit die Plasmaentladung im Ω-Modus betrieben wird. Wird die angelegte Spannung auf 1040 $V_{pp}$ erhöht, so wird der COST-Jet im reinen Penning-Modus betrieben, wie an den beiden Anregungsmaxima in den Randschichtbereichen vor den Elektroden zu erkennen ist.

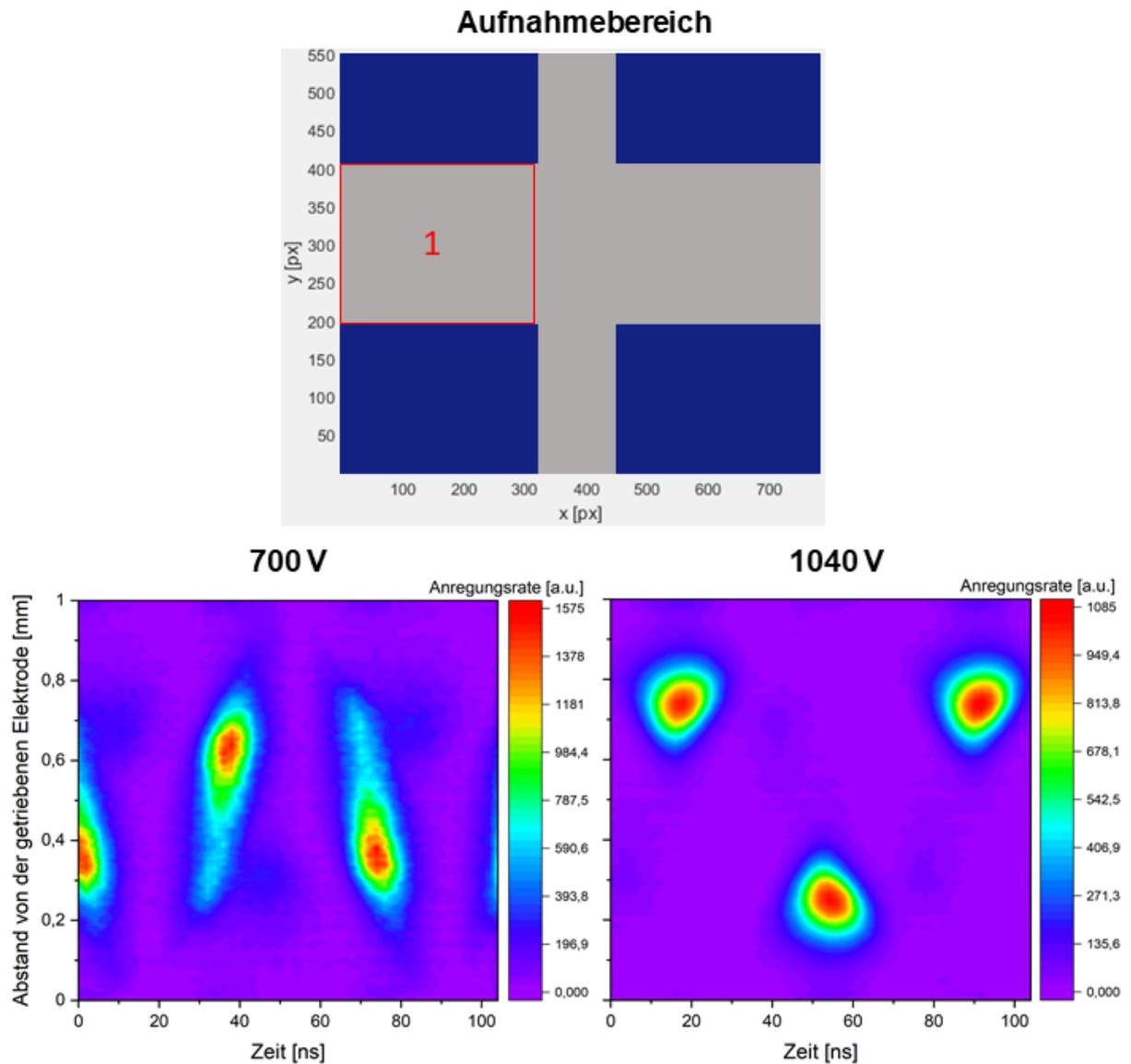

Abbildung 4.19: Raum- und phasenaufgelöste Plots der Elektronenstoßanregung einer beidseitig rechteckig strukturierten Elektrodenanordnung mit einer Strukturbreite von 0,5 mm in Abhängigkeit der Peak-to-Peak-Spannung. Der Gasfluss beträgt von 1 slm He und 5 sccm $O_2$.

## 4.1.5 Einseitig runde und dreieckig strukturierte Elektrodenanordnung aus Edelstahl

Abschließend werden eine runde und eine dreieckig strukturierte Elektrodenanordnung aus Edelstahl getestet. Die Elektrodenanordnung ist so gewählt, dass die getriebene Elektrode eine planare Elektrode ist und die geerdete Elektrode runde oder dreieckige Strukturen mit einer Breite von 1 mm besitzt. Der Gasfluss beträgt konstant 1 slm He und 0,5 sccm $O_2$ und es ist die maximal angelegte Peak-to-Peak-Spannung im Penning-Mode (hier 800 $V_{pp}$) gewählt.

Die Abbildung 4.20 a) - b) zeigt die Fotos der Plasmaentladungen für die verschiedenen Strukturformen. Die Fotos zeigen eine inhomogene Leuchtemission in dem planparallelen Bereich, die typischerweise im Penning-Modus auftritt. Innerhalb einer dreieckigen Struktur erkennt man eine helle punktförmige Leuchtemission. In einer runden Struktur erkennt man eine halbkreisförmige Leuchtemission.

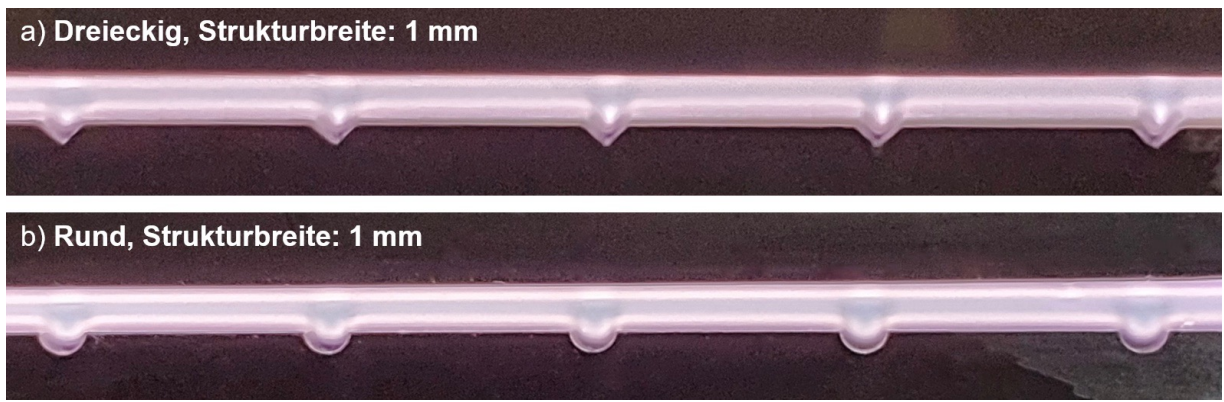

Abbildung 4.20: Fotos der Plasmaentladungen für eine dreieckig beziehungsweise rund strukturierten Elektrodenanordnung mit einer Strukturbreite von 1 mm bei einem Gasfluss von 1 slm He und 0,5 sccm $O_2$ und einer Peak-to-Peak-Spannung von 800 $V_{pp}$.

Die Abbildung 4.21 a) - f) zeigt die raumaufgelöste Emission der 706,5 nm He I-Linie für ausgewählte Zeitpunkte innerhalb einer RF-Periode für die verschiedenen Strukturtypen. In Abbildung 4.21 a) ist die Emission für den Zeitpunkt t = 5 ns innerhalb einer runden Struktur gezeigt. Zu diesem Zeitpunkt ist die Randschicht vor der geerdeten Elektrode expandiert und die Emission befindet sich in der runden Struktur. Abbildung 4.21 d) zeigt die Emission für den Zeitpunkt t = 5 ns innerhalb einer dreieckigen Struktur. Vergleicht man die Emission für diesen Zeitpunkt für die verschiedenen Strukturformen, so erkennt man, dass die Emission in der dreieckigen Struktur stärker in einem Punkt fokussiert wird. Dieser Effekt kann dadurch erklärt werden, dass an der Spitze der dreieckigen Struktur die benachbarten Randschichten und die dort auftretenden lokalen elektrischen Felder stärker überlagert werden. Die Abbildung 4.21 b) und e) zeigt die Emission für den Zeitpunkt t = 13 ns für die verschiedenen Strukturformen. Für diesen Zeitpunkt beginnt die Randschicht vor der geerdeten Elektrode zu kollabieren und die getriebene Elektrode zu expandieren. Die Abbildung 4.21 c) und f) zeigt die Emission für den Zeitpunkt t = 59 ns, für den die Randschicht vor der geerdeten Elektrode vollständig kollabiert ist. Nun ist die Randschicht vor der getriebenen Elektrode expandiert und man erkennt vor dieser Elektrode ein Emissionsmaximum.

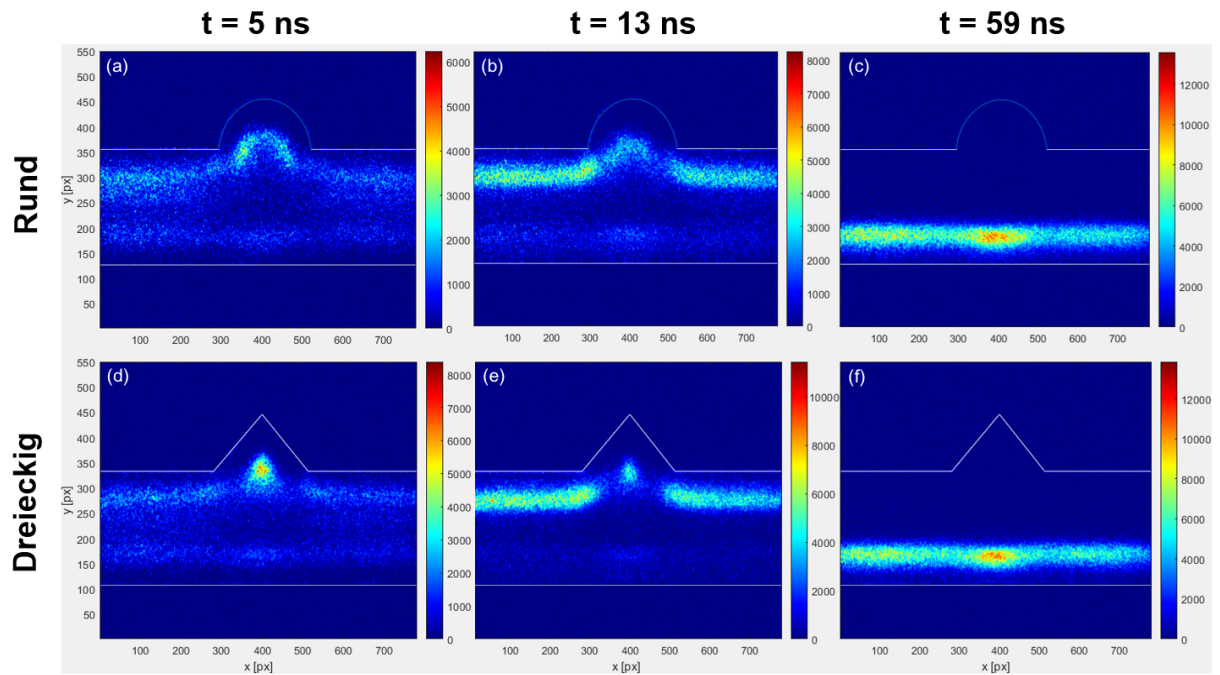

Abbildung 4.21: Raumaufgelöste Plots der Emission der 706,5 nm He I-Linie einer a) dreieckigen Struktur und b) runden Struktur mit der Breite 1 mm für ausgewählte Zeitpunkte innerhalb einer RF-Periode (74 ns). Der Gasfluss beträgt 1 slm He und 0.5 sccm $O_2$ und die Peak-to-Peak-Spannung beträgt 800 $V_{pp}$.

## 4.2 Auswertung der TALIF-Spektroskopie

In diesem Kapitel wird die atomare Sauerstoffdichte mit Hilfe der Zweiphotonen laserinduzierten Fluoreszenzspektroskopie für die Elektrodenanordnung untersucht, die aus zwei rechteckig strukturierten Edelstahlelektroden mit einer Strukturbreite von 0,5 mm besteht und zuvor in Kapitel 4.1.4 vorgestellt wurde. Als Erstes wird die Edelgaskalibrierung durchgeführt, um später aus den gemessenen TALIF-Signalen die absoluten atomaren Sauerstoffdichten zu berechnen.

Die Kalibrierung mit dem Edelgas Xenon ermöglicht die Berechnung der absoluten atomaren Sauerstoffdichten aus den gemessenen TALIF-Signalen. Hierfür wird zunächst die Vakuumkammer abgepumpt, da diese während der Sauerstoffmessungen geöffnet ist und so Verunreinigungen hinein gelangen können. Der COST-Jet befindet sich dabei in der Vakuumkammer. In die gesamte abgepumpte und abgedichtete Vakuumkammer wird dann über ein Nadelventil Xenon in einem Druckbereich von 1 Pa bis 10 Pa gefüllt. Das TALIF-Signal wird dann innerhalb des Entladungskanals des COST-Jets gemessen. Der Kalibirerungsfaktor $\chi$ (Gleichung 2.20) ist unter anderem von dem gemessenen TALIF-Signal und der dazugehörigen Teilchenzahl abhängig. Um den statistischen Fehler zu reduzieren, wird die Messung mit verschiedenen Drücken durchgeführt. Das relative TALIF-Signal $S_{Xe}$ erhält man aus der Steigung der Regressionsgerade der Messpunkte für einen bestimmten Druck. Die verschiedenen Werte von $S_{Xe}$ werden anschließend gegen die Xenon-Teilchendichte aufgetragen, die sich über das ideale Gasgesetz berechnen lässt. Die Werte der restlichen Parameter von dem Kalibrierungsfaktor $\chi$ werden nach [41] ermittelt.

Die Abbildung 4.22 zeigt das bei der TALIF-Methode verwendete Koordinatensystem und die untersuchte Struktur. Das Ende des Entladungskanals liegt bei z = 0 mm. Die Position einer Struktur kann aufgrund der begrenzten Auflösung der TALIF-Methode nicht exakt ermittelt werden. Die ausgewählte Struktur liegt zwischen z = -8 mm und z = -6 mm.

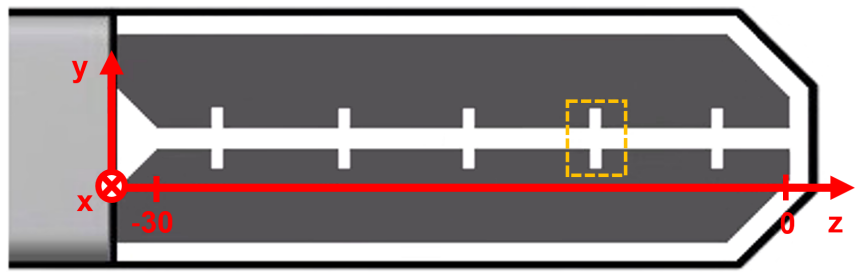

Abbildung 4.22: Veranschaulichung des verwendeten Koordinatensystems der TALIF-Messungen und schematische Darstellung der beidseitig rechteckig strukturierten Elektrodenanordnung mit einer Strukturbreite von 0,5 mm. Der gelbe Bereich zeigt die untersuchte Struktur.

Die Struktur wurde ausgewählt, da frühere Arbeiten zeigen, dass sich im Entladungskanal des standardisierten COST-Jets eine atomare Sauerstoffdichte entlang des Gasflusses exponentiell aufbaut und somit am Ende des Entladungskanals die größte atomare Sauerstoffdichte beob-achtet werden kann [47]. Für die folgenden Messungen am COST-Jet beträgt der Gasfluss 1 slm He und 5 sccm $O_2$. Der Gasfluss wird verwendet, da frühere Arbeiten ebenfalls zeigen, dass bei einer Beimischung von 5 sccm $O_2$ die höchste atomare Sauerstoffdichte im COST-Jet erzeugt wird [41, 48]. Es wird eine Peak-to-Peak-Spannung von 700 $V_{pp}$ und 1040 $V_{pp}$ gewählt. Wie bereits in Kapitel 4.1.4 gezeigt, befindet sich die Plasmaentladung für 700 $V_{pp}$ im $\Omega$-Modus und für 1040 $V_{pp}$ im Penning-Modus. Da der Laserstrahl einen Einfalls- und Ausfallswinkel von 45° an dem Entladungskanal des COST-Jets aufweist, kann das Signal nicht in einer Struktur gemessen werden. Somit kann mit der TALIF-Methode nur die absolute Dichte des atomaren Sauerstoffs im Bereich zwischen den gegenüberliegenden Strukturgräben gemessen werden, die eine Breite von 1 mm aufweisen.

Die Abbildung 4.23 a) und b) zeigt die zweidimensional raumaufgelöste absolute Dichte des atomaren Sauerstoffs zwischen den Elektroden (y-Richtung) und entlang der z-Achse von z = -8 mm bis z = -6 mm (zwischen den gegenüberliegenden Strukturgräben) für die verschiedenen Peak-to-Peak-Spannungen. In Abbildung 4.23 a) erkennt man für eine Peak-to-Peak-Spannung von 700 $V_{pp}$ ein symmetrisches Dichteprofil entlang der y-Achse. Erhöht man die Peak-to-Peak-Spannung auf 1040 $V_{pp}$, so erkennt man nun zwei Maxima der atomaren Sauerstoffdichte in der Nähe der Strukturöffnungen, wie in Abbildung 4.23 b) zu erkennen ist. Außerdem ist der maximale Wert der atomaren Sauerstoffdichte höher als in der Abbildung 4.23 a).

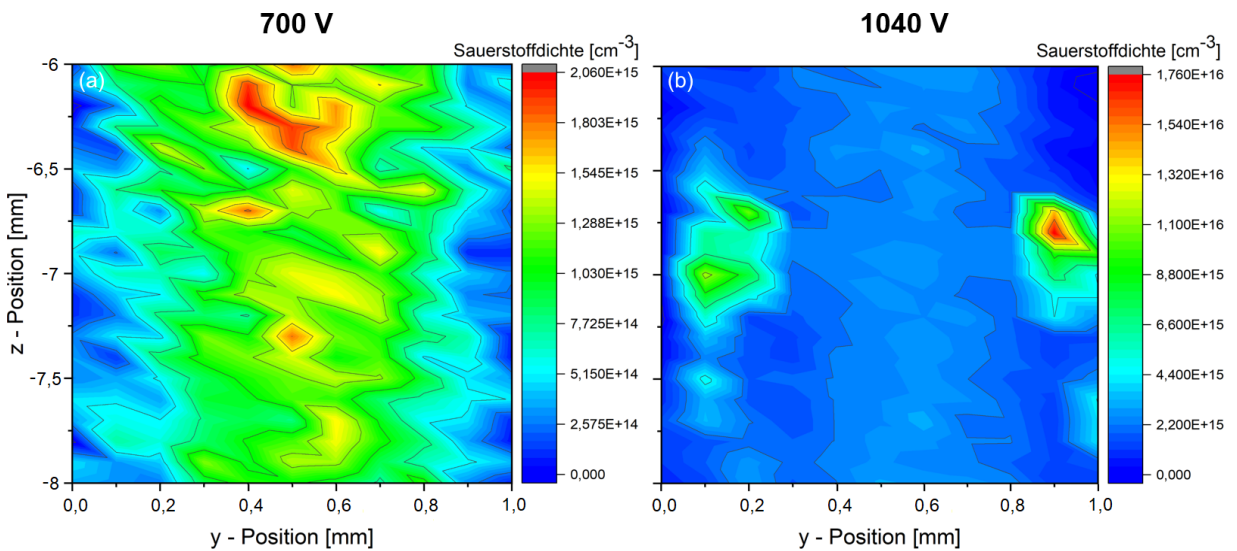

Abbildung 4.23: Zweidimensional raumaufgelöste Karte der absoluten Dichte des atomaren Sauerstoffs zwischen den Elektroden (y-Richtung) und entlang der z-Achse von z = -8 mm bis z = -6 mm (zwischen den gegenüberliegenden Strukturgräben) für die angelegte Peak-to-Peak-Spannung von a) 700 $V_{pp}$ und b) 1040 $V_{pp}$. Der Gasfluss beträgt 1 slm He und 5 sccm $O_2$.

Insgesamt zeigt sich somit, dass im Penning-Modus die atomare Sauerstoffdichte durch eine Struktur lokal beeinflusst wird. Außerdem zeigt sich in der Nähe der Strukturöffnungen eine höhere atomare Sauerstoffdichte im Vergleich zu den Messungen, die außerhalb der Struktur durchgeführt wurden. Wird der COST-Jet im Ω-Modus betrieben, so wird die atomare Sauerstoffdichte nicht erkennbar beeinflusst.

Die Abbildung 4.24 a) - d) zeigt die zweidimensional raumaufgelöste absolute Dichte des atomaren Sauerstoffs zwischen den Elektroden (y-Richtung) und den Quarzwänden (x-Richtung) für verschiedene Positionen in z-Richtung. Die Abbildung 4.24 a) und c) zeigt die zweidimensional raumaufgelöste absolute Dichte des atomaren Sauerstoffs außerhalb der ausgewählten Struktur für die verschiedenen Peak-to-Peak-Spannungen. Die Position entlang der z-Achse wird zu z = -10 mm gewählt. Die untersuchte absolute atomare Sauerstoffdichte zwischen der ausgewählten Struktur für die verschiedenen Peak-to-Peak-Spannungen ist in Abbildung 4.24 b) und d) gezeigt. Die Position wird hierfür zu z = -6,8 mm gewählt.

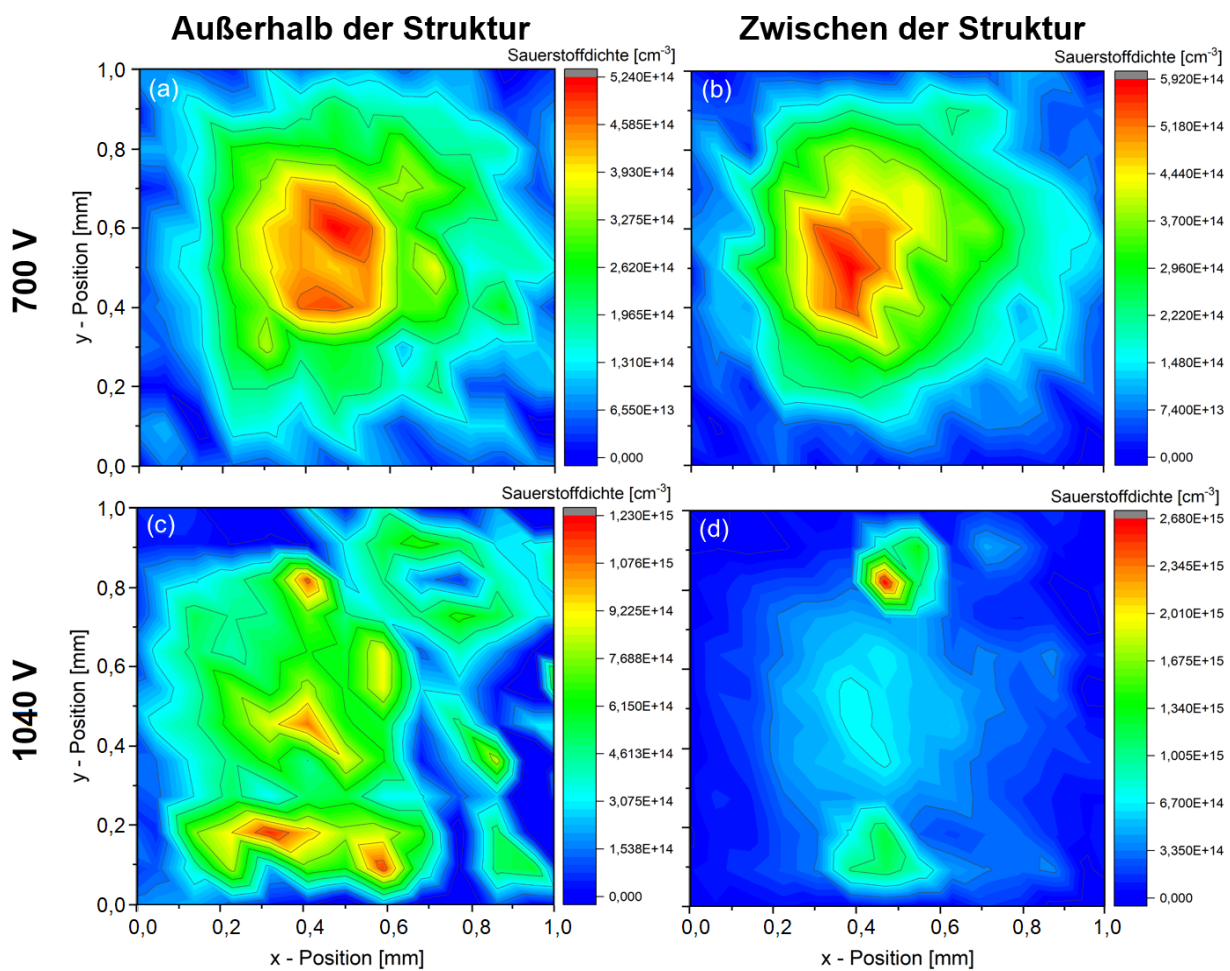

Abbildung 4.24: Zweidimensional raumaufgelöste Karte der absoluten Dichte des atomaren Sauerstoffs zwischen den Elektroden (y-Richtung) und den Quarzfenstern (x-Richtung) in Abhängigkeit der Position in z-Richtung und der angelegten Peak-to-Peak-Spannung. Der Gasfluss beträgt 1 slm He und 5 sccm $O_2$.

Bei einer Betrachtung der Karten in Abbildung 4.24 a) - b), bei denen eine Peak-to-Peak-Spannung von 700 $V_{pp}$ angelegt ist, wird deutlich, dass die absolute Dichte des atomaren Sauerstoffs sowohl für eine Position außerhalb der Struktur als auch zwischen den beiden Strukturgräben symmetrisch zwischen den Elektroden und den Quarzfenstern verteilt ist. Das jeweilige Maximum der absoluten Dichte des atomaren Sauerstoffs liegt im Zentrum der Karte und der Wert des Maximums ist für die unterschiedlichen Positionen sehr ähnlich. Dies liegt daran, dass sich die Plasmaentladung für die ausgewählten Kontrollparameter im $\Omega$-Modus befindet und die Elektronen durch das hohe elektrische Driftfeld im Plasmabulk beschleunigt werden. Anschließend können die Elektronen Dissoziationsstöße ausführen, die zu der Produktion von atomarem Sauerstoff führen. Da der $\Omega$-Modus unabhängig von den Oberflächenparametern ist, wird dieser nicht von der Struktur beeinflusst und somit ergibt sich hier für eine angelegte Peak-to-Peak-Spannung von 700 $V_{pp}$ kein Unterschied in dem Profil der absoluten Dichte des atomaren Sauerstoff für die verschiedenen Positionen.

Wird nun eine Peak-to-Peak-Spannung von 1040 $V_{pp}$ angelegt, so befindet sich der COST-Jet im Penning-Modus. Für diesen Betriebsmodus ist die Karte der absoluten Dichte des atomaren Sauerstoffs für die verschiedenen Positionen in Abbildung 4.24 c) - d) dargestellt. Vergleicht man nun die Karten in Abhängigkeit der Position, so erkennt man nun einen deutlichen Unterschied in dem Dichteprofil des atomaren Sauerstoffs. Für eine Position außerhalb der Struktur erkennt man ein Dichteprofil mit einem überlagerten Rauschen, welches aufgrund der Instabilität des Lasersystems entsteht. Hier ist jedoch auch eine Verteilung der absoluten Dichte des atomaren Sauerstoffes zu erwarten, die symmetrisch zwischen den Elektroden und den Quarzfenstern verteilt ist. Für eine Position zwischen den Strukturgräben erkennt man eine deutliche Asymmetrie der absoluten Dichte und das Maximum ist nun nicht mehr im Zentrum. Es befinden sich nun zwei Maxima an den äußeren Positionen der y-Achse, das heißt, sie befinden sich vor den Strukturöffnungen. Der Wert der Maxima zwischen den Elektroden ist nun auch größer als der Wert des Maximums außerhalb der Struktur. Somit beeinflussen die Strukturen die lokale atomare Sauerstoffproduktion im COST-Jet, jedoch nur, wenn dieser im Penning-Modus betrieben wird. Ein Grund für die Maxima der atomaren Sauerstoffdichte vor den Strukturöffnungen ist die erhöhte Elektronenstoßanregung, die durch die Überlagerung verschiedener lokaler elektrischer Felder entsteht und die vor den Strukturen beobachtet werden kann.

Das TALIF-Messverfaren bringt auch Nachteile bei der Untersuchung des COST-Jets mit einer strukturierten Elektrodenanordnung mit sich. Durch die TALIF-Messmethode können zum Beispiel keine Aussagen darüber gemacht werden, wie sich die atomare Sauerstoffdichte in einem Strukturgraben verhält, da der Laser nicht in die Struktur eindringen kann. Außerdem reicht die Auflösung dieses Messverfahrens nicht aus, um die exakte Position einer Struktur zu ermitteln. Somit sollte hier ein zusätzlicher Vergleich der experimentellen Ergebnisse mit PIC/MCC-Simulationen durchgeführt werden.

# 5 Zusammenfassung und Ausblick

Ziel dieser Arbeit war die Untersuchung von Effekten unterschiedlicher maßgeschneiderter Elektroden auf die Elektronenheizungsdynamik eines kapazitiv gekoppelten Radio Frequenz Mikroplasmajets (COST-Jet). Die maßgeschneiderten Elektroden können dabei aus verschiedenen Elektrodenmaterialien bestehen und verschiedene Strukturformen enthalten.

Mit Hilfe der phasenaufgelösten optischen Emissionsspektroskopie (PROES) wurde die Elektronenstoßanregungsrate der 706,5 nm He I-Linie für verschiedene maßgeschneiderte Elektrodenanordnungen, Gasmischungen und angelegten Spannungen untersucht.

Zunächst wurden die PROES-Messungen für planparallele Elektrodenanordnung aus verschiedenen Elektrodenmaterialien durchgeführt. Es wurden Elektroden aus Edelstahl, Kupfer und Aluminium verwendet. Wurde für die geerdete und die getriebene Elektrode das gleiche Material gewählt, so konnte man für die verschiedenen Elektrodenmaterialien nur einen geringen Unterschied in der Elektronenheizungsdynamik feststellen. Somit wurden zusätzlich Elektrodenanordnungen entworfen, die einen direkten Vergleich zwischen den Elektrodenmaterialien ermöglichten. So bestand die getriebene Elektrode aus Kupfer oder Aluminium und die geerdete Elektrode jeweils aus Edelstahl. Die raum- und phasenaufgelösten Plots der Elektronenstoßanregung zeigten für die verschiedenen Gasmischungen, dass die Heizungsdynamik des Penning-Modus von den verschiedenen Elektrodenmaterialien beeinflusst wird. So ergab sich eine deutliche Asymmetrie in der Intensität der Anregungsmaxima innerhalb der Randschichten. Wurde die Plasmaentladung im Ω-Modus betrieben, so konnte kein Einfluss des Elektrodenmaterials auf die Heizungsdynamik beobachten werden, da dieser nicht von Oberflächeneffekten abhängig ist. Es konnte gezeigt werden, dass die Elektrodenmaterialien den indirekten Kanal des Penning-Modus unterschiedlich beeinflussen, da dieser unter anderem von der Sekundärelektronenemission abhängt.

Anschließend wurden die PROES-Messungen für verschiedene rechteckig strukturierte Elektrodenanordnungen aus Edelstahl wiederholt. Hier zeigte sich, dass die Strukturbreite einen Einfluss auf die Heizungsdynamik hat. Für eine Strukturbreite von 0,5 mm konnte beispielsweise eine Fokussierung der Elektronenstoßanregung innerhalb der Struktur festgestellt werden, die durch die Überlagerung verschiedener lokaler elektrischer Felder entsteht. Die PROES-Messungen wurden auch mit einer rechteckig strukturierten Elektrodenanordnung aus Kupfer und Aluminium durchgeführt, jedoch ergaben sich hier nur geringe Unterschiede. Somit sollten für zukünftige Untersuchungen Elektrodenmaterialien verwendet werden, die zum Beispiel größere Unterschiede in den Oberflächenparametern aufweisen. Außerdem können auch zukünftig dielektrische Elektrodenmaterialien getestet werden, da diese möglicherweise das Abfließen von Ladungsträgern an den Wänden verhindern. Außerdem wurden PROES-Messungen für eine runde und dreieckig strukturierte Elektrodenanordnung aus Edelstahl durchgeführt. In einem dreieckigen Graben konnte das Emissionsmaximum deutlich besser fokussiert werden als in einer runden Struktur. Es ergab sich jedoch kein nennenswerter Unterschied in der Fokussierung eines Emissionsmaximums im Vergleich zu einer rechteckigen Struktur.

Abschließend wurden die absoluten Dichten von atomarem Sauerstoff mittels der Zweiphotonen laserinduzierten Fluoreszenzspektroskopie (TALIF) für eine ausgewählte rechteckig strukturierte Elektrodenanordnung untersucht. Insgesamt zeigte sich, dass nur im Penning-Modus die atomare Sauerstoffdichte durch die Strukturen lokal beeinflusst wird. Außerdem zeigt sich in der Nähe der Strukturöffnungen eine höhere atomare Sauerstoffdichte im Vergleich zu den Messungen, die außerhalb der Struktur durchgeführt wurden. Es ergaben sich auch Nachteile des TALIF-Verfahrens bei der Untersuchung des COST-Jets mit einer strukturierten Elektrodenanordnung. So können zum Beispiel keine Aussagen darüber gemacht werden, wie sich die atomare Sauerstoffdichte in einem Strukturgraben verhält, da der Laserstrahl nicht in eine Struktur eindringen kann. Außerdem reicht die Auflösung nicht aus, um die exakte Position einer Struktur zu ermitteln.

Als zukünftige Arbeit empfiehlt es sich, die experimentellen Ergebnisse mit PIC/MCC-Simulationen zu vergleichen. Abschließend ist in dem Teilprojekt A4 des Sonderforschungsbereiches 1316 der Ruhr-Universität vorgesehen, die maßgeschneiderten Elektrodenanordnungen und das Voltage Waveform Tailoring zu kombinieren.

# Literaturverzeichnis